\documentclass[fleqn,usenatbib]{mnras}
\usepackage{fix-cm}

\usepackage{newtxtext,newtxmath}
\usepackage[T1]{fontenc}

\DeclareRobustCommand{\VAN}[3]{#2}
\let\VANthebibliography\thebibliography
\def\thebibliography{\DeclareRobustCommand{\VAN}[3]{##3}\VANthebibliography}

\usepackage{graphicx}	% Including figure files
\usepackage{amsmath}	% Advanced maths commands

\usepackage{xcolor}

\definecolor{mycolor}{HTML}{007900}

\title[PSF effect on IHL measurements]{Correcting for the effects of the point spread function in intra-halo light measurements and application to deep Hyper Suprime-Cam data}

\author[Garate-Nuñez et al.]{Lucía P. Garate-Nuñez,$^{1,2}$\thanks{E-mail: lucia.garatenunez@research.uwa.edu.au}
Aaron S. G. Robotham,$^{1,2}$ 
Sabine Bellstedt$^{1}$ and Luke J. M. Davies$^{1}$
\\
$^{1}$ICRAR, The University of Western Australia, 35 Stirling Highway, Crawley, WA 6009, Australia\\
$^{2}$Australian Research Council Centre of Excellence for All-Sky Astrophysics in 3 Dimensions (ASTRO 3D), Stromlo, ACT 2611, Australia}

\date{Accepted 2026 November 5. Received 2026 November 5; in original form 2025 May 30}

\pubyear{2024}

\begin{document}
\label{firstpage}
\pagerange{\pageref{firstpage}--\pageref{lastpage}}
\maketitle

\begin{abstract}
The intra-halo light (IHL) is the diffuse stellar component that surrounds galaxies, groups, and clusters. Its formation is linked to the hierarchical assembly of the system, making it a key tracer of galaxy evolution. However, the low surface brightness of the IHL makes it challenging to detect and distinguish from the point spread function (PSF) effect of the telescope. In this paper, we present \textcolor{black}{an innovative technique that corrects for PSF-scattering effects in astronomical images. We use this method to provide a statistically robust estimation of the IHL in galaxy groups and clusters by fitting an exponential model to this component using an MCMC optimiser algorithm.} \textcolor{black}{To test our methodology, we build 5440 mock observations of Galaxy And Mass Assembly (GAMA) groups using data from the Hyper Suprime-Cam Subaru Strategic Program Public Data Release 3 (HSC-SSP PDR3). Each mock contains an IHL component with a flux fraction ($f_{\mathrm{IHL}}$) ranging from 0.01 to 0.5.} \textcolor{black}{Our results demonstrate the importance of properly removing the PSF-scattered light, especially at lower $f_{\mathrm{IHL}}$. The IHL effective radius is overestimated by $\sim9.1$ and $\sim2$ at $f_{\mathrm{IHL}}=0.01$ and $0.5$, respectively, while the IHL flux requires correction for $f_{\mathrm{IHL}} < 0.4$, with an overestimate of $\sim84.4$ at $f_{\mathrm{IHL}} = 0.01$. After applying our PSF correction, the recovered flux and effective radius are much closer to their true values.} Finally, we apply our methodology to \textcolor{black}{HSC-PDR3 UD} real observations of the GAMA group G400138 in the \textit{g,r} and \textit{i}-bands, finding median IHL fractions of: \textcolor{black}{$f_{\mathrm{g,IHL}}$} $\sim$ 0.19$^{+0.09}_{-0.01}$, \textcolor{black}{$f_{\mathrm{r,IHL}}$} $\sim$ 0.08$^{+0.06}_{-0.02}$, and \textcolor{black}{$f_{\mathrm{i,IHL}}$} $\sim$ 0.06$^{+0.04}_{-0.02}$.

\end{abstract}


%%%%%%%%%%%%%%%%%%%%%%%%%%%%%%%%%%%%%%%%%%%%%%%%%%

%%%%%%%%%%%%%%%%% BODY OF PAPER %%%%%%%%%%%%%%%%%%

\section{Introduction}
\label{sec:Intro}
%The goal of this paper is to estimate how much of the total light of a galaxy group is actually in the intra-halo light component (i.e. the fraction of IHL or \textcolor{black}{$f_{\mathrm{IHL}}$}). The first step we take is to remove the PSF-scattered flux from the image. \textcolor{red}{This extra flux can be confused with real IHL and lead to inaccurate estimations.} Secondly, we

%Estos papers comparan los dos metodos: \citep{2011ApJ...732...48R, 2024MNRAS.528..771B}. Decir surface brightness cut method y composite model porque despues en IHL technie los nombroy esta bueno ya haberlos introduced y citado

 The most massive galaxies form at the centre of groups and clusters during the hierarchical assembly process (e.g. \citealt{2007MNRAS.375....2D, 2009ApJ...697.1290B, 2009ApJ...699L.178N, 2010Natur.468..940V}). Mainly through the accretion of smaller objects and merging events, a diffuse stellar light component is also developed surrounding these systems, known as intra-halo light (IHL, \citealt{2002ApJ...575..779F, 2007ApJ...666...20P, 2020ApJS..247...43K}). As a result, this faint component is a fossil record of the assembly history of the group or cluster and can be used as a tool to analyse the dynamical state of the system (e.g. \citealt{1984ApJ...276...26M, 2021Galax...9...60C, 2022NatAs...6..308M}).

 Since it was first observed \citep{1937ApJ....86..217Z}, the IHL has been widely studied as an important stellar component of the overall optical luminosity of a cluster. However, its low surface brightness (LSB) makes it both challenging to detect and significantly inhibits our ability to analyse the properties of these stellar remnants and determine their origin. From a theoretical point of view, cosmological simulations provide insight into the mechanisms of IHL formation. The three most important mechanisms are mergers between satellites and the brightest cluster galaxy (BCG) \citep{2006ApJ...652L..89M,2007MNRAS.377....2M}, stellar stripping of satellite galaxies \citep{2006ApJ...648..936R,2011ApJ...732...48R, 2024ApJ...969..142C} and preprocessing of IHL from other haloes \citep{2005ApJ...631L..41M, 2014MNRAS.437.3787C}. In addition, thanks to the availability of phase-space information in simulations, several works have studied the relationships between the IHL and other global cluster properties, such as dynamical state \citep{2023ApJ...958...72C,2024ApJ...965..145Y,2025arXiv250320857K}, morphology \citep{2024MNRAS.527.2624P}, total halo mass \citep{2017Galax...5...49R,2018MNRAS.475..648P,2020MNRAS.494.4314C} and dark matter distribution \citep{2020MNRAS.494.1859A,2021MNRAS.500.4181D,2024A&A...683A..59C,2025MNRAS.539.2279B}.

 From an observational point of view, modern CCD imaging techniques have made it possible to conduct deep studies of Milky Way-like systems \citep{2008ApJ...689..184M, 2015ApJ...800L...3W,2016ApJ...830...62M}, and small samples of galaxy groups and clusters \citep{1991ApJ...369...46U, 2000ApJ...536..561G, 2004ApJ...617..879L, 2005MNRAS.364.1069D, 2005ApJ...631L..41M,2008MNRAS.388.1433D,2014MNRAS.443.1433D}. Only over the last ten years, we have seen a remarkable increase in the sensitivity of instruments and telescopes, as well as improvements in image-processing techniques, which allows us a better understanding of the IHL formation and evolution \citep{2016ApJ...823..123T, 2018MNRAS.474..917M,2020A&A...639A..14S,2022ApJ...940L..51M, 2023A&A...670L..20R,2024arXiv240513503K,2024A&A...689A.306S,2025MNRAS.538..622G}. Despite these great improvements, there is still no observational or theoretical consensus on the fraction of light that contributes to the IHL component (\textcolor{black}{$f_{\mathrm{IHL}}$}), nor on its dependence on redshift or cluster mass \citep{2017ApJ...846..139M, 2019ApJ...874..165Z, 2021Galax...9...60C,2024AJ....167....7C}. Surveys with a wide field of view (FoV) are crucial for addressing these challenges, as large samples of groups and clusters are needed to make more robust estimations. Examples of such surveys include the Dark Energy Survey (DES; \citealt{2005astro.ph.10346T, 2023MNRAS.521..478G,2025MNRAS.538..622G}), the Hyper Suprime-Cam Subaru Strategic Program (HSC-SSP; \citealt{2018PASJ...70S...8A}), European Space Agency’s \textit{Euclid} Wide Survey (EWS; \citealt{2011arXiv1110.3193L, 2022A&A...662A.112E, 2024arXiv240513503K}), as well as the upcoming Roman's High Latitude Wide Area Survey for Low Surface Brightness Astronomy (HLWAS; \citealt{2023arXiv230609414M}) and Vera Rubin Observatory Legacy Survey of Space and Time (LSST, \citealt{2019ApJ...873..111I, 2020arXiv200111067B}). These surveys will be crucial in determining the \textcolor{black}{$f_{\mathrm{IHL}}$} for representative samples of groups and clusters at a given epoch.

 %However, the halo dynamical state is also an ill-defined quantity in both observations and simulations, since each of the many parameters used refers to a peculiar condition of relaxed/disturbed systems, see (e.g. Rasia et al. 2013; Cui et al. 2017; Haggar et al. 2020; De Luca et al. 2021; Capalbo et al. 2021; Zhang et al. 2022, for research along this line)

%From a theoretical point of view, the current generation of cosmological hydrodynami cal simulations \citep{}

Even with high-quality imaging data, isolating the IHL component from the BCG and the satellite galaxies is also a challenging task, where multiple methods have been proposed in the literature (see \citealt{2024MNRAS.528..771B}
for an overview of the different methods). The most widely used technique is to apply a threshold in the surface brightness (SB) and consider all light fainter than this limit to be part of the IHL \citep{2004ApJ...609..617F, 2005MNRAS.358..949Z, 2006ApJ...648..936R, 2021MNRAS.502.2419F}. The advantage of this method is its easy implementation. However, it is highly dependent on the data, the selected photometric band, and the choice of the SB cut. In addition, \textcolor{black}{by definition}, it does not account for the IHL flux that has a higher SB than the threshold. Another typical method is to use functional forms to fit both the BCG and the IHL, such as double Sérsic \citep{1963BAAA....6...41S}, exponential, or composite profiles \citep{2005ApJ...618..195G, 2015MNRAS.451.2703C, 2021ApJS..252...27K, 2021ApJ...910...45M, 2023MNRAS.518.3685A}. This method successfully captures the IHL flux overlapping with the BCG, but does not account for the overlapped flux with satellite galaxies. Additionally, depending on the choice of functional forms, this method might become degenerate. The approach that successfully captures the IHL flux overlapped with BCG + satellites consists of modelling and subtracting all the galaxies present in the image, but this method can be computationally expensive \citep{2014ApJ...781...24G, 2017ApJ...846..139M, 2023MNRAS.518.1195M}.

 Another requirement for accurately estimating the IHL properties in groups and clusters is a well-behaved and extended Point Spread Function (PSF; \citealt{1999prop.book.....B}) model. Numerous studies have shown that the PSF-scattered flux has a small but non-negligible effect at large radii \textcolor{black}{from the sources centres} \citep{2008MNRAS.388.1521D,2014PASP..126...55A,2023ApJ...953....7L}. Consequently, if the scattered light from stars and extended objects is not properly subtracted from the image, the estimations of low surface brightness features can be biased (e.g., \citealt{2024MNRAS.531.2517G}). %The impact of the PSF on astronomical images is further discussed in Sec.\ref{sec:PSF analysis}.

\textcolor{black}{Throughout this work, we define the IHL as an extended component that is structurally distinct from the BCG and follows an exponential profile, extending into the central regions of the system and spatially overlapping with the BCG. Based on the results of \citealt{2024MNRAS.527.2624P}, there is a spatial overlap between the different stellar components of galaxies. Defining the IHL as a non-overlapping component would therefore exclude a fraction of the total IHL mass and potentially bias estimates of IHL properties.
}

We present two \textcolor{black}{complementary} techniques specially designed to analyse LSB flux. The first one aims to correct for the PSF-scattering effect in astronomical images. Following advanced source extraction routines, our technique first rescales the flux of each detected source to account for the lost flux when being observed, and then removes the scattered flux that is contaminating the background of the image. The second technique consists of fitting an exponential model to the IHL component of groups and clusters using a Markov Chain Monte Carlo (MCMC) algorithm. By using our IHL estimation technique, we model the IHL light projected over the BCG and the satellite galaxies, which is not possible with a SB threshold approach. Moreover, modelling a single component is computationally more efficient than modelling all the components present in the image. The aim is to apply the IHL detection technique to images that have previously been PSF-corrected following our PSF-scattered flux removal technique.

By applying our methodology to Hyper Suprime-Cam Subaru Strategic Program Public Data Release 3 (HSC-SSP PDR3; \citealt{2022PASJ...74..247A}) mock observations of Galaxy And Mass Assembly (GAMA, \citealt{2011MNRAS.413..971D}) groups, we analyse the impact of PSF-scattered flux on measurements of the IHL component in galaxy groups and clusters. We demonstrate the importance of accounting for this effect, particularly when the IHL component is less prominent. The structure of this work is as follows. In Section~\ref{sec:PSF analysis}, we present our technique to correct for the PSF-scattered flux in an image, and in Section~\ref{sec:IHL analysis}, we present our technique to estimate the IHL component in galaxy groups and clusters. Section~\ref{sec:Data} is dedicated to explaining how we construct the dataset of mock observations that we use to test both techniques. The results of these tests are discussed in Section~\ref{sec:results}. We additionally show the results of applying our techniques to estimate the \textcolor{black}{$f_{\mathrm{IHL}}$} on a GAMA group using real HSC-SSP PDR3 UD data in Section~\ref{sec:application}. We finally summarise the results of this work in Section~\ref{sec:Conclusions}.

%\textcolor{mycolor}{In Section \ref{sec:non-circular} we demonstrate the robustness of our technique in modelling different IHL configurations}

We adopt throughout a standard $\mathrm{\Lambda}$CDM cosmology model with \textcolor{black}{$H_{\mathrm{0}}$} =  70 $ \, \mathrm{km}\,\mathrm{s}^{-1} \,\mathrm{Mpc}^{-1}$, \textcolor{black}{$\Omega_{\mathrm{\Lambda_0}}$} = 0.7 and $\mathrm{\Omega_{\mathrm{m_0}}}$ = 0.3. All magnitudes are in the AB magnitude system and the HSC-SSP PDR3 zero point magnitude for all the five filters is \textcolor{black}{\textit{mag}$_\mathrm{zero}$} = 27 mag.

\section{PSF-scattered flux removal technique}
\label{sec:PSF analysis}

The PSF characterises how the flux of a source is scattered in astronomical images as a consequence of the atmosphere and all the optical elements of a telescope (instruments, detectors, mirrors, lenses, detectors, etc.). The central region of the PSF is produced by atmospheric turbulence \citep{1941DoSSR..30..301K} and can be described by a Moffat profile \citep{1969A&A.....3..455M, 1996PASP..108..699R}. However, the outer PSF remains less well understood due to its faintness and the challenges associated with accurately measuring the extended wings. As mentioned in Sec.~\ref{sec:Intro}, accurate estimation of \textcolor{black}{any low surface brightness feature} requires an extended PSF model to remove all PSF-scattered flux that could otherwise be mistaken for genuine stellar light \citep{2014A&A...567A..97S, 2015A&A...577A.106S, 2016ApJ...823..123T,2019A&A...629A..12M, 2023MNRAS.518.1195M, 2024MNRAS.531.2517G}.

%This part of the radial profile is mostly fitted using a power-law profile with a power index ranging from 1.6 to 3 \citep{2005ApJ...618..195G, 2007ApJ...666..663B,2009PASP..121.1267S}.

%However, the outer part of the PSF is less well understood and caused by scattering by atmospheric aerosols and dust, micro-ripples and dust on optical surfaces and effects of diffusion and reflection within the instrument. include scattering by atmospheric aerosols and dust, as well as micro-ripples and dust on optical surfaces (for example, van de Hulst 1948; Deirmendjian 1957, 1959), and effects of diffusion and reflection within the instrument (Hasan & Burrows 1995; Racine 1996; B07; Slater et al. 2009, hereafter SHM09). 

%As mentioned in Sec.\ref{sec:Intro}, in order to make accurate estimations of any LSB feature, having an extended PSF model is crucial to remove all of the PSF-scattered flux that can be confused with real stellar light 

\textcolor{black}{In this section, we build on source extraction routines to \textcolor{black}{estimate and remove} the PSF-scattered flux from any astronomical image, regardless of the presence of LSB features. During the process, we make use of a source identification code to find the sources in the image and an extended PSF model to maximise the removal of scattered flux at large distances from the centre of the sources.}

\textcolor{black}{This technique involves modifying the flux within and outside the identified sources to compensate for the PSF effect. The process is divided into two parts, corresponding to adjustments inside and outside the sources. We first modify the flux inside each detected source to compensate for the scattered flux lost due to the PSF convolution (Section \ref{Source PSF-scattered flux recovery}). The aim is for the sources to recover \textcolor{black}{an approximation of} their intrinsic flux (unaffected by the PSF). However, the PSF-scattered flux outside the sources remains physically present in the image. In the second step, we use the output from the first step to subtract this excess flux from the surrounding regions (Section \ref{Background PSF-scattered flux subtraction}). \textcolor{black}{It is important to note that this technique does not model PSF-scattered light from pixels outside the detected sources.} The final output of the technique is the PSF-corrected image. \textcolor{black}{A schematic representation of the steps involved in this technique is shown in Fig.~\ref{PSF removal}}.}

%nor does it subtract PSF-scattered flux from within the sources, as the subtraction is performed exclusively on unmasked pixels.}

%\textcolor{mycolor}{The technique consists of two main steps. First, we modify the flux of each detected source in the image to compensate for the lost scattered flux as a consequence of the PSF convolution (Section \ref{Source PSF-scattered flux recovery}). The aim is for the sources to recover the intrinsic flux (unaffected by the PSF). However, the PSF-scattered flux outside the segments remains physically present in the image. Secondly, we remove this extra flux that contaminates the LSB estimations using the product of the first step (Section \ref{IHL PSF-scattered flux removal}).}

%to mimic the PSF convolution that an image suffers when being observed. This way we ensure that only the scattered flux produced in the convolution between the sources and the PSF is being removed from the surroundings of each segment.

%original group image with the PSF and identify the scattered flux \textcolor{mycolor}{to remove it later in Sec.~\ref{IHL}}.

%\begin{figure}
%\centering  %\qquad
% \includegraphics[width=1\linewidth]{Paper2/Figures/2ndVersion.png}
%    \caption{}
%    \label{convolution 2}
%\end{figure}

%VER DE PONER IMAGEN MAYOR 

\subsection{Source PSF-scattered flux recovery}
\label{Source PSF-scattered flux recovery}
\begin{figure*}
\centering  %\qquad
 \includegraphics[width=.77\linewidth]{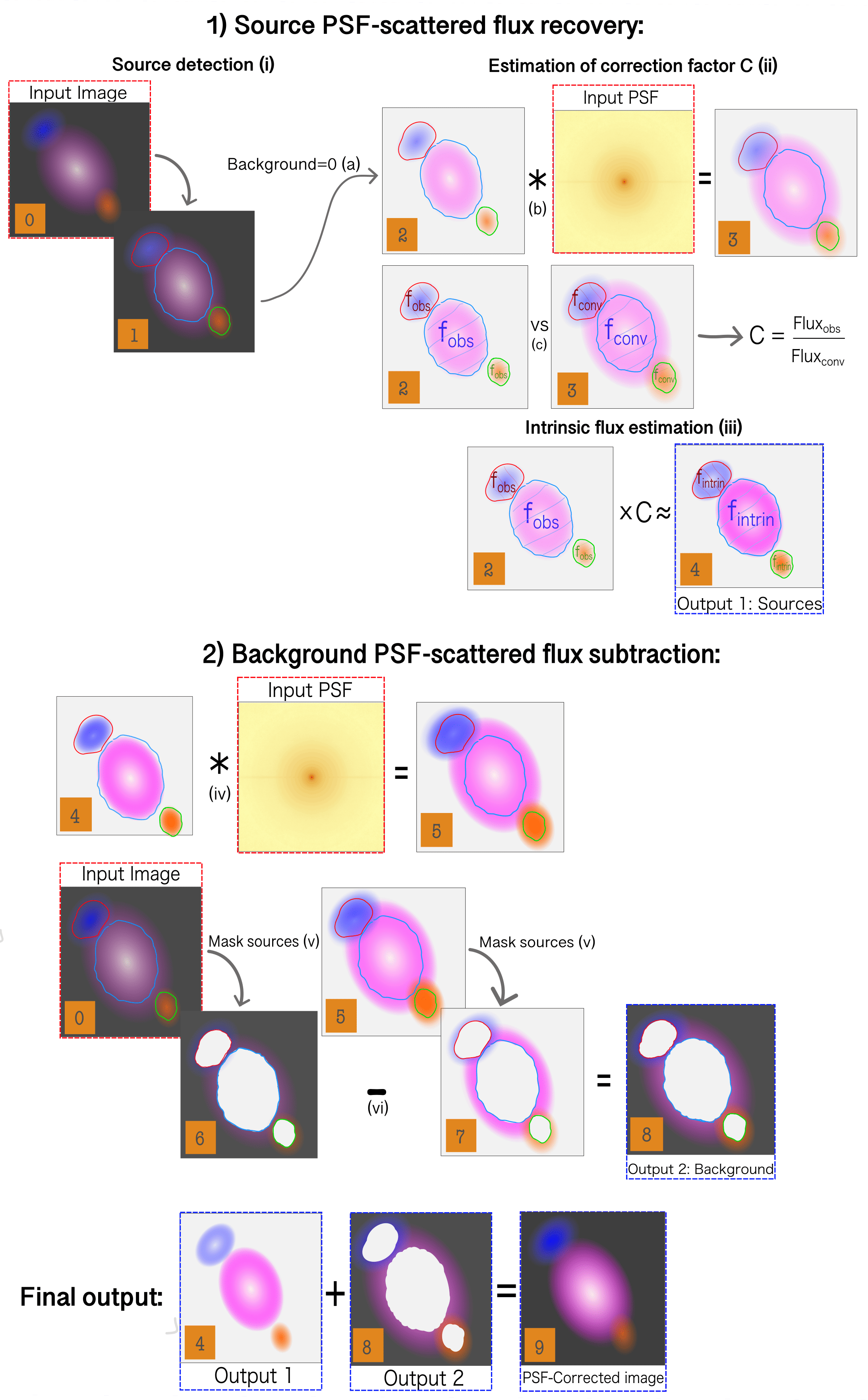}
    \caption{Schematic view of the processing steps outlined in Sec.~\ref{sec:PSF analysis} to remove the PSF scattered light from an astronomical image. The first part of the technique covers Steps i-iii, which are explained in Sec.~\ref{Source PSF-scattered flux recovery}. The second part covers Steps iv-vi, explained in  Sec.~\ref{Background PSF-scattered flux subtraction}. Throughout both parts of the technique, we make reference to the orange numbered labels to clarify the intermediate and final outputs of each step. \textcolor{black}{The grey background of the panels represents the sky, while the white background corresponds to pixels that have been set to zero. The colours of the sources and the segments are intended for qualitative visualisation only.} Solid dark grey lines indicate transitions between steps. \textcolor{black}{This figure shows only galaxy sources (and excludes point sources) for illustrative purposes.}} %The sources identified by \textsc{ProFound} in (ii) are delimited by multi-colored segments. , whereas dashed light grey lines indicate repeated panels.
    \label{PSF removal}
\end{figure*}

\textcolor{black}{In this part, we estimate a correction factor for each source that scales its observed flux towards an approximation of its intrinsic flux.} \textcolor{black}{This process is explained here, with references to the orange-numbered labels shown in Fig.~\ref{PSF removal}}:

  \begin{enumerate}
  \item \textcolor{black}{\textbf{Source detection:} Run the source extraction code to detect sources in the Input image and measure their fluxes ($Flux_\mathrm{{obs}}$)\textcolor{black}{.}}
  \vspace{2mm}

  \item \textcolor{black}{\textbf{Estimation of correction factor ($C$):}}
    \begin{enumerate}
      \item \textcolor{black}{Reset all pixel values outside the sources (background) to zero\textcolor{black}{.}}
      \item \textcolor{black}{Convolve the image with background zero with the PSF model to reproduce the PSF effect\textcolor{black}{.}}
      \item \textcolor{black}{Compute the fraction of flux scattered outside each source due to the PSF. To do this, compare the flux per source before ($Flux_\mathrm{{obs}}$) and after ($Flux_\mathrm{{conv}}$) the PSF-convolution from (b). This results in a correction factor per source: $C$ = $\frac{Flux_\mathrm{{obs}}}{Flux_\mathrm{{conv}}}$\textcolor{black}{.}}
      %\[
      %C = \frac{\mathrm{Flux_{obs}}}{\mathrm{Flux_{conv}}}.
      %\]
    \end{enumerate}
    \vspace{2mm}

  \item \textcolor{black}{\textbf{Intrinsic flux estimation:} Calculate an approximation of the intrinsic flux within each source ($Flux_\mathrm{{intrin}}$) by rescaling the observed \textcolor{black}{integrated} flux with $C$: $Flux_{\mathrm{intrin}}$ $\approx$ $Flux_{\mathrm{obs}}$ $\cdot$ $C$} \textcolor{black}{(Output 1, bounded by the blue box).}
\end{enumerate}

\textcolor{black}{The following paragraphs provide a detailed explanation of each step listed above. After detecting the sources in the image in Step i, we proceed to estimate the PSF-correction factor $C$. In this estimation, Step ii-a is essential because it prevents PSF-scattered flux from contaminating the sources during the ii-b convolution, which would otherwise bias $C$. In Step ii-b, we convolve the observed flux with the PSF (producing Label 3), ignoring the fact that the source flux is already PSF-affected, since our goal is solely to quantify the fraction of light scattered outside the sources due to the PSF. However, because the image is already PSF-convolved, the factor $C$ does not exactly represent the value needed to recover the true intrinsic flux of each source. This point is discussed better in Section \ref{cavetas psf}. For clarification, the twice PSF-convolved image (Label 3) is used exclusively to estimate $C$, and $C$ is then used to estimate the intrinsic flux, but the twice PSF-convolved image itself is not used in any subsequent steps. In Step ii-c, we estimate $C$, which is then used in Step iii to approximate the intrinsic flux per source. The output of the first part of the PSF-scattered flux removal technique is an image where background pixels are zero and source regions contain the estimated intrinsic flux.}

\subsection{\textcolor{black}{Background PSF-scattered flux subtraction}}
\label{Background PSF-scattered flux subtraction}
\textcolor{black}{In the first part of the technique we focus on modifying the flux per source to recover the flux unaffected \textcolor{black}{by} the PSF convolution. In this part, we use Output 1 to correct the background pixels, which are still contaminated by PSF-scattered flux. The procedure is as follows:}

\begin{enumerate}
 \item[(iv)] \textcolor{black}{\textbf{Estimation of PSF flux outside the sources:} Convolve Output 1 with the PSF to identify the scattered light that needs to be removed from the Input image\textcolor{black}{.}}

 \item[(v)] \textcolor{black}{\textbf{Masking source regions:}} \textcolor{black}{Mask all pixels inside the sources in the convolved version of Output 1, and apply the same mask to the Input image.}

 \item[(vi)] \textcolor{black}{\textbf{Background correction:} Subtract the estimated PSF flux outside the sources from the \textcolor{black}{masked version} of the Input image (Output 2, bounded by the blue box).}   
\end{enumerate}

\textcolor{black}{Output 1 plays a key role in accurately subtracting the PSF-scattered flux from the background for two reasons. First, as it contains the intrinsic flux of the sources rather than the observed flux, the convolution in Step \textcolor{black}{iv} provides a better approximation of the PSF-scattered flux to be subtracted\footnote{\textcolor{black}{If the steps from Section~\ref{Source PSF-scattered flux recovery} are skipped (i.e. the intrinsic flux is not} \textcolor{black}{estimated) and the observed flux is convolved with the PSF in Step iv, the total flux is underestimated, leading to a corresponding underestimation of the PSF-scattered light in the outskirts and thus less flux being subtracted from the background.}}. Second, since Output 1 has zero-valued pixels in the background, the flux to be removed from the background represents ideally purely PSF-scattered light\textcolor{black}{, with no sky contribution except in source segments where sky pixels overlap.}}

%the flux outside the sources would be redistributed even farther toward the outskirts, resulting in less flux being subtracted from the Input image in Step viii.

%to create a PSF-convolved version of the image, which will be later subtracted from the Input image. This results in Output 2, which contains only the IHL of the system mostly unaffected by the PSF-scattered flux. The procedure is as follows:

%With our first step we  rescale the flux per source inside the segments to account for the PSF-scattered flux, but the scattered light that surrounds the segments is still present in the original image. To remove it, the second part of the technique applies the following steps:

%To achieve this, we use Output 1 to reproduce the PSF flux (Step vi), which is then subtracted from the flux outside the segments in the input image (Step viii).

%In Label 5, we make sure that the flux outside the segments comes purely from the sources, 
%The key to this second part of the PSF-correction is removing the PSF-scattered flux that contaminates the IHL. To achieve this, we use Label 4 to reproduce the PSF flux that we want to remove (Step vi) from the flux outside the segments in Label 0 (Step viii), using the masks from Step vii. In Label 7, the flux outside the segments comes purely from the sources, whereas the flux outside the segments in Label 6 is, in principle, a combination of diffuse stellar light and PSF flux. Label 8 (bounded by the blue box) is then the IHL component unaffected by the PSF-scattered flux. 

The final output of the technique is the PSF-corrected image (Label 9, bounded by the blue box), obtained by combining Outputs 1 and 2. \textcolor{black}{With this approach, the net effect is to reduce PSF-scattered light in the background while increasing the flux within the sources to account for the flux lost due to PSF scattering.}

%we recover most of the intrinsic flux within the sources while substantially reducing the PSF-scattered flux outside the segments.}

%CORTE MANUALMENTE EL PLOT IHL_pipeline.png para que no queden bordes grandes y ver lo d width .8

\subsection{Caveats}
\label{cavetas psf}

  %The first caveat of the technique is that \textcolor{mycolor}{the details of how segmentation maps are defined by the source detection code influence the resulting flux redistribution, leading to variations in the PSF-corrected image.}

  The first caveat of the technique is that \textcolor{black}{the results depend on how the segmentation maps are defined by the source detection algorithm. Different segmentation definitions assign different sets of pixels to sources and background, which in turn changes how flux is partitioned between these components. Since the PSF correction relies on this partition, the resulting PSF-corrected image can vary depending on the adopted segmentation.} \textcolor{black}{In some cases, as shown in Appendix \ref{B}, the \textsc{ProFound} segmentation map may fragment bright sources into multiple smaller objects, requiring careful manual masking to correctly model the PSF-scattered light in the subsequent steps.} \textcolor{black}{A related limitation is that our technique only takes into account the PSF light that is scattered from the unmasked pixels, which contain the galaxies and stars present in the image. That is, the technique does not model the PSF-scattered light that might be produced on the intra-halo light (IHL) component of the system. However, in a typical cluster field, the PSF-scattered light is expected to be dominated by the brightest sources, primarily point sources such as foreground stars and, to a lesser extent, the extended BCG \citep{2020ApJS..247...43K, 2024MNRAS.531.2517G}. Therefore, although our technique does not explicitly model the PSF contribution from the IHL itself, this contribution is expected to be subdominant.}

  %In addition, as mentioned above, the PSF-scattering correction is applied exclusively to unmasked pixels. Consequently, PSF-scattered flux from surrounding sources that falls within the footprint of a detected source is not corrected. This unmodelled contribution may introduce a systematic uncertainty in the recovered source fluxes.}

  %the estimation of the intrinsic flux per source depends on how we define the segmentation maps around each source in \textcolor{mycolor}{the source detection code}. The amount of flux enclosed depends on our choice of input values for each parameter in \textsc{ProFound}. Additionally, some galaxy light will always remain unaccounted for, as it is not fully enclosed within the segments. \textcolor{mycolor}{This unaccounted flux is therefore not considered in the factor $C$ used to estimate the intrinsic flux.}

The \textcolor{black}{next} caveat\textcolor{black}{, as previously mentioned, concerns the estimation of  the correction factor $C$.} Since the sources \textcolor{black}{are already affected by the PSF}, the flux inside the segments has already been scattered, modifying the shape of the source. This modification of the profile implies that our estimation of $C$ will never exactly match the true value required to recover the intrinsic flux \textcolor{black}{of the sources}.  \textcolor{black}{That is}, \textcolor{black}{while the factor $C$ allows us to estimate the fraction of flux redistributed by the convolution, our technique cannot fully recover the true intrinsic flux or profile of the sources.} 

%the remaining IHL in Label 8 will be slightly underestimated compared to the real IHL. 

%\textcolor{mycolor}{In our PSF-correction technique,} we slightly overestimate $Flux_{\mathrm{intrin}}$, which causes an overestimation of \textcolor{mycolor}{the PSF flux to be subtracted from the background of the Input image. As a consequence,} \textcolor{mycolor}{the background of the PSF-corrected image is slightly oversubtracted.} \textcolor{mycolor}{This conclusion follows from the test described below, which compares the true correction factor ($C_{\mathrm{true}}$) that should be used in our PSF correction technique to recover the real intrinsic flux of each source, versus the estimated factor ($C_{\mathrm{est}}$) we actually use.}

\textcolor{black}{To assess the accuracy of our PSF-correction technique, we compare the true correction factor ($C_{\mathrm{true}}$), which should be used to recover the intrinsic flux of each source, with the estimated factor ($C_{\mathrm{est}}$) that we actually use in our method. This comparison shows that $C_{\mathrm{est}}$ slightly overestimates the correction needed to recover the intrinsic flux. As a result, the PSF flux subtracted from the background of the Input image is also slightly overestimated, leading to a mild oversubtraction of the background in the PSF-corrected image.}

%o demonstrate how our PSF correction slightly overestimates the intrinsic flux per source, we calculate the true correction factor ($C_{\mathrm{true}}$) that should be used in our PSF correction technique to recover the real intrinsic flux of each source, and compare it to the estimated factor ($C_{\mathrm{est}}$) we actually use. 

\textcolor{black}{To do this test, we use a mock image of a galaxy group, in order to have the non-PSF convolved version. The comparison of $C_{\mathrm{true}}$ versus $C_{\mathrm{est}}$ could be done with any data, but we take advantage of the HSC-PDR3 mock dataset used in this work and select one mock group to make this comparison. The mock generation process is further explained in Section \ref{sec:Data}, but briefly, we select the coordinates of a real group from the GAMA Galaxy Group Catalogue (G$^3$Cv10; \citealt{2011MNRAS.416.2640R}) and use real HSC-PDR3 data to create the galaxies in the mock image. This is done by running our source extraction code \textsc{ProFound}\footnote{\url{https://github.com/asgr/ProFound}}\citep{2018MNRAS.476.3137R} in the real HSC group image and creating a new image where all pixels outside the \textcolor{black}{identified segmentation maps} are masked. We then \textcolor{black}{analytically} inject the foreground stars at their true coordinates as single pixels, convolve the image with our HSC-PDR3 PSF model \textcolor{black}{\citep{2024MNRAS.531.2517G}}, add Gaussian noise, and finally \textcolor{black}{analytically} inject an exponential Sérsic profile modelling the IHL component of the group. This approach allows our mocks to capture the full complexity and clustering of sources, as well as the realistic stellar foreground.}

%During this process, we aim to avoid the IHL pixels, as

%to generate this mock, where the generation process is better explained in Section \ref{sec:Data}. Briefly, we run \textsc{ProFound} in a real group image and create a new image where all pixels outside the segments are masked. We then inject the foreground stars at their true coordinates as single pixels, convolve the image with our HSC-PDR3 PSF model, and add Gaussian noise. Finally, we inject a circular exponential IHL component. This approach allows our mocks to capture the full complexity and clustering of sources, as well as the realistic stellar foreground (features that are not achievable with hydrodynamical simulations). Although hydrodynamical simulations are useful, they are limited by stellar mass resolution and often fail to accurately reproduce LSB structures.}

To estimate $C_{\mathrm{true}}$, we compare the flux inside each \textsc{ProFound} segment in the non-PSF convolved version and in the PSF-convolved version of the mock image. \textcolor{black}{This is possible because the intrinsic, pre-PSF-convolved flux of each source is known by construction in the mock images. In real observations, however, the intrinsic flux of each source before PSF convolution is fundamentally unknown, and therefore $C_{\mathrm{true}}$ cannot be directly determined or used in the PSF-correction pipeline. Thus, $C_{\mathrm{true}}$ serves purely as a theoretical validation metric in our mock tests, allowing us to assess how accurately the estimated correction factor $C_{\mathrm{est}}$ recovers the correction that would be required if the intrinsic source flux were known}. To calculate $C_{\mathrm{est}}$, following the Step \textcolor{black}{ii} outlined in the technique, we compare the flux inside each segment in the PSF-convolved version and in the twice PSF-convolved version of the mock image.

\begin{figure}
\centering  %\qquad
 \includegraphics[width=1\linewidth]{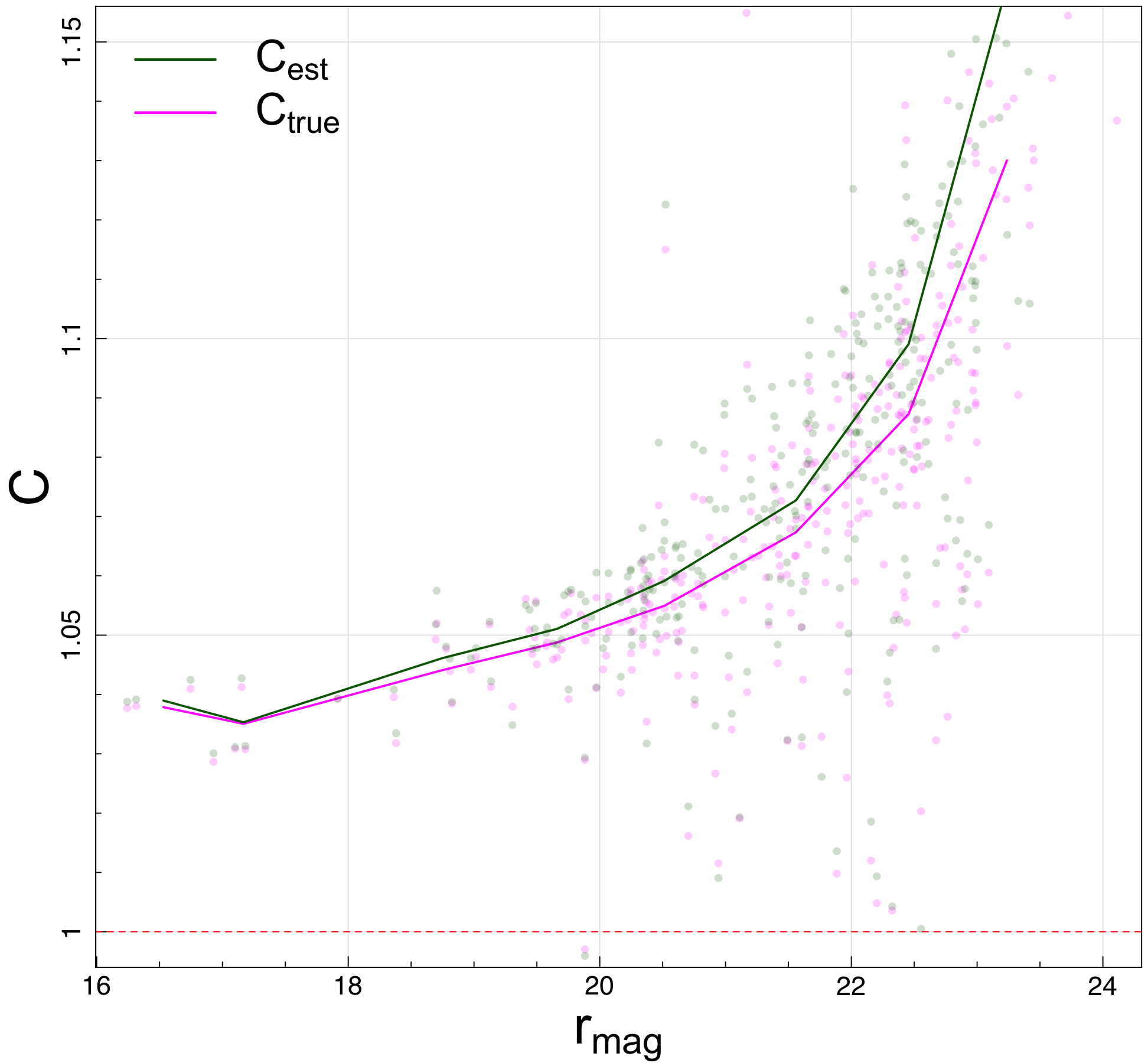}
    \caption{\textcolor{black}{Comparison between the true PSF-correction factor ($C_{\mathrm{true}}$) and the estimated PSF-correction factor using our technique ($C_{\mathrm{est}}$) as a function of the magnitudes of the sources present in the group mock image. At fainter magnitudes, $C_{\mathrm{est}}$ is slightly larger than $C_{\mathrm{true}}$, suggesting that our PSF-scattered flux removal method tends to marginally overestimate the intrinsic flux when the observed flux is rescaled using $C_{\mathrm{est}}$}.} %The sources identified by \textsc{ProFound} in (ii) are delimited by multi-colored segments.
    \label{PSF percentage}
\end{figure}

\textcolor{black}{Fig.~\ref{PSF percentage} illustrates the difference between $C_{\mathrm{est}}$ and $C_{\mathrm{true}}$ as a function of the magnitude of each source. This plot provides the reader with a sense of the impact of convolving the image twice and identifying which sources are most affected. Ideally, in the PSF-correction we should use $C_{\mathrm{true}}$ to re-estimate the intrinsic flux per source; instead, we perform the correction with $C_{\mathrm{est}}$. \textcolor{black}{We first observe a trend of increasing correction factor $C$ towards fainter sources, reflecting the fact that for bright sources the SB remains above the detection threshold over a sufficiently large area for the segmentation map to enclose most of the flux. In contrast, for faint sources only the high signal-to-noise central region is typically detected, while a larger fraction of the low SB outskirts lies outside the segments. As a result, PSF scattering redistributes a greater fraction of their total flux beyond the segmentation boundaries, leading to a larger relative correction. We also observe that $C_{\mathrm{est}}$ is consistently larger than $C_{\mathrm{true}}$, meaning that we systematically overestimate the intrinsic flux of sources across all magnitudes. However, } $C_{\mathrm{est}}$ is more accurate for the brighter objects, whereas for the fainter ones we overestimate the intrinsic flux by up to 4\%. This overestimation of the source flux leads to an overestimation of the PSF-scattered flux that is subtracted from the original image, resulting in a slightly oversubtracted background. This issue is discussed in more detail in Section \ref{sec:results}.} \textcolor{black}{To provide context, a typical correction factor of $C \approx 1.15$ indicates that the observed flux underestimates the intrinsic flux by $\sim$15\% due to PSF effects, implying that the residual bias in our method (up to $\sim$4\%) corresponds to approximately 27\% of the total PSF-induced effect.}  %4/15 ~0.25

\textcolor{black}{Another caveat is the dependence of our PSF-correction technique on the sky-subtraction performed during the primary data reduction. The extended PSF wings may either be partially removed as part of the sky subtraction or contribute to the local sky estimate, potentially leading to an over-subtraction of LSB flux. Since our technique operates on the reduced image, any PSF-scattered flux removed during the sky-subtraction process cannot be recovered by our correction.}

%The upper panel shows the true PSF correction factor as a function of the magnitude of the detected segments in an HSC-PDR3 mock image. The median and 1$\sigma$ range are shown in solid and dashed pink respectively. The medium left panel shows the ratio between the true and estimated correction factors $\left( \frac{C_{\mathrm{true}}}{C_{\mathrm{est}}}\right)$ also as a function of the segment magnitude. The median indicates that $C_\mathrm{{est}}$ is larger than $C_\mathrm{{true}}$ towards fainter magnitudes, showing that our PSF-scattered flux removal technique slightly overestimates the intrinsic flux of the sources. The medium right panel shows $\frac{C_{\mathrm{true}}}{C_{\mathrm{est}}}$ as a function of the true correction factor. The lower panel shows $\frac{C_{\mathrm{true}}}{C_{\mathrm{est}}}$ escalated by the flux of each segment, as a function of the segment magnitude.

\subsection{\textcolor{black}{Comparison with existing PSF-correction methods}}
\label{Comparison with existing PSF-correction methods}
\textcolor{black}{\textcolor{black}{A common approach for removing PSF-scattered light from an image is to use image deconvolution algorithms.} There are several methods to deconvolve an image, including the classical Richardson–Lucy algorithm, Wiener filtering, wavelet-based techniques, linear regularised methods, and more recent deep-learning–based approaches that have shown promising results in recovering faint extended structures (e.g., \citealt{wiener1949,1972JOSA...62...55R,1974AJ.....79..745L,2002PASP..114.1051S, 2017A&A...601A..86K,2017MNRAS.467L.110S,2025A&A...697A.136A}). However, deconvolution approaches can be computationally expensive due to regularisation and iterative steps, highly sensitive to noise, \textcolor{black}{and can amplify residual uncertainties in the PSF model, which may lead to unstable or unreliable results if the PSF is not accurately characterised.} \textcolor{black}{In App. \ref{D}, we further explore the Wiener deconvolution method by comparing the results obtained from applying our PSF correction method and the Wiener algorithm to a group image, demonstrating that our method produces a more stable PSF-corrected version and is less sensitive to small numerical variations in the PSF.}}

%computationally expensive due to regularisation and iterative steps,

\textcolor{black}{Other approaches used in the literature that do not employ a deconvolution technique involve generating 2-D PSF-convolved models of the individual sources and subtracting them from the original image \citep{2016ApJ...823..123T,2017MNRAS.470..427P,2023MNRAS.518.1195M}. However, the scattered light from sources that are not modelled can contaminate the image. In addition, this method requires the assumption of a specific profile for each source \textcolor{black}{present in the image}. Our method offers a more straightforward and computationally efficient way to estimate the intrinsic flux of all sources present in the image. \textcolor{black}{The efficiency of the method mainly depends on the size of the image and of the PSF model. In the case discussed in Sec.~\ref{cavetas psf}, for an HSC-SSP PDR3 mock group image of 2000 $\times$ 2000 pix$^2$ and a PSF model of $\sim 4000 \times 4000$ pix$^2$, the PSF-correction procedure takes $\sim$ 15 sec.} \textcolor{black}{By performing two convolutions with the PSF model and estimating the source flux through a computationally efficient procedure}, we retrieve the intrinsic flux of the sources within an error of up to $\sim$ 4\%, as demonstrated in Fig.~\ref{PSF percentage}. That is, our method does not require modelling individual sources and performs a PSF correction to all the sources present in the image, rather than only to a selected subset.}

\section{IHL estimation technique}
\label{sec:IHL analysis}
In this section, we present our technique to create a two-dimensional \textcolor{black}{exponential} model  of the intra-halo light component in images of galaxy groups and clusters. \textcolor{black}{We use a 2D galaxy-profiling tool that includes a Sérsic template suitable for describing the IHL, such as \textsc{ProFit} \citep{2017MNRAS.466.1513R}, \textsc{galfit} \citep{2010AJ....139.2097P}, \textsc{imfit} \citep{2015ApJ...799..226E} and \textsc{autoprof} \citep{2021MNRAS.508.1870S}.}

%\textcolor{black}{with circular symmetry}

%\textcolor{mycolor}{Although this technique can also be applied to model non-circular IHL components, we describe it here for the circular case for simplicity.}

%To model the IHL we make use of \textsc{ProFit}\footnote{\url{https://github.com/asgr/ProFit}}, a fully Bayesian 2-D galaxy profiling tool \citep{2017MNRAS.466.1513R}. \textsc{ProFit} offers flexible modelling with built-in profiles (Sérsic, Moffat, exponential, etc.), support for multiple components, various likelihood and optimisation methods -- including Markov Chain Monte Carlo (MCMC) and genetic algorithms -- fitting in linear or logarithmic space, and the ability to apply complex priors and parameter constraints (e.g.,\citealt{2019MNRAS.490.4060C, 2022MNRAS.516..942C}). 

\begin{figure*}
\centering  %\qquad
 \includegraphics[width=.85\linewidth]{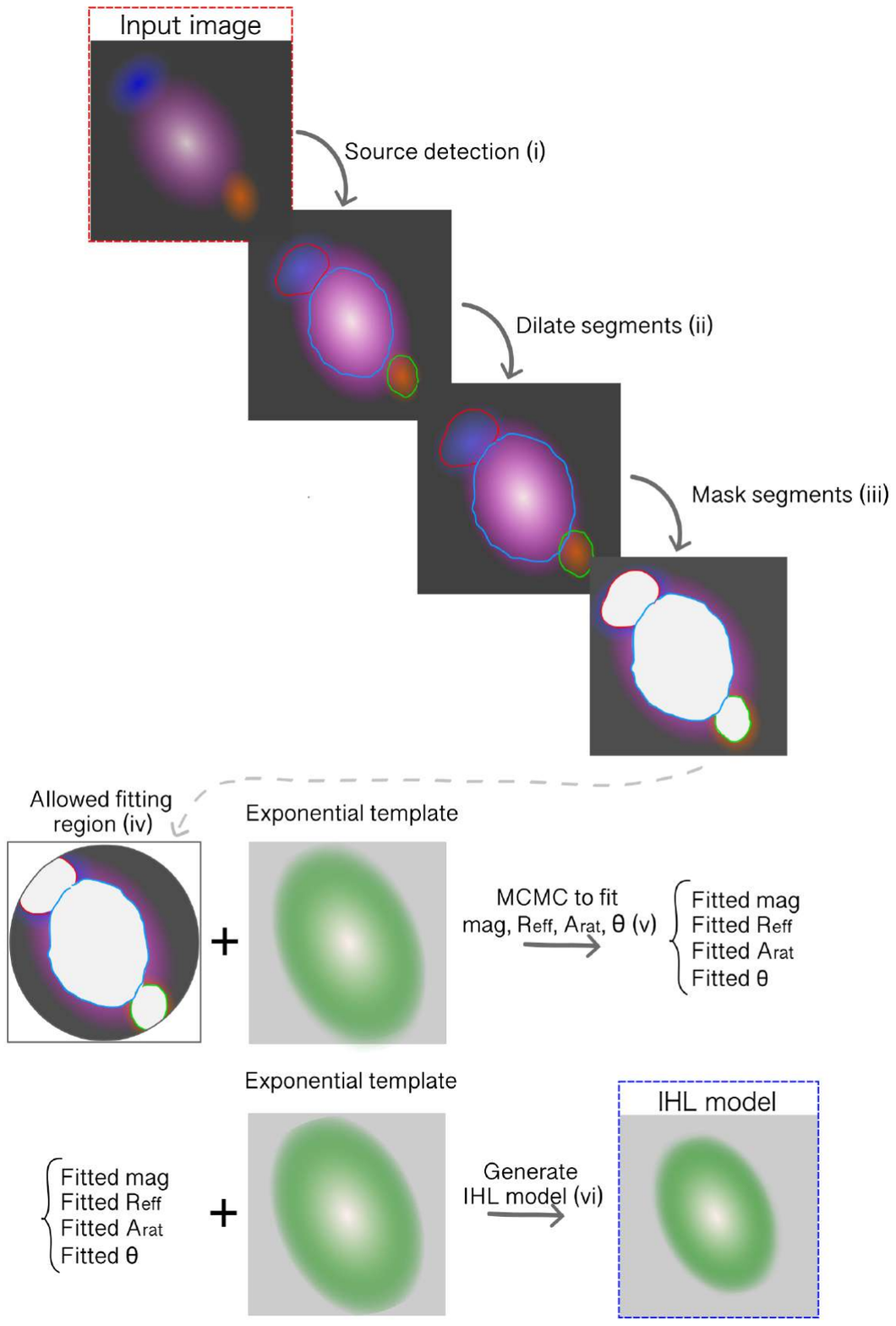}
    \caption{Schematic view of the processing steps outlined in Sec.~\ref{sec:IHL analysis} to model the IHL component of a system using a 2D exponential template. The steps shown correspond to those outlined in that section. Solid dark grey lines indicate transitions between steps, whereas dashed light grey lines indicate repeated panels.}
    \label{ihl model}
\end{figure*}

The process to model the IHL component is explained in detail here, with a schematic representation of the steps shown in Fig.~\ref{ihl model}:

%We also show the results of testing this technique in our sample of HSC mock observations of GAMA groups introduced in Sec.~\ref{mock section}.

\begin{enumerate}
 % \item apply the PSF-scattered flux removal technique presented in Sec.~\ref{sec:PSF analysis} to eliminate PSF-scattered flux;
  
  %We first start by applying our PSF-scattered flux removal technique presented in Sec.~\ref{sec:PSF analysis} to our sample of astronomical images. To do this, we need a PSF extended model of the telescope used to collect the data, which is required in steps iii and iv. After removing the PSF-scattered flux that contaminates the estimation of low surface brightness features in the image, we proceed to model the IHL component using \textsc{ProFound} and \textsc{ProFit} \citep{2017MNRAS.466.1513R} in each mock image. 

 \item Run the source extraction code to detect all sources in the image. 

%\textcolor{mycolor}{As the PSF correction redistributes the flux in the image, a second run of the source extractor is required\textcolor{black}{.}}

 %\item apply the \texttt{profoundMakeSegimDilate} function to dilate the segments and \textcolor{mycolor}{ensure the galaxy flux is fully captured\textcolor{black}{.}} This function depends on the \textit{size} parameter, which in this case we use \textit{size} = 21.

 %\item \textcolor{mycolor}{apply the \texttt{profoundMakeSegimDilate} function to dilate segments and fully capture galaxy flux, which depends on the parameter \textit{size} = 21\textcolor{black}{.}}

      %\item \textcolor{mycolor}{apply the \texttt{profoundMakeSegimDilate} function to dilate segments and fully capture galaxy flux\textcolor{black}{.} this function depends on the \textit{size} parameter, set to 21 in this case\textcolor{black}{.}}

            %\item \textcolor{mycolor}{apply \texttt{profoundMakeSegimDilate} with the \textit{size} parameter set to  21 to dilate segments and fully capture galaxy flux\textcolor{black}{.}}

        \item \textcolor{black}{Apply mask dilation tool to expand the segments \textcolor{black}{in order to include pixels containing significant intrinsic source emission.}}

 %capturing any extra flux potentially contaminating the IHL\textcolor{black}{.}

 \item Mask all the detected sources using the dilated segments\textcolor{black}{.}  

%After creating segmenation maps for all detected sources, we apply the \texttt{profoundMakeSegimDilate} function to dilate the segments, ensuring that any extra flux potentially contaminating the IHL modelling is effectively masked.

%\textsc{ProFound} is source finding and photometry analysis package that detects sources using dilated segments (ishophotal outlines) of arbitrary shape (rather than elliptical apertures) and measures statistics such as flux, size, and ellipticity.

%\item \textcolor{mycolor}{restrict the \textsc{ProFit} fitting region to all unmasked pixels within a circle to avoid distortion problems around the image borders\textcolor{black}{.}}

%\item \textcolor{mycolor}{restrict the \textsc{ProFit} fitting region to all unmasked pixels within a circle of radius $ \textcolor{black}{$R_{\mathrm{fit}}$} to avoid distortion} problems near the image borders, where
 % \begin{equation}
 % \mathrm{R_{fit}} = \mathrm{R_{cutout}} - 25,
 % \end{equation}
 % \textcolor{mycolor}{where $\mathrm{R_{cutout}}$ is the radius of the mock group cutout in pixels, from which we subtract 25 pixels to mitigate distortion effects near the image edges\textcolor{black}{.}}

\item \textcolor{black}{Restrict the fitting region to the unmasked pixels within a circle of radius}  \textcolor{black}{$R_{\mathrm{fit}}$} = \textcolor{black}{$R_{\mathrm{cutout}}$} - 25 \textcolor{black}{pix} \textcolor{black}{to avoid edge distortions, where  \textcolor{black}{$R_{\mathrm{cutout}}$} is \textcolor{black}{the distance in pixels from the centre of the cutout to its edge.}}

%cutout radius is defined as the distance from the centre of the cutout to its edge (i.e. half the cutout size)

%\item use \textsc{ProFit} to fit a 2D circular exponential model to the IHL component, adopting a Sérsic template with Sérsic index $\textit{n}$ = 1. This template is fully characterised by eight parameters, two of which, magnitude \textcolor{mycolor}{(overall brightness of the IHL component)} and effective radius \textcolor{mycolor}{(radius enclosing half of the total light)}, are left as free parameters. \textcolor{mycolor}{In this step, we fit these two free parameters within specified limits in the fitting range\textcolor{black}{.}}

\item \textcolor{black}{Use \textcolor{black}{a} 2D galaxy-profiling tool to fit a 2D exponential model to the IHL component, adopting a Sérsic profile with Sérsic index $\textit{n}$ = 1. The Sérsic model contains eight parameters, of which the magnitude (overall brightness), the effective radius (radius enclosing half of the total light), \textcolor{black}{the major-axis angle (orientation of the major axis), and the axis ratio (ratio of minor to major axis)} are left free. These \textcolor{black}{four} parameters are fitted within specified limits\textcolor{black}{.}}

\item Use again an exponential Sérsic template with Sérsic index $\textit{n}$ = 1 with the fitted \textcolor{black}{parameters} from Step \textcolor{black}{v} to generate the image model of the IHL component.

%We use the Sérsic template to model a circular IHL following an exponential profile. This approach is supported by cosmological simulations, which have proved the necessity of an extra Sérsic or exponential profile to fit the stars forming the diffuse BCG component (e.g., \citealt{2005ApJ...618..195G}, \citealt{2010MNRAS.405.1544D}, \citealt{2017A&A...603A..38S}). Three key parameters are used to characterise the IHL exponential template: Sérsic index, magnitude, and effective radius. We fix the Sérsic index \texttt{n} equal to 1, while the magnitude and effective radius are left as free parameters and optimized using a Markov Chain Monte Carlo (MCMC) algorithm. In order to avoid distortion problems around the borders of the images, we are restricting the fitting region to a circular shape. 

\end{enumerate}

%\textcolor{mycolor}{The following paragraphs provide a detailed explanation of each step listed above}. To model the IHL, we first remove the PSF-scattered flux in Step i. \textcolor{mycolor}{In Step ii, we run \textsc{ProFound} for the second time on each image, as it was previously run to correct for the PSF-scattered flux. The PSF correction increases the flux within the segments but also removes flux outside them, which implies that the second run of \textsc{ProFound} will be slightly different from the first run. This second run is used to detect the sources, dilate the segmentations maps (Step iii), mask them (Step iv), and create the fitting region for \textsc{ProFit} (Step v).

 %The \texttt{profoundMakeSegimDilate} function expands out the image segments already identified by \textsc{ProFound}. This function depends on the \textit{size} parameter, which in this case we use \textit{size} = 21. }

%The fitting region is delimited by all the unmasked pixels within a circle of radius $ \textcolor{black}{$R_{\mathrm{fit}}$} equal to:}

%\begin{equation}    
%\\
%\\
%\\
% \mathrm{R_{fit}} =  \mathrm{R_{cutout}} - 25,\\
%\end{equation}

%\textcolor{mycolor}{where $\mathrm{R_{cutout}}$ is the radius of the mock group cutout in pixels, from which we subtract 25 pixels to mitigate distortion effects near the image borders}

\textcolor{black}{In this technique, we use both operational modes of the 2D galaxy-profiling tool: the fitting mode (\textcolor{black}{Step v}), and the 2D image-generation mode (\textcolor{black}{Step vi}). In \textcolor{black}{Step v}, we fit the magnitude, the effective radius, \textcolor{black}{the major-axis angle, and the axis ratio} of the IHL component. As mentioned, the Sérsic template is fully characterised by eight parameters: xcen (\textcolor{black}{$x_{\mathrm{cen}}$}), ycen (\textcolor{black}{$y_{\mathrm{cen}}$}), magnitude ($\textit{mag}$), effective radius (\textcolor{black}{$R_{\mathrm{eff}}$}), Sérsic index ($\textit{n}$), major-axis angle ($\textit{$\theta$}$), minor to major axis ratio (\textcolor{black}{$A_{\mathrm{rat}}$}), and boxiness ($\textit{B}$). In both Steps v and vi, in order to recreate an exponential profile, \textcolor{black}{the next} parameters are fixed: \textcolor{black}{$x_{\mathrm{cen}}$} and \textcolor{black}{$y_{\mathrm{cen}}$} are set to the centre of the mock, $\textit{n}$ = 1, and $\textit{B}$ = 0. We use an optimisation algorithm to fit $\textit{mag}$, \textcolor{black}{$R_{\mathrm{eff}}$, $\textit{$\theta$}$, and $A_{\mathrm{rat}}$}. Finally, in \textcolor{black}{Step vi}, we use the output parameters from \textcolor{black}{Step v} together with the remaining fixed parameters to recreate the IHL model of the group.}

 %We use \textsc{Highlander} to fit $\textit{mag}$ and \textcolor{black}{$R_{\mathrm{eff}}$}, which is an optimisation algorithm that combines a genetic algorithm and MCMC to explore local extrema and identify the global maximum (e.g., \citealt{2021MNRAS.505..540T}).

\subsection{\textcolor{black}{Technique details}}

            Selecting a Sérsic template with fixed Sérsic index equal to 1 is supported by cosmological simulations, which have proved the necessity of an extra Sérsic or exponential profile to fit the stars forming the diffuse BCG component (e.g., \citealt{2010MNRAS.405.1544D}). \textcolor{black}{Separating the IHL from the BCG using two components has also been performed in observational works}, \textcolor{black}{with the IHL often described using profiles motivated by theoretical expectations, such as de Vaucouleurs or exponential profiles \citep{2005ApJ...618..195G,2007AJ....134..466K,2017A&A...603A..38S,2018MNRAS.474..917M}}. Further evidence for our choice of an exponential model is provided in Appendix~\ref{C}.

%\textcolor{mycolor}{Our motivation for modelling a circular (exponential) IHL component is establishing an ideal scenario in which differences between fitted IHL models can be attributed to the PSF effect when comparing results with and without our PSF correction applied. For this reason, in Section \ref{sec:Data}, we create mock observations of galaxy groups with an analytical circular IHL component, which we then fit using our circular IHL model. This approach allows us to minimise fitting-related complications and attribute any remaining deviations to PSF effects.}

 \textcolor{black}{2D galaxy-profiling tools can be used to fit both circular and elliptical exponential IHL components, where the reliability of these fits depends on the signal-to-noise ratio (S/N) of the data and the complexity of the model. Most haloes, especially those that are dynamically unrelaxed, show significant low surface brightness (LSB) substructures in their IHL, associated with ongoing or recent galaxy disruption \citep{2015MNRAS.451.2703C}. In these cases, a circular IHL model will not be the best choice.} \textcolor{black}{In Sec.~\ref{sec:results}, we explore in detail the performance of our galaxy fitting tool \textsc{ProFit} when modelling an intrinsically circular IHL component, and in App.~\ref{sec:non-circular}, an intrinsically elliptical IHL component.}

 %when modelling an intrinsically elliptical IHL component using a non-circular template.}

%The choice of a circular IHL model is further motivated by future research based on} image median stacking of a large sample of galaxy groups and clusters, where the average IHL distribution will approximate a circular shape given the random orientations on the sky.

 \textcolor{black}{The main advantage of the IHL estimation technique} is that we are able to model the IHL flux projected over the BCG and the satellite galaxies, which is \textcolor{black}{excluded} when using a surface brightness threshold method.  \textcolor{black}{Our technique is also} computationally efficient compared to the composite model method, which models the BCG and IHL separately, and \textcolor{black}{is especially efficient compared to} the multi-galaxy fitting method, which models all components in the image, as we are modelling only a single component.

 %On the other hand, the main disadvantage of our technique is the lack of flexibility to model non-circular and non-exponential IHL distributions. Although an exponential profile is widely used in the literature to fit the diffuse IHL component, some systems, especially the unrelaxed ones, might exhibit the presence of significant substructures originated by disrupting or recently disrupted galaxies \citep{2015MNRAS.451.2703C}. In these cases, a circular IHL model will not be the best choice. \textcolor{mycolor}{However, the purpose of this simple technique is to establish a scenario in which we can analyse the effect of the PSF on the IHL component. For this reason, in Section \ref{sec:Data}, we create mock observations of galaxy groups with an analytical circular IHL component. This approach allows us to minimise fitting-related complications and attribute any remaining deviations to PSF effects.} In addition, \textsc{ProFit} can also fit elliptical exponential IHL components, where the reliability of these fits depends on the signal-to-noise ratio (S/N) of the data and the complexity of the model. In Sec.~\ref{caveats results}, we explore in detail the impact of using a circular versus an elliptical model to fit an intrinsically elliptical IHL component with \textsc{ProFit}.

%CORTE MANUALMENTE EL PLOT IHL_pipeline.png para que no queden bordes grandes y ver lo d width .8

%This might be expected, as more relaxed galaxy clusters have had more time to disrupt the accreted galaxies and add their stellar content to the BCG and/or the ICL. Note, however, that there is scatter in this relation so clusters that are currently assembling with large BCG+ICL fractions and vice versa also exist.

%resto de parametros?

%We use \textsc{ProFit} as a 2-D galaxy fitting tool where we select a Sérsic template with fixed Sérsic index n equal to 1, while the magnitude and effective radius are left as free parameters and optimized using a Markov Chain Monte Carlo (MCMC) algorithm. 

\section{Data: HSC-PDR3 Wide mock observations of GAMA groups}
\label{sec:Data}
\textcolor{black}{To test our methodology for measuring the IHL component corrected for PSF flux, we generate a dataset of 5440 \textit{r}-band Hyper Suprime-Cam Subaru Strategic Program (HSC-SSP; \citealt{2018PASJ...70S...8A}) Wide PDR3 mock images based on real groups and clusters. \textcolor{black}{As the goal of this work is} to quantify the impact of the PSF \textcolor{black}{on the IHL component}, we establish the ideal scenario where our dataset of mock group images contains an injected circular exponential IHL component, which is then fitted with a circular exponential template using our IHL estimation technique.} \textcolor{black}{ This allows us to isolate the effects of the PSF from modelling uncertainties associated with profile mismatch.}

\textcolor{black}{We use real HSC data to replicate the real spatial configuration of both galaxy group members and \textcolor{black}{foreground and background} galaxies \textcolor{black}{in our mock dataset}. However, we use analytical models to inject the stars (following their real spatial configuration) and the IHL component of the groups in the mocks.} \textcolor{black}{This combination between real data for galaxies and analytical models for both stars and IHL component allows us to recreate realistic group configurations while having full control over the IHL and} \textcolor{black}{over the stellar contribution to the PSF-scattered light.} \textcolor{black}{For stars, using HSC star segments would be problematic for bright objects, as PSF effects in these segments are too strong and the segmentation maps can become fragmented, resulting in unreliable source identifications.}

\textcolor{black}{To select the sample of real groups and clusters used to create the HSC mock dataset we make use \textcolor{black}{of the Galaxy And Mass Assembly (GAMA, \citealt{2011MNRAS.413..971D}) survey, which overlaps with the HSC-SSP survey}. In particular, we use the GAMA Galaxy Group Catalogue (G$^3$Cv10; \citealt{2011MNRAS.416.2640R}),} \textcolor{black}{which gives us information about the GAMA groups and allows us to restrict our group selection to systems with ten or more members (\textcolor{black}{$N_{\mathrm{fof}}$} $\geq 10$),} \textcolor{black}{as groups below this threshold are less reliably identified in the G$^3$Cv10 catalogue.} \textcolor{black}{During the mock generation process, we also make use of the Wide Area VISTA Extragalactic Survey (WAVES, \citealt{2019Msngr.175...46D}),} \textcolor{black}{which is also overlapped with HSC-SSP.} \textcolor{black}{In particular, we use the extracted photometry from WAVES to classify the different sources in the HSC data \textcolor{black}{cutouts of the GAMA selected groups,} enabling us to separate the galaxies from the stars.}

%These groups are located using the coordinates from the Galaxy And Mass Assembly (GAMA, \citealt{2011MNRAS.413..971D}) Galaxy Group Catalogue (G$^3$Cv10; \citealt{2011MNRAS.416.2640R}). In addition, during the mock generation process, we also use the photometry from the Wide Area VISTA Extragalactic Survey (WAVES, \citealt{2019Msngr.175...46D}) to classify the different sources in the real HSC data which will allow us t

 HSC is a state-of-the-art camera that reaches deep optical imaging with a wide FoV and \textcolor{black}{features a well-characterised PSF with relatively limited extended wings, reducing the level of scattered light.} \textcolor{black}{\citep{2015ApJ...807...22M,2018PASJ...70S...1M, 2018PASJ...70S...2K,2018PASJ...70...66K}}. Attached at the prime focus of the 8.2-m Subaru Telescope (Mauna Kea, Hawaii) and developed by the National Astronomical Observatory of Japan, HSC covers a FoV of \textcolor{black}{$\sim$} 1.7 deg$^2$ with a pixel scale of 0.168 arcsec. The HSC team carried out the HSC-SSP, a three-layered survey (Wide, Deep \& Ultradeep) with a coverage of 1400 deg$^2$ in five different broad-bands (\textit{griZY}), as well as four narrow filters. Due to its depth and coverage, HSC-SSP has been used in recent studies to analyse the IHL component in galaxy groups and clusters \citep{2021MNRAS.502.2419F,2023MNRAS.518.1195M,2025ApJ...980..245C}. The most recent public data release of HSC-SSP, PDR3, features increased sky coverage reaching the \textcolor{black}{nominal target depths of the survey} across all five filters\textcolor{black}{: $\textit{r}_{\mathrm{mag}}$ $\le$ 26 for the WIDE layer, $\textit{r}_{\mathrm{mag}}$ $\le$ 27 for the Deep layer, and $\textit{r}_{\mathrm{mag}}$ $\le$ 28 for the UltraDeep layer.} \textcolor{black}{These survey depths are defined as the 5$\sigma$ PSF magnitude limit for isolated point sources, with photometric errors determined by the HSC imaging pipeline \citep{2018PASJ...70S...5B}.} Additionally, the implementation of a new global sky subtraction algorithm has improved the overall data quality \textcolor{black}{by reducing the oversubtraction of extended LSB emission, thereby better preserving the extended wings of the PSF.} For these reasons, we use the HSC-SSP PDR3 data to generate our dataset and evaluate the performance of our IHL detection technique.

%To identify the galaxy group members and generate the HSC group cutouts, we use the coordinates of the galaxy members from the GAMA Galaxy Group Catalogue G$^3$Cv10

\textcolor{black}{GAMA is a wide area spectroscopic programme conducted with the Anglo-Australian Telescope (AAT), targeting five fields (G02, G09, G12, G15, and G23), and covering 330542 galaxies over a total sky area of 230 deg$^2$. This survey is complemented by panchromatic imaging across 20 bands, spanning the far-infrared to the far-ultraviolet. GAMA has been conducted in three phases: GAMA I \citep{2010MNRAS.404...86B}, GAMA II \citep{2015MNRAS.452.2087L} and GAMA III \citep{2020MNRAS.496.3235B,2022MNRAS.513..439D}. For the purpose of this work, we make use of GAMA II, which has a spectroscopic redshift completeness of 98 \% above a magnitude limit of $\textit{r}_{\mathrm{mag}}$ $\le$ 19.8, and its data products were released in the third data release (DR3\footnote{\url{https://www.gama-survey.org/dr3/schema/}}). \textcolor{black}{As mentioned before,} we make use of the GAMA Galaxy Group Catalogue G$^3$Cv10, which provides information about the spectroscopically confirmed group members}. \textcolor{black}{In particular, G$^3$Cv10 provides the coordinates used to perform the HSC cutouts around the selected groups. We additionally use the G$^3$Cv10 information on the total flux of each group and the halo mass to inject the analytical IHL component into each mock group image.}

\begin{figure*}
\centering  %\qquad
 \includegraphics[width=.9\linewidth]{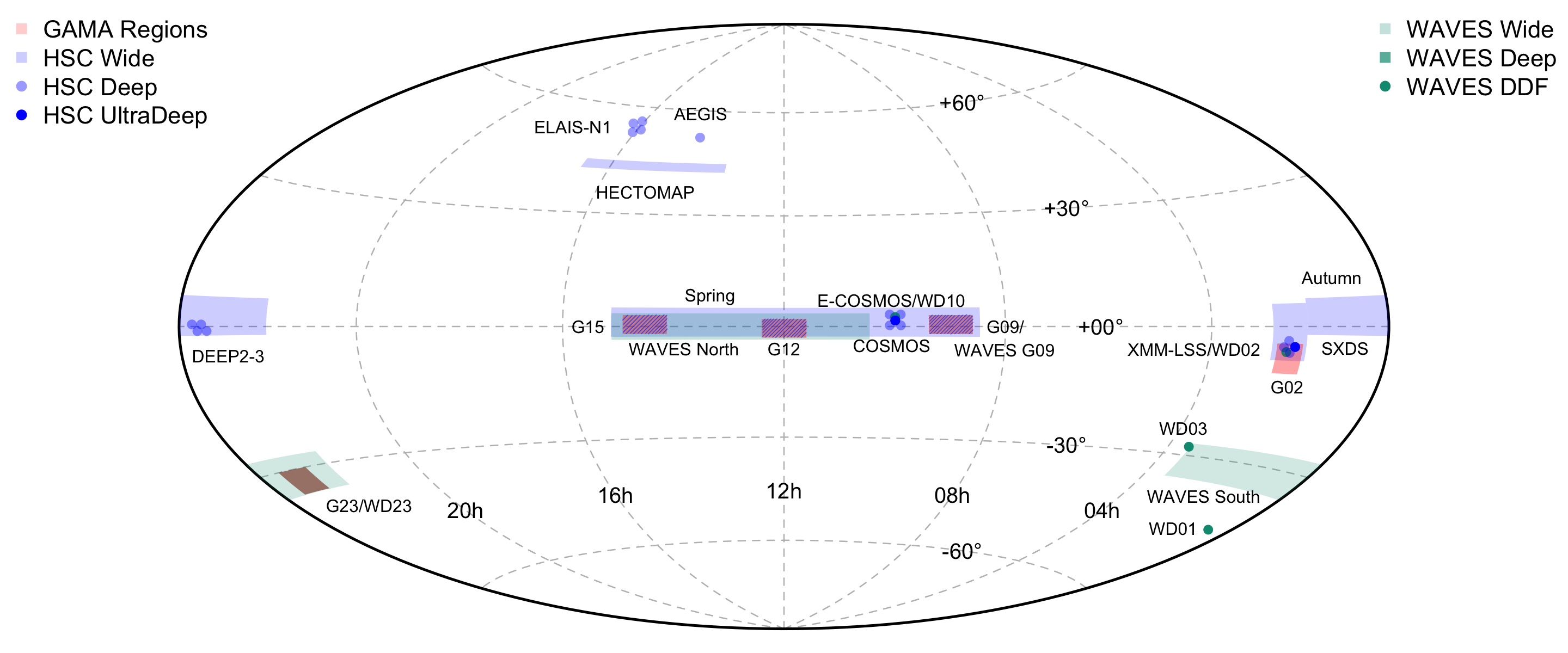}
    \caption{\textcolor{black}{Celestial map showing the \textcolor{black}{fields where} the 544 groups and clusters used to generate our HSC-SSP PDR3 mock observations \textcolor{black}{are located}, corresponding to the overlap region between HSC, GAMA, and WAVES (G09, G12, and G15, \textcolor{black}{hatched regions}). Our mocks are based on the real spatial configuration and data of these groups. During the mock generation process, we use HSC data to model the galaxies, GAMA to provide the group information, and WAVES to identify the sources in the HSC data.}}
    \label{Map}
\end{figure*}

\textcolor{black}{WAVES is a deep spectroscopic survey conducted with the VISTA telescope, complemented by high-quality photometric data, providing a comprehensive view of galaxy properties over wide areas.} \textcolor{black}{As WAVES overlaps with both HSC-SSP and GAMA (see \textcolor{black}{the hatched regions in} Fig.~\ref{Map}), we take advantage of the extracted photometry from WAVES\footnote{\textcolor{black}{WAVES photometry data has already been observed and processed (Bellstedt et al. in prep), whereas the spectroscopic observations are planned to be carried out in 2026.}} to identify the galaxies in the real HSC data and generate the HSC mocks. Using the WAVES photometry allows us to avoid the time-consuming process of identifying the sources in the HSC data, a necessary step in the way we construct our mock observations. The source identification process on WAVES data was carried out in a similar way to that used for GAMA (\citealt{2020MNRAS.496.3235B}), but WAVES reaches a fainter \textcolor{black}{photometric} limit of $\textit{Z}_{\mathrm{mag}}$ $\le$ \textcolor{black}{23}, making it more appropriate than GAMA for this purpose.}\footnote{\textcolor{black}{As a comparison, HSC-SSP reaches its full depth of $\sim$ 26 mag (at 5$\sigma$) in all five bands of the Wide layer.}}

\subsection{Mock generation process}

\textcolor{black}{There are 544 groups \textcolor{black}{with ten or more members} identified by G$^3$Cv10 in the overlapping area between HSC-SSP, GAMA, and WAVES (i.e., G09\footnote{Formally, the G09 field is not part of the WAVES-Wide survey footprint, but photometry in G09 was also generated following the same procedures as for the WAVES regions to support photo-\textit{z} training. As a result, WAVES photometric data are also available for the G09 GAMA field.}, G12 \& G15). Our dataset of mock observations replicates the real configuration of these 544 groups using \textcolor{black}{\textit{r}-band} HSC real data. That is, we first create 1 $\times$ 1 Mpc${^2}$ HSC-PDR3 Wide cutouts centred\footnote{The group centre is defined using the iterative method described in Section 4.2.1 of \citet{2011MNRAS.416.2640R}.} on the groups using the group coordinates from G$^3$Cv10. In parallel, we create 1 $\times$ 1 Mpc${^2}$ WAVES cutouts centred on the same GAMA groups. \textcolor{black}{As we have the star/galaxy separation from the WAVES photometry,} we mask all pixels in each WAVES cutout except for the galaxies (both galaxy group members and \textcolor{black}{foreground and background} galaxies). We then warp them into the HSC cutouts, which is equivalent to geometrically transform the WAVES cutouts to match the coordinate system, pixel scale, and orientation of the HSC cutouts. At this stage, we have 544 HSC-PDR3 cutouts of GAMA groups with real HSC data inside the galaxy segments and zero-valued pixels outside them. To recreate the real Milky Way stellar foreground in our HSC mock cutouts, we manually add the stars as single pixels using their real magnitudes and coordinates from the WAVES catalogue.} We then convolve the HSC cutouts (with extended segments as galaxies and single pixels as stars) with our \textit{r}-band HSC-SSP PDR3 extended PSF model presented in \citealt{2024MNRAS.531.2517G}. Finally, we add image noise with a Gaussian distribution with 0 mean and standard deviation equal to \textcolor{black}{the median of the derived HSC background root mean square map (skyRMS $\sim$ 0.0498).}

 \textcolor{black}{As we ideally only select the HSC galaxy-light pixels delimited by the segments to generate the galaxies in our mocks, our dataset of 544 HSC–GAMA mock groups does not have an IHL component by construction}. This is important as the next step involves manually adding an \textcolor{black}{analytical} IHL component following a circular exponential profile (further evidence for our choice of an exponential model is provided in Appendix~\ref{C}). \textcolor{black}{This exponential profile is the same analytical model we use in our IHL estimation technique to fit this component.} For each of the 544 mock groups, we generate ten versions with identical GAMA group configurations but different injected IHL components, where each version has a different \textcolor{black}{$f_{\mathrm{IHL}}$} value selected from the following set: 0.01, 0.02, 0.03, 0.04, 0.05, 0.1, 0.2, 0.3, 0.4, and 0.5. \textcolor{black}{With this wide range, we explore IHL fractions spanning values reported in the literature from observations of galaxy groups and clusters at $z$$\sim$ 0, reflecting both differences between measurement techniques and intrinsic variations between haloes (e.g. \citealt{2022NatAs...6..308M}). We sample a discrete set of values across this interval without favouring any particular value, as these are not intended to represent preferred models or specific definitions of the IHL, but rather to uniformly sample the observationally motivated range of $f_{\mathrm{IHL}}$.} 
 
 Considering the ten different versions of each of the 544 different groups, our final dataset consists of 5440 HSC-GAMA mock groups. 
%This range of IHL fractions is found in observations of galaxy groups and clusters at \textcolor{black}{$z$}$\sim$ 0\textcolor{black}{, reflecting both differences between measurement techniques and intrinsic variations between haloes} (e.g. \citealt{2022NatAs...6..308M})

\textcolor{black}{To inject the IHL component in our set of HSC-GAMA mocks we use the 2-D galaxy fitting tool \textsc{ProFit}\footnote{\url{https://github.com/asgr/ProFit}}, which offers flexible modelling with built-in profiles (Sérsic, Moffat, exponential, etc.), support for multiple components, various likelihood and optimisation methods -- including Markov Chain Monte Carlo (MCMC) and genetic algorithms -- fitting in linear or logarithmic space, and the ability to apply complex priors and parameter constraints (e.g.,\citealt{2019MNRAS.490.4060C, 2022MNRAS.516..942C}). We use \textsc{ProFit} in the 2D image-generation mode, where all the template parameters are fixed.} We select a circular exponential model, characterised by the eight parameters of the Sérsic template: \textcolor{black}{$x_{\mathrm{cen}}$}, \textcolor{black}{$y_{\mathrm{cen}}$}, $\textit{n}$, $\textit{$\theta$}$, \textcolor{black}{$A_{\mathrm{rat}}$}, $\textit{B}$, $\textit{mag}$, and \textcolor{black}{$R_{\mathrm{eff}}$}. We set \textcolor{black}{$x_{\mathrm{cen}}$} and \textcolor{black}{$y_{\mathrm{cen}}$} to the centre of the mock, $\textit{n}$ = 1, $\textit{$\theta$}$ = 0, \textcolor{black}{$A_{\mathrm{rat}}$} = 1 and $\textit{B}$ = 0. \textcolor{black}{To estimate the $\textit{mag}$ and \textcolor{black}{$R_{\mathrm{eff}}$} parameters of the IHL model, we use the real information of the selected group \textcolor{black}{obtained} from G$^3$Cv10.} To estimate $\textit{mag}$, we first calculate the flux in the IHL component (\textcolor{black}{$F_{\mathrm{IHL}}$}) with the \textcolor{black}{following} equation:

%(also applied in the Step \textcolor{mycolor}{vii} of Sec.~\ref{sec:IHL analysis})

\begin{equation}  
\label{ecuacion}
\\
\\
\\
\textcolor{black}{F_{\mathrm{IHL}}} = \textcolor{black}{F_{\mathrm{group}}} \frac{\textcolor{black}{f_{\mathrm{IHL}}}}{1-\textcolor{black}{f_{\mathrm{IHL}}}}, \\
\end{equation}
\\
\noindent where \textcolor{black}{$F_{\mathrm{group}}$} is the total group light calculated by summing the fluxes from the confirmed members using G$^3$Cv10, and \textcolor{black}{$f_{\mathrm{IHL}}$} is the desired IHL fraction \textcolor{black}{assigned to the model}, ranging from 0.01 to 0.5. \textcolor{black}{To estimate \textcolor{black}{$R_{\mathrm{eff}}$}, we use the relations presented in \citealt{2024MNRAS.527.2624P}. Their figure 10 shows a correlation between \textcolor{black}{$M_{\mathrm{200}}$} and the IHL transition radius \textcolor{black}{$R_{\mathrm{IHL}}$}\footnote{The authors found this relation using Gaussian mixture methods to decompose the stellar halo of galaxies in the flagship (100 Mpc)$^3$ \textsc{Eagle} simulation into three components (i.e. a disc, a bulge, and IHL) according to their kinematic properties. \textcolor{black}{$M_{\mathrm{200}}$} is defined as the total mass within the spherical radius \textcolor{black}{$r_{\mathrm{200}}$} that encloses a mean density of 200 $\times$ $\mathrm{\rho_{crit}}$, where the critical density is given by $\rho_\mathrm{{crit}} = 3 H(z)^2/8\pi G $, $H(z)$ is the Hubble parameter and $G$ is the gravitational constant.}, which characterises the radial scale at which the IHL begins to dominate the stellar mass of a galaxy. By using the \textcolor{black}{$M_{\mathrm{200}}$} value of the group from G$^3$Cv10, we first infer \textcolor{black}{$R_{\mathrm{IHL}}$} from this relation. We then compare these values to the stellar mass–size relation shown in their figure 9, which links stellar mass \textcolor{black}{$M_{\mathrm{\star}}$} to an effective radius. By mapping halo mass to stellar mass and comparing the corresponding radii in both relations, we find that the IHL transition radius and the effective radius are of similar magnitude over the relevant mass range. Based on this consistency, we approximate the effective radius of the IHL Sérsic model with \textcolor{black}{the IHL transition radius}. For example, for our most massive haloes (\textcolor{black}{$M_{\mathrm{halo}}$} $\sim 10^{15}~\mathrm{M}_\odot$, corresponding to \textcolor{black}{$M_{\mathrm{\star}}$}  $\sim 10^{12.5}~\mathrm{M}_\odot$), both relations yield characteristic radii of $R$ $\sim$ 300 kpc. Although the IHL transition radius and the effective radius are not strictly equivalent, this approximation is sufficient for our purposes, as our focus is on assessing the impact of the PSF on the IHL component rather than modelling its structure perfectly.}

After estimating the effective radius and magnitude for each \textcolor{black}{combination of} GAMA group configuration and IHL fraction, we inject the corresponding \textcolor{black}{analytical} IHL component into each mock group. \textcolor{black}{By definition, a distinct value of $R_{\mathrm{eff}}$ is estimated for each of the 544 different GAMA group configurations,} while $\textit{mag}$ changes depending on both the specific group and the selected \textcolor{black}{$f_{\mathrm{IHL}}$}.

%By definition, \textcolor{black}{$R_{\mathrm{eff}}$} is the same for each of the 544 different GAMA group configurations,

%Finally, we discard mock groups where the distance to the most distant member of the group is smaller than half the size of the box (500 Kpc), or where the size of the box exceeds 5000 pixels. With the first restriction, we exclude groups that are too compact and the IHL detection may be challenging, while the second restriction avoids large images that could make our IHL fitting code computationally expensive. 

%This results in a dataset of 544 HSC-SSP Wide PDR3 GAMA groups, with 10 mock images created for each group configuration. Each mock image includes a circular exponential IHL component with \textcolor{black}{$f_{\mathrm{IHL}}$} ranging from 0.01 to 0.5. Considering all versions of each group configuration, our final mock dataset consists of 5440 images. 

\begin{figure*}
\centering  %\qquad
 \includegraphics[width=1\linewidth]{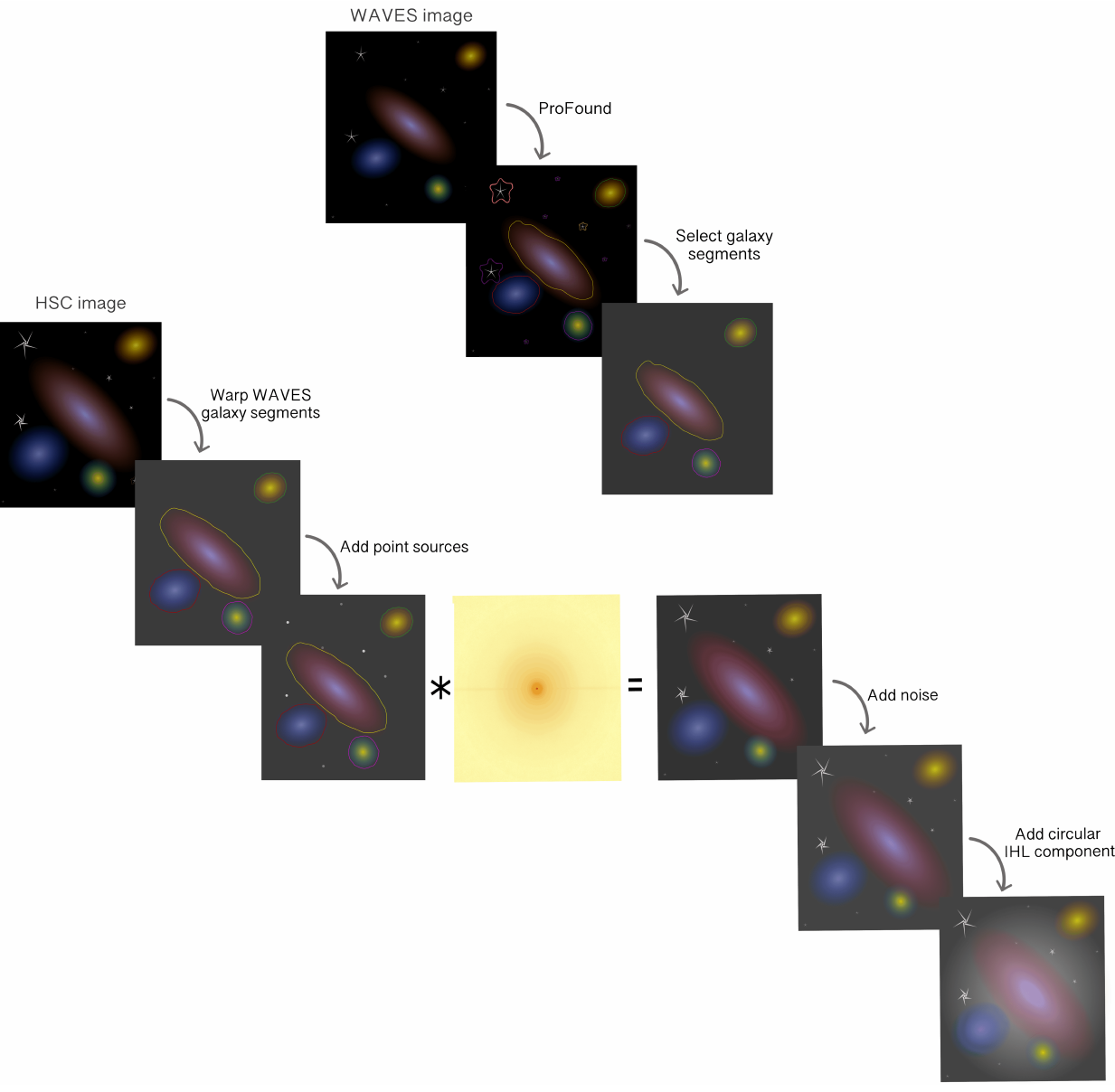}
    \caption{Schematic view of the processing steps outlined in Sec.~\ref{sec:Data} to construct our set of HSC-SSP PDR3 mock observations.}
    \label{mock generation}
\end{figure*}

This way, our set of 5440 HSC-SSP Wide PDR3 mock observations of GAMA groups incorporates a circular exponential IHL component with known characteristics, which we will later estimate in Sec.~\ref{sec:results}. A schematic view of the construction process of our HSC-GAMA mock dataset is shown in Fig.~\ref{mock generation}.

%as well as the realistic stellar foreground—features that are not achievable with hydrodynamical simulations. While hydrodynamical simulations are useful, they are limited by stellar mass resolution and often fail to accurately reproduce low surface brightness structures.}

%Celestial map highlighting the surveys that were involved in the process to generate our datset of HSC-SSP PDR3 mock observations: GAMA, WAVES, and HSC-SSP PDR3.

%with N$_{\mathrm{{fof}}}$ $\geq$ 10

\begin{figure}
\centering  %\qquad
 \includegraphics[width=1\linewidth]{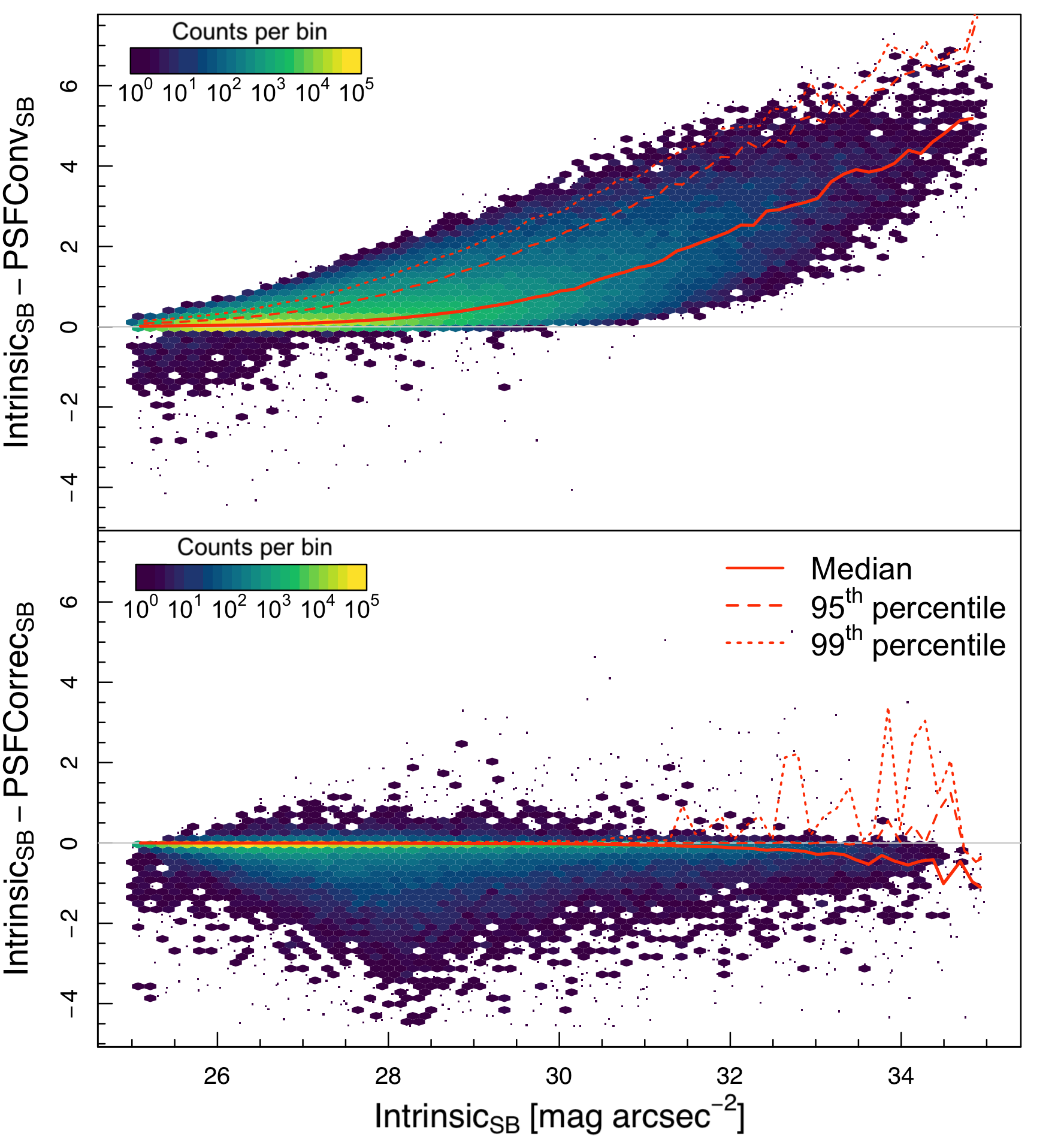}
    \caption{Comparison between the intrinsic (non-PSF-convolved), PSF-convolved, and PSF-corrected versions of the sample of 544 HSC-SSP Wide PDR3 GAMA groups with \textcolor{black}{$f_{\mathrm{IHL}}$} = 0.1. All sources in the different versions of the mock images have been masked to emphasise the impact of the PSF correction on the fainter pixels. The upper panel shows the difference of SB between \textcolor{black}{each pixel of} each intrinsic mock image and its PSF-convolved version as a function of SB. The lower panel shows the difference of SB between \textcolor{black}{each pixel of} each PSF-convolved mock image and its PSF-corrected version as a function of SB. The orange solid line indicates the median of the distribution and the two orange dashed lines indicate the 95$^{\mathrm{th}}$ and 99$^{\mathrm{th}}$ percentiles. Fainter pixels are the \textcolor{black}{ones} most affected by the PSF convolution. However, after applying our PSF correction, the majority of this extra flux is successfully removed, bringing the median of the lower panel closer to the zero line. The lower panel also shows an excess of pixels with $\Delta_{\mathrm{SB}} < 0$, which reflects the slight overestimation of the PSF-scattered flux we are aware of.}
    \label{SB grid}
\end{figure}

In Fig.~\ref{SB grid} we show the comparison between the intrinsic (non-PSF-convolved), the PSF-convolved and the PSF-corrected versions of the sample of 544 HSC-SSP Wide PDR3 GAMA groups with \textcolor{black}{$f_{\mathrm{IHL}}$} = 0.1.  All sources have been masked in each version of the mock images to emphasise the impact of the PSF correction on the dimmer pixels. In the upper panel, we show the 2D histogram of the difference in SB \textcolor{black}{per pixel} between each intrinsic mock image and its PSF-convolved version, as a function of the SB of each pixel in the intrinsic image. In the lower panel, we show the 2D histogram of the difference in SB between each PSF-convolved mock image and its PSF-corrected version, again as a function of the SB of each pixel in the intrinsic image. The median of each distribution is indicated with an orange solid line, and the two orange dashed lines indicate the 95$^{\mathrm{th}}$ and 99$^{\mathrm{th}}$ percentiles. \textcolor{black}{By looking at the top panel, the faint end is the most affected by the PSF convolution, where positive pixel values correspond to regions that become brighter due to the PSF convolution, and negative values indicate regions that become fainter. This effect is  mainly driven by} the extended wings of the \textcolor{black}{bright point sources \citep{2024MNRAS.531.2517G}}.  \textcolor{black}{The HSC-SSP PDR3 Wide layer reaches a depth of $\mu_i^{lim}$ $\sim$ 28.5 mag arcsec$^{-2}$ in the \textit{i}-band, indicating that the majority of the affected pixels are brighter than this surface brightness level, where PSF-induced flux redistribution is relatively small compared to the faintest regime.} However, \textcolor{black}{as seen in the lower panel}, we manage to reduce this excess of flux that contaminates the IHL component after applying our PSF-scattered flux correction technique, which successfully brings the median closer to the zero line across all surface brightness values. \textcolor{black}{The cloud of pixels with $\Delta \mathrm{SB} < 0$ at $\mathrm{SB} \gtrsim 32$ mag arcsec$^{-2}$, indicating that these pixels become fainter than their intrinsic surface brightness after the correction, is a consequence of the slight overestimation of the intrinsic flux of each source. As mentioned in Sec.~\ref{cavetas psf}, this overestimation leads to a corresponding overestimate of the PSF-scattered flux to be removed outside the sources after the convolution performed in Step (iv) of the PSF correction technique. Consequently, slightly more flux than the actual scattered contribution is subtracted from the extended outskirts of the source segments, which correspond to LSB pixels (as all sources have been masked in this plot), producing the observed $\Delta \mathrm{SB}<0$ values.}

\subsection{\textcolor{black}{Motivation for our mock-generation method}}

\textcolor{black}{There are several reasons why we use real HSC data to model the galaxies in our mocks rather than relying on hydrodynamical simulations or analytical models. Our choice is primarily a pragmatic, empirical one: using real galaxy cutouts preserves the observed distribution of galaxy morphologies, clustering, interactions, and environmental complexity, while naturally incorporating the observational characteristics of the HSC data. These include the background properties, PSF effects, instrumental artefacts, and the LSB features present in the observations. In contrast, hydrodynamical simulations and their associated mock-image generation frameworks provide a complementary approach, in which these observational effects can be modelled and added to simulated galaxies, but this generally requires additional modelling assumptions and choices. Similarly, analytical models provide useful controlled scenarios but necessarily impose simplified descriptions of galaxy structure and environment. In addition, by injecting stars at their observed positions and magnitudes, we directly reproduce the foreground stellar population present in the HSC data, including its observed spatial distribution and brightness distribution. This is useful for our purpose because stellar contamination can significantly affect the estimation of the IHL component.}

\textcolor{black}{However, we model both the stars and the IHL component using analytical profiles, which is essential for this study. Since the first step in our methodology for fitting the IHL component is applying our PSF correction technique, injecting the stars as point sources allows us to more effectively remove their extended wings during the PSF correction. This therefore represents an ideal scenario for comparing the resulting fitted IHL with and without the PSF correction applied, under the assumption that in real images, residual flux from imperfectly subtracted stars might have a higher impact.}

\textcolor{black}{In the case of the IHL, injecting this component using an analytical model allows full control over the properties fitted by our IHL estimation technique. This therefore also represents an idealised scenario for comparing the resulting fitted IHL components with and without the PSF correction applied.}

%\textcolor{mycolor}{However, we use analytical models for the stars and for the IHL component, which are crucial for this study. In the case of the stars, injecting them as point sources allows us to more effectively remove their extended wings during the PSF correction. This therefore represents an ideal scenario for comparing the resulting fitted IHL with and without the PSF correction applied, under the assumption that in real images, residual flux from imperfectly subtracted stars might have a higher impact.}

%\textcolor{mycolor}{In the case of the IHL, our IHL estimation technique fits this component with a circular exponential model, which is the same analytical model injected during the mock generation process. The goal behind this ideal scenario is to compare the fitted IHL with and without the PSF correction applied, and attributing the differences between the results mainly to the PSF scattering effect. While a circular exponential IHL model represents the simplest scenario to measure the PSF effect, we acknowledge that, in reality, the IHL distribution is more complex, and the impact of the PSF is likely to be even stronger than in our idealised case.}

\subsection{\textcolor{black}{Caveats}}
\label{caveats mocks}
\textcolor{black}{We acknowledge that, by using the real HSC galaxy pixels to model the galaxies in our mock observations, these pixels are already slightly blurred due to PSF scattering, and that our additional PSF convolution further spreads this flux.} \textcolor{black}{This introduces a small systematic over-convolution in our mock observations and therefore represents a limitation of our validation tests.} \textcolor{black}{However, we prefer this approach, as mentioned above, because using real observational data preserves the observed diversity of galaxy morphologies, SB profiles, and clustering properties without requiring additional modelling assumptions.}

%which effectively captures the diversity of morphologies, SB profiles, and clustering properties present in the observed universe. In contrast, analytical Sérsic models or \textcolor{black}{hydrodynamical} simulations do not reproduce these features with the same level of realism.} 

\textcolor{black}{We also acknowledge that using the real HSC galaxy pixels of the group members to construct the mocks inevitably includes IHL light overlapping with the BCG. However, this diffuse component is expected to dominate primarily in the outskirts of the system \textcolor{black}{(with the precise radius and contribution depending on the mass of the host halo)}, and our approximation of neglecting this light in the galaxy pixels is reasonable. \textcolor{black}{As explained above, the main IHL component is implemented in a later step using an analytical model with controlled parameters, allowing direct comparison between the injected and recovered IHL properties.}}

%As explained above, the full implementation of the IHL component is performed in a later step, where it is analytically modeled with all properties explicitly defined.

\section{Results}
\label{sec:results}
In this section, we test both the PSF-scattered flux removal technique (Sec.~\ref{sec:PSF analysis}) and the IHL estimation technique (Sec.~\ref{sec:IHL analysis}) on \textcolor{black}{our dataset of} HSC-SSP PDR3 Wide mock images of GAMA groups.

\textcolor{black}{While both techniques can be implemented using different photometric tools, we adopt \textsc{ProFound}\footnote{\url{https://github.com/asgr/ProFound}}\citep{2018MNRAS.476.3137R} as our source extraction code and \textsc{ProFit} (previously described in Section \ref{sec:Data}) as our 2D galaxy-profiling tool to model the IHL.} \textcolor{black}{\textsc{ProFound}} \textcolor{black}{uses irregular apertures that adapt to the observed morphology of each source, rather than assuming simplified analytical or elliptical apertures. This approach is particularly useful for extended or irregular LSB structures, where simplified aperture definitions may fail to capture the full extent of the emission.} It also estimates pseudo-total fluxes by gradually dilating segments until flux convergence by employing a watershed deblending method that prevents flux overlap between sources. \textsc{ProFound} has been used to make the multiband catalogs for the GAMA Survey \citep{2020MNRAS.496.3235B} and the Deep Extragalactic VIsible Legacy Survey (DEVILS; \citealt{2021MNRAS.506..256D}). The \textsc{ProFound} input parameters used to detect the sources in the HSC-PDR3 mock images are described in Table \ref{tab:image_params}. In Step ii of the PSF correction technique, we use the \texttt{profoundMakeSegimDilate} function to dilate the segments. \textcolor{black}{This depends on the \textit{size} parameter, which defines the size of the dilation kernel in pixels, set to \textit{size} = 21 in this case.} \textcolor{black}{For the IHL modelling, \textsc{ProFit} uses \textsc{Highlander} as the optimisation algorithm to determine the best-fitting parameters. This combines a genetic algorithm and MCMC to explore local extrema and identify the global maximum (e.g., \citealt{2021MNRAS.505..540T}).}

%As our 2D galaxy-profiling tool to model the IHL, we use \textsc{ProFit} (previoulsy described in Section \ref{sec:Data}), with \textsc{Highlander} as the optimisation algorithm to find the best parameters of the fit. This software combines a genetic algorithm and MCMC to explore local extrema and identify the global maximum (e.g., \citealt{2021MNRAS.505..540T}).}

%\textcolor{black}{In order to be as general as possible, we do not mention any specific photometric tool in the description of both techniques. We now explicitly state the used tools applied to test both techniques. As our source extraction code, we use \textsc{ProFound}\footnote{\url{https://github.com/asgr/ProFound}}\citep{2018MNRAS.476.3137R}.}

\begin{table}
\centering
\begin{tabular}{lcccccc}

\textcolor{black}{\textit{mag}$_\mathrm{zero}$} & \textit{pixscale} & \textit{skyRMS} & \textit{box} & \textit{skycut} & \textit{roughpedestal}\\
\hline
~~~~27.0 & 0.168 & 0.05 & 300 & 1 & TRUE\\

\end{tabular}
\caption{\textcolor{black}{\textsc{ProFound} input parameters used to detect the sources in the HSC mock images. \textcolor{black}{\textit{mag}$_\mathrm{zero}$} and \textit{pixscale} are the photometric zero point and pixel scale of HSC-PDR3 respectively, while \textit{skyRMS} is the median of the derived HSC background root mean square map \textcolor{black}{and is expressed in ADU, which are the units of the HSC mock images.} \textit{box} sets the size (in pixels) of the local background estimation box used by \textsc{ProFound}, and \textit{skycut} defines the detection threshold in units of \textit{skyRMS}. Finally, \textit{roughpedestal} = TRUE forces the initial sky estimate to use a single median pedestal value, helping prevent over-subtraction in images dominated by large objects.}}
\label{tab:image_params}
\end{table}

\textcolor{black}{We use our IHL estimation technique, which assumes an exponential model, but in this case we use circular symmetry to fit the IHL component of the HSC-PDR3 mock images. Our motivation for adopting a circular model in this case is that, as described in Section~\ref{sec:Data}, the mock galaxy groups include an analytically injected circular IHL component. Therefore, modelling them with a circular profile establishes an idealised scenario in which differences between the recovered IHL models can be attributed to PSF effects when comparing results with and without the PSF correction applied. This approach allows us to minimise fitting-related complications and attribute any remaining deviations to PSF effects\footnote{\textcolor{black}{We note that PSF-scattered light from neighbouring sources, as well as asymmetries in the PSF itself, can introduce non-circular features into the IHL distribution. These morphology-related effects are therefore not captured by the circular model adopted here.}}. \textcolor{black}{In this case, only two parameters of the circular exponential IHL model are left free during the optimisation: magnitude ($\textit{mag}$) and effective radius ($R_{\mathrm{eff}}$), as the major-axis angle ($\theta$) is fixed to 0$^\circ$ and the minor-to-major axis ratio ($A_{\mathrm{rat}}$) is fixed to 1}. \textcolor{black}{We impose bounds on these two parameters, restricting the magnitude to 12 mag $<$ $\textit{mag}$ $<$ 25 mag, and the effective radius to 20 kpc $<$ \textcolor{black}{$R_{\mathrm{eff}}$} $<$ 2000 kpc. These limits are selected in order to be sufficiently broad to avoid artificially constraining the fit, while excluding clearly unphysical solutions.}}

%Our motivation for modelling

We first focus on the IHL estimation technique to demonstrate the flexibility of \textsc{ProFit} to robustly fit the IHL profile with automated Bayesian modelling on a particular mock group image in Sec.~\ref{case-study}. We then test our PSF-scattered flux removal technique on our full dataset of 5440 mock group images in Sec.~\ref{full dataset}. %We finally comment on the caveats of both techniques in Sec.~\ref{caveats results}.

\subsection{\textcolor{black}{IHL profile fitting with \textsc{ProFit}}}
\label{case-study}
In this section, we show the results of applying our IHL estimation technique on a particular HSC-GAMA mock group image to demonstrate its robustness. This mock image was generated using the G$^3$Cv10 catalogue information of the GAMA group identified by ID G200195, with a manually injected exponential circular IHL component of \textcolor{black}{$f_{\mathrm{IHL}}$} = 0.5. \textcolor{black}{G200195 is a group of 10 observed members, with halo mass $\mathrm{M_{halo}}$ $\sim$  1.7 $\times 10^{14}\,\mathrm{M_\odot}$ and redshift $z$ $\sim$ 0.26.} \textcolor{black}{The image used in this example has not been PSF-corrected, as the purpose of this section is to present an example in which PSF-related uncertainties are not dominant, allowing us to demonstrate the performance of the fitting procedure under conditions where the recovered IHL component can be robustly trusted.} 

%G200195 is a group of 10 observed members, with halo mass $\mathrm{M_{halo}}$ $\sim$  1.7 $\times 10^{14}\,\mathrm{M_\odot}$ and redshift $z$ $\sim$ 0.26. 

\begin{figure*}
\centering  %\qquad
 \includegraphics[width=1\linewidth]{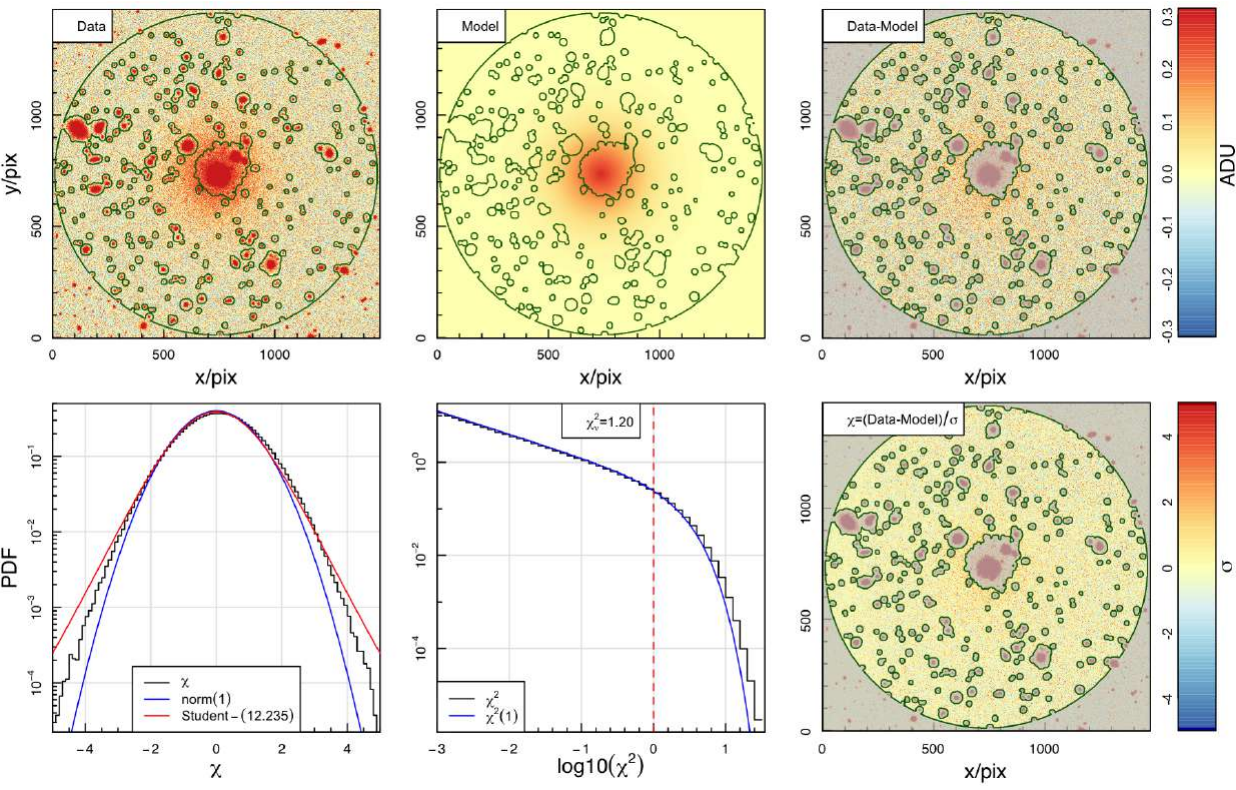}
    \caption{Example of a galaxy group mock fitted with \textsc{ProFit}. Top panels show the mock image data (left), the \textsc{ProFit} IHL model (middle) and the result of subtracting the model from the data (right).  It is worth highlighting that the fitting region comprises all pixels located outside the green mask boundaries that enclose the sources. \textcolor{black}{The bottom-left and bottom-middle panels show the histogram distributions of the residuals in units of $\sigma$ ($\chi$), constructed using the unmasked pixels and compared with different statistical distributions. In the bottom-left panel, $\chi$ is compared with a standard Normal distribution (blue line) and with a Student-T distribution (red line), where all curves are expected to overlap if the residuals were equal to 0.  Deviations from this expected shape —such as skewness, heavy tails, or a shifted mean— indicate potential issues with the model fit or with the error estimation. In this case, the deviation is expected given the observed residuals in the top-right panel. \textcolor{black}{These positive residuals around the central regions mainly arise from limitations of the assumed parametric IHL model in reproducing the complex light distribution of the inner regions. In particular, the central regions have higher surface brightness and may contain departures from the smooth parametric profile adopted here, resulting in local differences between the model and the mock data. The masking procedure may also contribute to minor residual structure near bright sources. These residuals are spatially confined to the central regions and do not affect the LSB outskirts relevant for the IHL measurement.} We focus on the positive tail of the $\chi$ distribution in the bottom-middle panel, where the black line shows the $\chi^2$ distribution of the data and the blue line represents the analytic expectation for a $\chi^2$ distribution with one degree of freedom. As before, the two curves are expected to overlap if the residuals were centred on zero. The low value of $\chi^2$ ($\chi^2_{\nu} = 1.2$) indicates that our model fits reasonably well the IHL component, \textcolor{black}{and that the residuals in the central regions do not significantly affect the overall goodness of fit.} The bottom-right panel shows the 2-D residuals expressed in terms of $\sigma$ significance, with no strong evidence for significant tension. For clarity, all galaxy sources that are not included in the fit or in the histogram distributions are masked with a semi-transparent grey overlay in the top-right and bottom-right panels.}}
    \label{image1}
\end{figure*}

\begin{figure}
\centering  %\qquad
 \includegraphics[width=1\linewidth]{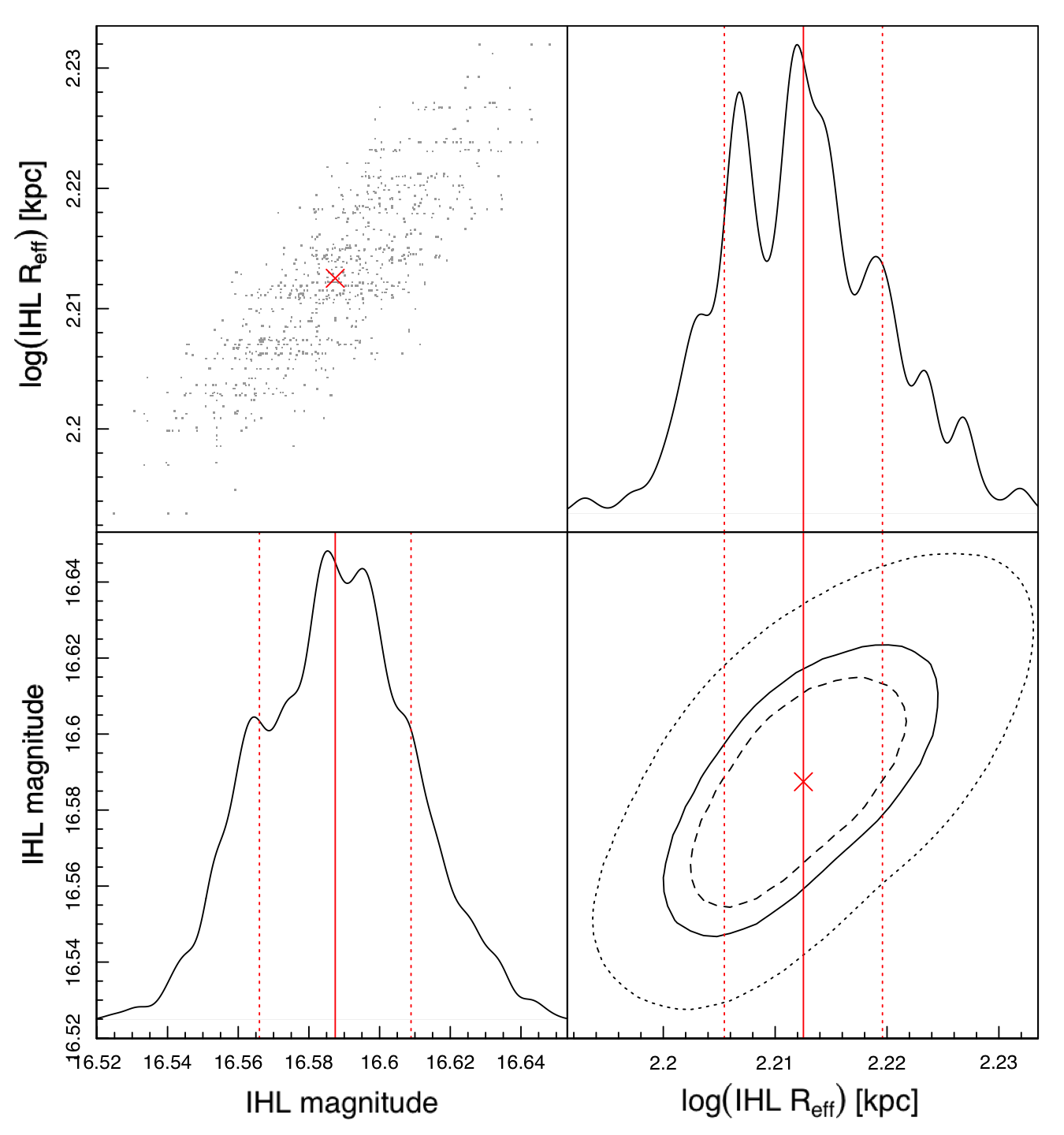}
    \caption{\textcolor{black}{Corner} plot of the stationary MCMC chains for the model fit of \textcolor{black}{a galaxy group mock}. Raw and contoured sample distributions are shown in the 1\textsuperscript{st} and 4\textsuperscript{th} panels respectively (\textcolor{black}{centres are marked by a red cross}), while the other diagonal panels present the marginalised one-dimensional density plots for each fitted parameter \textcolor{black}{(centres are marked by a solid red line and 1-$\sigma$ intervals by dashed red lines)}. \textcolor{black}{A correlation between magnitude and effective radius is observed in some modelling cases, reflecting a parameter degeneracy in which different combinations of these parameters can produce similarly good fits. This behaviour is not consistently present across all cases and therefore does not appear to be an intrinsic property of the Sérsic model. Such parameter correlations do not necessarily indicate a poor fit, provided that the covariance between the parameters is properly accounted for when estimating their uncertainties.}}
    \label{image2}
\end{figure}

By applying the IHL estimation technique to the G200195 mock, we obtain the magnitude and effective radius of the circular exponential template, which are used to later construct the IHL model. In Figs.~\ref{image1} and \ref{image2}, we show the \textsc{ProFit} output plots for this case. In the top panels of Fig.~\ref{image1}, we show the group mock image (left), the IHL model fitted by \textsc{ProFit} (middle), and the result of subtracting the model from the image (right). \textcolor{black}{It is worth highlighting that the fitting region comprises all pixels located outside the green mask boundaries that enclose the sources, which are shown as a mask in the top-right panel.} The bottom panels show the histogram of residuals of the unmasked pixels in units of $\sigma$ compared to a reference Normal and Student-T distribution (left), the $\chi^2$ data distribution compared to a $\chi^2$ analytical distribution with one degree of freedom (middle), and the 2-D residuals expressed in \textcolor{black}{units of $\sigma$.} \textcolor{black}{The histogram distributions would overlap with the analytical distribution in the case of zero residuals. While this is not strictly the case}\textcolor{black}{, the observed deviation from the zero-residual case is expected and remains minor.} \textcolor{black}{The bottom-right panel shows no evidence of significant deviations at the level of the pixel uncertainties (i.e. in units of $\sigma$), indicating that our model provides a reasonable fit to the IHL component of the G200195 mock.} \textcolor{black}{This is reflected in the reduced chi-squared value, $\chi^2_{\nu} = 1.2$, which is slightly above unity, indicating that the model reproduces the data within the expected level of uncertainty.} Fig.~\ref{image2} shows the \textcolor{black}{corner} plot of the stationary MCMC chains for the model fit of the G200195 mock. The top-left panel of the figure shows the raw samples, the bottom-right shows the contoured version of the samples, with dashed/solid/dotted lines containing 50\%$-$68\%$-$95\% of the samples. The diagonal one-dimensional density plots show the marginalised distributions for each of the two fitted parameters. \textcolor{black}{In this particular case, we see a strong correlation between $\textit{mag}$ and \textcolor{black}{$R_{\mathrm{eff}}$}, reflecting a covariance between these parameters in the fit. However, this covariance is not present in all modelling cases, indicating that it arises from the specific constraints of individual fits rather than being an intrinsic property of the Sérsic model. Such parameter correlations do not necessarily indicate a problematic fit, as ProFit's posterior sampling accounts for the covariance when estimating the parameter uncertainties.}

%\textcolor{mycolor}{We observe slightly more scatter in the data than expected from the assumed errors, probably due to a slight underestimation of the background noise, as this effect appears consistently across different mock groups.}

\textcolor{black}{In summary, we demonstrate that the performance of \textsc{ProFit} is robust when fitting the IHL of a mock group with high \textcolor{black}{$f_{\mathrm{IHL}}$}.}

 %digo esto porque si en porfitmakeplots multiplico a sigma por 1.1 en este caso particular se mejora

%ojo que esto no compara con los valroes reales que yo inyecte, esto solo analiza cuan bueno es el ajute, osea profit

 %In the left half of the plot, the first panel shows the mock image of 200043 with no IHL component, whereas in the second panel, we have included an exponential IHL component of \textcolor{black}{$f_{\mathrm{IHL}}$} = 0.5. The third panel shows the exponential model generated using the magnitude and effective radius fitted from the second panel after applying the PSF-correction. In the fourth panel, we have subtracted the IHL fitted model and the sources from the group image (2\textsuperscript{nd} panel - (3\textsuperscript{rd} panel + 1\textsuperscript{st} panel)). We use the same colour scale with fixed upper and lower limits across the first three panels, but the limits have been modified for the forth. The residuals are 

%As mentioned in Sec.\ref{sec:IHL analysis}, using an exponential template to fit the IHL component represents the best-case scenario, as systems with recent accretion history exhibit significant substructures. Following this, the results of our simple tests provide a lower limit for the impact of the PSF-scattered light on IHL measurements. We expect this impact to be higher in more complicated systems, where irregular structures and abrupt variations in surface brightness further amplify the effects of PSF contamination. esto esta mal creo

 %\textcolor{black}{These positive residuals around the central regions mainly arise from limitations of the assumed parametric model in the high surface-brightness inner regions, where PSF convolution effects can amplify small modelling inaccuracies. The masking procedure may also contribute to minor residual structure near bright sources. These residuals are spatially confined and do not affect the LSB outskirts relevant for the IHL measurement.}

    %\textcolor{black}{A strong correlation between magnitude and effective radius is observed in some modelling cases, although this behaviour is not consistently present across all cases. This indicates that the correlation is not intrinsic to the Sérsic model, but can emerge in certain fits due to parameter degeneracies.}

%hablar de profit aplocado a la psf corrected image, como para porcentajes bajos es malaso pero mayores es bueno y la imporntancia de emvoer la scattered light faint que contamina y predice higher fractions of ihl

\subsection{Effect of PSF-scattered light on IHL estimations}
\label{full dataset}
\textcolor{black}{In Section \ref{case-study}, we demonstrate that \textsc{ProFit} performs well in modelling the IHL component of a particular HSC/GAMA group at \textcolor{black}{$f_{\mathrm{IHL}}$}\textcolor{black}{= 0.5}. In this section, we analyse the impact of the PSF-scattered flux on IHL estimations at all \textcolor{black}{$f_{\mathrm{IHL}}$} values by comparing the injected and fitted IHL values for the full dataset of mocks. To do so, we apply our IHL estimation technique to two versions of the same dataset of 5440 HSC–GAMA mock groups: the original dataset, which has not been PSF-corrected \textcolor{black}{and for which a simple masking technique is applied}, and the PSF-corrected version, where the PSF-scattered flux removal technique has been applied.} \textcolor{black}{This way, we analyse how the fitted IHL model changes across the different versions of the same group image as a result of the PSF-correction.}

%, regardless of whether the fitted flux and size closely match the true values used to inject the IHL component. This result is important because our PSF correction slightly oversubtracts the IHL flux, meaning that even when the S/N of the IHL is sufficiently high, the fitted flux will be accurate but will not necessarily match the injected value.

%We apply the IHL estimation technique to both the original and the PSF-corrected datasets. 

\textcolor{black}{As mentioned above, the only two parameters that we fit are the magnitude and the effective radius, as the remaining six parameters required to fully characterise a Sérsic template are the same for both the injected IHL and the fitted IHL model. Thus, for each group image in each dataset, we \textcolor{black}{have} the injected $mag$ and $R_\mathrm{eff}$ and, after applying the IHL estimation technique to each image, we \textcolor{black}{obtain} the fitted $mag$ and $R_\mathrm{eff}$.}

\textcolor{black}{The injected $mag$ and $R_\mathrm{eff}$ are identical in both versions of each group image (original and PSF-corrected), but the recovered parameters differ. This is because, after applying our PSF correction, we aim to remove the scattered flux that contaminates the IHL component, thereby modifying the flux distribution used by \textsc{ProFit} to model the IHL and resulting in different recovered values. As demonstrated in Sec.~\ref{case-study}, \textsc{ProFit} recovers the flux of the IHL component with good accuracy at large $f_\mathrm{IHL}$. However, because our PSF correction slightly oversubtracts the IHL, the fitted flux in the PSF-corrected image does not necessarily match the flux originally injected into the mock, regardless of the IHL fraction. Instead, it matches the PSF-corrected IHL component, i.e. the quantity actually present in the image.}

\begin{figure*}
\centering  %\qquad
 \includegraphics[width=1\linewidth]{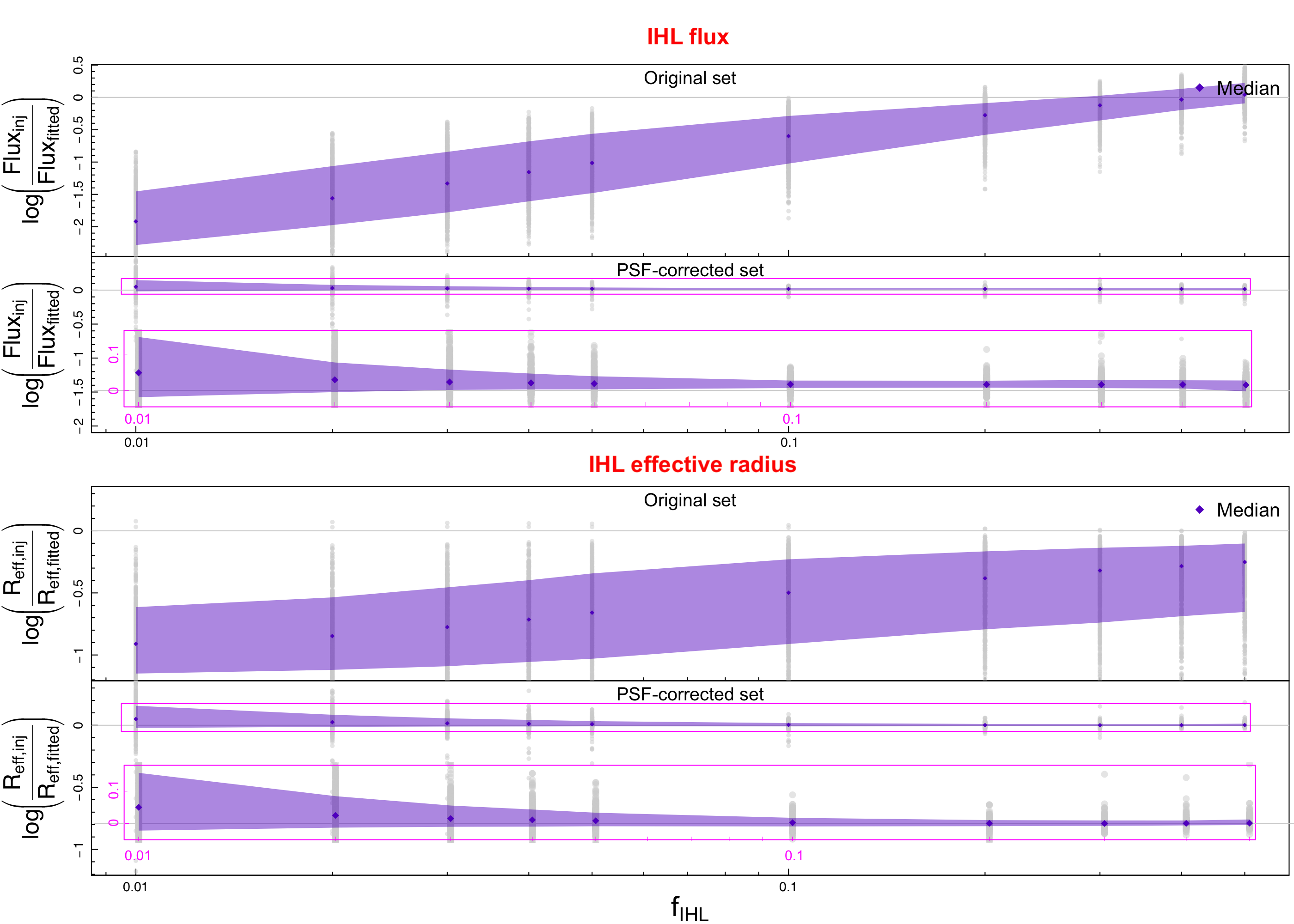}
    \caption{Comparison between the injected versus fitted parameters for both original and PSF-corrected datasets. \textcolor{black}{The} upper half of the plot corresponds to the flux, \textcolor{black}{and the} lower half to the effective radius. \textcolor{black}{The} upper panel of each half corresponds to the original dataset and \textcolor{black}{the} lower panel to the PSF-corrected one. \textcolor{black}{In the original dataset, the fitted parameters systematically exceed the injected values, whereas in the PSF-corrected dataset, they systematically fall below them. However, the PSF-corrected results are overall much closer to the injected values,} \textcolor{black}{with better performance towards brighter IHL components.}}
    \label{Difference}
\end{figure*}

Fig. \ref{Difference} compares the injected and fitted parameters for each group mock in both datasets. The upper half of the plot corresponds to the flux (derived from the magnitude), while the lower half corresponds to the effective radius. In each half, the upper panel was generated using the original dataset, and the lower panel using the PSF-corrected dataset. The purple diamonds represent the median of each distribution, with the 1$\sigma$ range shaded in light purple. \textcolor{black}{To better visualise the results of Fig. \ref{Difference}, we show in Table \ref{tab:results} the median ratio between the fitted and injected values of both parameters and datasets at \textcolor{black}{all} selected IHL fractions. This indicates how much larger or smaller the fitted values are compared to the injected ones.}

\begin{table*}

\centering
\begin{tabular}{ccccc}

 \textcolor{black}{$f_{\mathrm{IHL}}$} & $\frac{\mathrm{Original~Flux_{{\mathrm{fitted}}}}}{\mathrm{Flux_{\mathrm{inj}}}}$ & $\frac{\mathrm{PSF-corrected~Flux_{{\mathrm{fitted}}}}}{\mathrm{Flux_{\mathrm{inj}}}}$ & $\frac{\mathrm{Original~{R_\mathrm{{eff, fitted}}}}}{\mathrm{R_\mathrm{{eff, inj}}}}$  & $\frac{\mathrm{PSF-corrected~{R_\mathrm{{eff, fitted}}}}}{\mathrm{R_\mathrm{{eff, inj}}}}$\\

\hline
0.01 & \textcolor{black}{84.38}  & \textcolor{black}{0.89}  & \textcolor{black}{9.13}  & \textcolor{black}{0.89} \\

\textcolor{black}{0.02} & \textcolor{black}{38.02}  & \textcolor{black}{0.92}  & \textcolor{black}{7.91}  & \textcolor{black}{0.94} \\

\textcolor{black}{0.03} & \textcolor{black}{22.18}  & \textcolor{black}{0.94}  & \textcolor{black}{6.68}  & \textcolor{black}{0.97} \\

\textcolor{black}{0.04} & \textcolor{black}{15.08}  & \textcolor{black}{0.95}  & \textcolor{black}{5.82}  & \textcolor{black}{0.98} \\

\textcolor{black}{0.05} & \textcolor{black}{10.79}  & \textcolor{black}{0.95}  & \textcolor{black}{5.09}  & \textcolor{black}{0.99} \\

0.1 & \textcolor{black}{4.03}  & \textcolor{black}{0.95}  & \textcolor{black}{3.55}  & \textcolor{black}{1.00} \\ 

\textcolor{black}{0.2} & \textcolor{black}{1.93}  & \textcolor{black}{0.95}  & \textcolor{black}{2.68}  & \textcolor{black}{1.01} \\

\textcolor{black}{0.3} & \textcolor{black}{1.30}  & \textcolor{black}{0.95}  & \textcolor{black}{2.37}  & \textcolor{black}{1.01} \\

\textcolor{black}{0.4} & \textcolor{black}{1.05}  & \textcolor{black}{0.95}  & \textcolor{black}{2.17}  & \textcolor{black}{1.01} \\

0.5 & \textcolor{black}{0.87}  & \textcolor{black}{0.95}  & \textcolor{black}{2.01}  & \textcolor{black}{1.01} \\

\end{tabular}

\caption{\textcolor{black}{Median ratio between the fitted and injected values of flux and effective radius for the original and PSF-corrected datasets, at \textcolor{black}{all} IHL fractions. The values indicate how much larger or smaller the fitted quantities are compared to the injected ones. In all cases, the PSF-corrected IHL estimations are more accurate than the original ones.}}
\label{tab:results}

\end{table*}

\textcolor{black}{In the upper panel, the original Flux${_\mathrm{fitted}}$ is brighter than Flux${_\mathrm{inj}}$ at $f_{\mathrm{IHL}} < 0.4$, as the PSF-scattered light contributes additional flux, making the IHL appear brighter and leading to higher fitted values. This effect is particularly pronounced for the least prominent IHL components (i.e. at lower \textcolor{black}{$f_{\mathrm{IHL}}$})}. In contrast, the PSF-corrected Flux${_\mathrm{fitted}}$ shown in the \textcolor{black}{lower panel of the upper half} is significantly more accurate and closely matches Flux${_\mathrm{inj}}$. Still, Flux$_{\mathrm{fitted}}$ is slightly fainter than Flux$_{\mathrm{inj}}$. Removing the PSF-scattered flux ideally leaves only the true IHL flux to be modelled. However, as mentioned in Sec.~\ref{cavetas psf}, we are aware that our technique oversubtracts PSF flux by slightly removing IHL flux as well, leading to fainter fitted fluxes at all \textcolor{black}{$f_{\mathrm{IHL}}$}. \textcolor{black}{This approach reduces PSF-induced contamination in the inferred IHL signal, at the cost of slightly reducing the recovered flux.}

%\textcolor{mycolor}{To give the reader a sense of the PSF-corrected IHL estimations}, at \textcolor{black}{$f_{\mathrm{IHL}}$} = 0.01, $\log\left(\mathrm{\frac{Flux_{inj}}{Flux_{fitted}}}\right)$ $\sim$ 0.048, which means that the fitted flux is around 9.6\% lower than the real flux. In the case of the effective radius, $\log\left(\mathrm{\frac{R_{eff,inj}}{R_{eff,fitted}}}\right)$ $\sim$ 0.05, which means that the fitted effective radius is around 12.2\% smaller than the real effective radius. At the other extreme, when \textcolor{black}{$f_{\mathrm{IHL}}$} = 0.5, $\log\left(\mathrm{\frac{Flux_{inj}}{Flux_{fitted}}}\right)$ $\sim$ 0.015, which means that the fitted flux is around 3.5\% lower than the real flux. In the case of the effective radius, $\log\left(\mathrm{\frac{R_{eff,inj}}{R_{eff,fitted}}}\right)$ $\sim$ 0.0012, which means that the fitted effective radius is around 0.28\% smaller than the real effective radius. 

 In the case of the effective radius, we overestimate the size of the IHL component when the PSF correction technique is not applied, \textcolor{black}{as seen in the upper panel of the lower half. This effect persists even at high \textcolor{black}{$f_{\mathrm{IHL}}$}, where the IHL flux itself is well recovered but the effective radius remains biased. This is because, although the extended PSF-scattered flux has a limited impact on the total integrated flux at high \textcolor{black}{$f_{\mathrm{IHL}}$} due to the high S/N of the IHL, it can still alter the spatial distribution of the light in the LSB outskirts and therefore bias the recovered \textcolor{black}{$R_{\mathrm{eff}}$}. When the PSF correction is applied (lower panel of lower half), we instead slightly underestimate the effective radius because the correction removes the extended wings of bright objects, which would otherwise make the IHL component look larger.}

 For both datasets, the uncertainties in the flux distributions are larger at lower \textcolor{black}{$f_{\mathrm{IHL}}$}, as the IHL component has a low S/N and is more difficult to detect and distinguish from the sky background and other contaminants. At higher \textcolor{black}{$f_{\mathrm{IHL}}$}, the IHL S/N increases, leading to better flux recovery and reduced uncertainties. \textcolor{black}{In contrast, the uncertainties in \textcolor{black}{$R_{\mathrm{eff}}$} remain large across the full range of \textcolor{black}{$f_{\mathrm{IHL}}$} in the original mocks, including at high \textcolor{black}{$f_{\mathrm{IHL}}$, where the recovered \textcolor{black}{$R_{\mathrm{eff}}$} can still be substantially overestimated, as mentioned above. After the PSF correction, the uncertainties in \textcolor{black}{$R_{\mathrm{eff}}$} decrease towards higher \textcolor{black}{$f_{\mathrm{IHL}}$}, reflecting the improved recovery of the spatial extent of the IHL component.}}

 %The uncertainties in \textcolor{black}{$R_{\mathrm{eff}}$} are equally large across all values of \textcolor{black}{$f_{\mathrm{IHL}}$} in the original dataset of mocks (2\textsuperscript{nd} panel), but after the PSF correction the uncertainties in \textcolor{black}{$R_{\mathrm{eff}}$} are reduced towards higher \textcolor{black}{$f_{\mathrm{IHL}}$} (4\textsuperscript{th} panel).

%Mencionar casos fallaidos por estrellas brillantes u objetos extendidos (poner imagenes?)

%HABLAR DE DEGENERACION EN EL MODELO DE SERSIC?
%HABLAR SOBRE CASOS DONDE NO SE CUMPLE LO QUE DIGO ARRIBA? LOS OUTLIERS ?

%\textcolor{mycolor}{This is expected as the intensity of the profile at $\textit{R}_\textit{e}$ ($\textit{I}_\textit{e}$) and $\textit{R}_\textit{e}$ are inversely related in a Sérsic model, and $\textit{I}_\textit{e}$ is inversely related with $\textit{mag}$ in an exponential way (see Sec. 2.1.1 of \citealt{2017MNRAS.466.1513R} for more details). This means that a dimmer $\textit{mag}$ implies a smaller $\textit{R}_\textit{e}$ and viceversa (check corner plots)}.

% Our results improve significantly after applying the PSF correction, with better performance towards brighter IHL components

Our results demonstrate the importance of removing the PSF-scattered flux of the telescope when estimating the IHL component of a group or cluster. \textcolor{black}{In this simplified scenario, where a circular exponential IHL component is fitted with a circular IHL exponential model \textcolor{black}{and the only background effect considered is PSF scattering,} we are able to clearly demonstrate the significant impact that PSF-scattered flux can have on the recovered IHL properties.}

\subsubsection{PSF-corrected IHL estimations}
\label{subsubsection PSF-corrected}
\textcolor{black}{In Section \ref{case-study}, we demonstrate that \textsc{ProFit} is reliable for fitting the IHL component when the S/N of the IHL is good (i.e. high \textcolor{black}{$f_{\mathrm{IHL}}$}) \textcolor{black}{by analysing the reduced $\chi^2$ value}. \textcolor{black}{We now} highlight the importance of removing the PSF-scattered light when measuring the IHL using both original and PSF-corrected datasets of mock groups. In this subsection, we analyse the systematics introduced by both the performance of \textsc{ProFit} when the \textcolor{black}{$f_{\mathrm{IHL}}$} of the analysed mock is low, and the oversubtraction issue associated with the PSF correction. To do so, we focus on the IHL estimations from the PSF-corrected mock dataset.}

\begin{figure*}
\centering  %\qquad
 \includegraphics[width=1\linewidth]{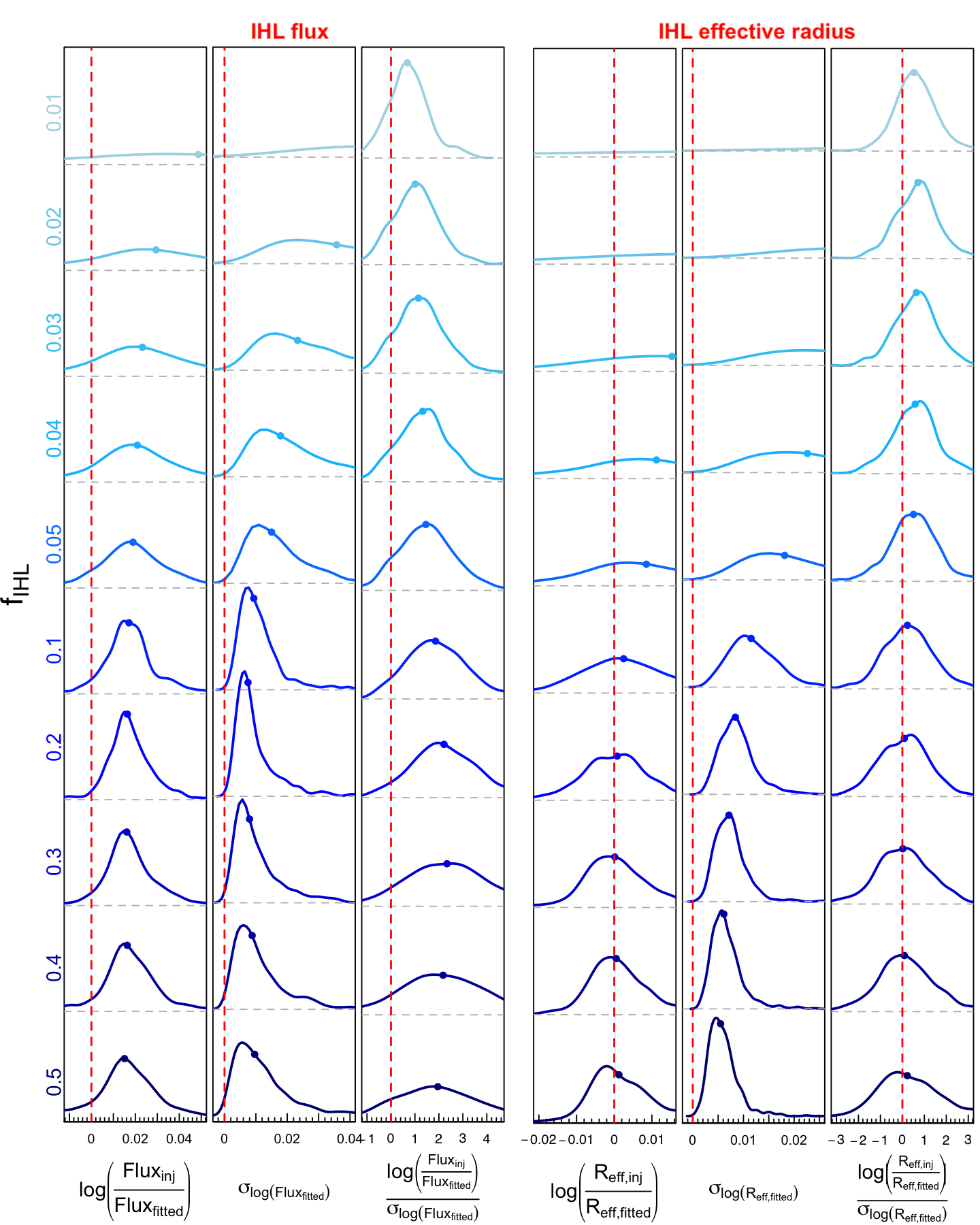}
    \caption{PSF-corrected normalised distributions of the ratio between injected and fitted values, the 1$\sigma$-quantile posterior, and the \textcolor{black}{scaled} ratio between injected and fitted values binned by \textcolor{black}{$f_{\mathrm{IHL}}$} in rows for Flux \textcolor{black}{(derived from $\textit{mag}$)} and \textcolor{black}{$R_{\mathrm{eff}}$}, where \textcolor{black}{distributions corresponding to} higher \textcolor{black}{$f_{\mathrm{IHL}}$} are indicated with darker blues. \textcolor{black}{That is, the \textcolor{black}{$f_{\mathrm{IHL}}$} values on the y-axis do not represent the actual y-values of the distributions; they are shown only to indicate the IHL fraction corresponding to each row of distributions.} The median of each distribution is indicated with a solid point. \textcolor{black}{The vertical dashed red lines indicate the zero line of the distributions.}}
    \label{Hist_triple}
\end{figure*}

\textcolor{black}{By comparing the injected and fitted IHL parameters for the PSF-corrected dataset, we construct Fig.~\ref{Hist_triple}, which presents three types of distributions for both flux and effective radius, binned by \textcolor{black}{$f_{\mathrm{IHL}}$} and arranged in rows. That is, the y-axis values indicate the \textcolor{black}{$f_{\mathrm{IHL}}$} corresponding to each row of distributions, \textcolor{black}{where darker shades of blue indicate a higher $f_{\mathrm{IHL}}$ (from 0.01 to 0.5)}. These distributions are}:

\begin{enumerate}

    \item \textcolor{black}{The ratio distributions between the injected and fitted values (1\textsuperscript{st} and 4\textsuperscript{th} columns)\textcolor{black}{.}}
    \item \textcolor{black}{The fitting uncertainties, estimated as the 1$\sigma$-quantile\footnote{Half of the range containing the central 68$\%$ of data} from each parameter posterior of each mock (2\textsuperscript{nd} and 5\textsuperscript{th} columns)\textcolor{black}{.}}
    \item \textcolor{black}{The ratio distributions \textcolor{black}{scaled} by the fitting uncertainties ($\frac{\textcolor{black}{X_\mathrm{inj}/X_\mathrm{fitt}}}{\sigma}$, 3\textsuperscript{rd} and 6\textsuperscript{th} columns).  \textcolor{black}{This quantifies the deviation relative to the reported uncertainties and therefore provides a diagnostic of whether those uncertainties are appropriately estimated, indicating whether the recovered parameters are statistically consistent with the injected values.} }

\end{enumerate}

 %\textcolor{black}{$\frac{\Delta}{\sigma}$ allows us to assess how significant the deviation is relative to the uncertainty, providing insight into the reliability of the fits. That is, whether the recovered parameters are statistically consistent with the injected values. 
 \textcolor{black}{All distributions are normalised such that the area under each curve equals one. We also indicate the median of each distribution with a solid blue point.}

\textcolor{black}{There are two factors that affect the results to different extents, depending on the injected IHL fraction of the analysed mock group: the S/N of the IHL component, which affects the \textcolor{black}{precision} of the fits, and the PSF correction, which slightly oversubtracts IHL flux. Focusing on the flux estimation, at \textcolor{black}{$f_{\mathrm{IHL}}$} \textcolor{black}{< 0.1}, the dominant contribution to the uncertainty in the recovered flux comes from the fitting, which is large due to the weak IHL S/N,  resulting in a broad scatter in the recovered values. In this regime, the scatter in the recovered flux is large compared to any systematic offset, so potential biases from the PSF correction are not clearly detectable.} This leads to on average \textcolor{black}{accurate} results (the median in \textcolor{black}{log($\textcolor{black}{Flux_\mathrm{inj}/Flux_\mathrm{fitt}}$)} is close to the zero line), but not \textcolor{black}{precise} at all (the spread in $\sigma$ is large). \textcolor{black}{At intermediate \textcolor{black}{$f_{\mathrm{IHL}}$}, both the IHL S/N (i.e. $f_{\mathrm{IHL}}$), and the PSF correction have a significant impact on the recovered parameters.} \textcolor{black}{The precision improves as the S/N is better, but} the impact of our PSF correction in the IHL flux estimation becomes more noticeable, causing \textcolor{black}{log$\left(\frac{\textcolor{black}{Flux_\mathrm{inj}/Flux_\mathrm{fitt}}}{\sigma}\right)$} to increase the positive deviation from the zero line. Going to larger \textcolor{black}{$f_{\mathrm{IHL}}$}, \textcolor{black}{the median of} log$(\textcolor{black}{Flux_\mathrm{inj}/Flux_\mathrm{fitt}})$ deviates up to 2$\sigma$ from the injected flux values. \textcolor{black}{These values are not accurate, as the most dominant factor in this IHL range is the PSF correction.} As the uncertainties in the fitting are low enough, the systematic biases of our correction are more prominent. \textcolor{black}{That is, while our PSF-corrected flux estimations are not entirely free of systematics introduced by the correction itself, they are significantly more accurate than those obtained without applying the PSF correction, as shown in Fig. \ref{Difference}.} \textcolor{black}{log$(\textcolor{black}{Flux_\mathrm{inj}/Flux_\mathrm{fitt}})$ deviating up to 2$\sigma$ also indicates that the uncertainties are not realistic, which can be a result of the MCMC algorithm occasionally becoming trapped in local minima while exploring the posterior distribution.} In addition, in our plots, $\sigma$ only captures the fitting uncertainties, while there are other sources that are not being considered but introduce errors in our results: background noise, the source detection technique in the PSF correction process, masking of sources in the IHL estimation technique, etc. \textcolor{black}{Accounting for these additional contributions would increase the scatter in the recovered fluxes and broaden the error bars.} \textcolor{black}{However, this would not remove the systematic bias observed in the PSF-corrected fluxes, which arises from the PSF correction itself.} Regarding the estimation of the IHL effective radius, \textcolor{black}{which is shown in the right half of Fig.~\ref{Hist_triple}}, at low \textcolor{black}{$f_{\mathrm{IHL}}$} the dominant source of uncertainty remains the fitting itself. However, at intermediate and high \textcolor{black}{$f_{\mathrm{IHL}}$}, once the extended wings of bright objects are removed through our PSF correction, the size estimates of the IHL component become more reliable, \textcolor{black}{as log$(\textcolor{black}{R_\mathrm{eff,inj}/R_\mathrm{eff,fitt}})$ and \textcolor{black}{log$\left(\frac{\textcolor{black}{R_\mathrm{eff,inj}/R_\mathrm{eff,fitt}}}{\sigma}\right)$} are closed to the zero line, indicating both small systematic offsets and appropriately estimated uncertainties.}

 \textcolor{black}{In summary, our technique is accurate when fitting the IHL flux if sufficient S/N is available. However, due to a slight oversubtraction introduced by the PSF correction, the fitted IHL flux tends to be slightly smaller than the real flux. Despite this, the PSF-corrected flux values are significantly more accurate than those obtained without the PSF correction at all \textcolor{black}{$f_{\mathrm{IHL}}$}. By comparing the IHL flux estimated with and without the PSF correction, we find that the true value likely lies between the two, but} \textcolor{black}{much closer to}  \textcolor{black}{the PSF-corrected estimation. Regarding the IHL size estimation, our PSF-corrected estimation is a good match to the true value \textcolor{black}{(if sufficient S/N is available)}.} 

 \textcolor{black}{Our results demonstrate that applying a masking technique to exclude all pixels affected by PSF-scattered light is insufficient to fully account for this contamination, as the extended PSF wings can distribute flux beyond the masked regions. Therefore, additional care is required when estimating the IHL to ensure that the contribution from PSF-scattered light is properly accounted for, both by removing its contribution to the IHL and also by recovering the scattered flux corresponding to the sources that produced it.}

 \textcolor{black}{Regarding the physical properties of the IHL, the presence of significant PSF-scattered light does not prevent the IHL from having a distinct intrinsic morphology. As demonstrated in Appendix~\ref{sec:non-circular}, our PSF-correction technique can recover the intrinsic shape of elliptical IHL models once the contribution from PSF-scattered light is removed. Furthermore, the impact of PSF-scattered light is expected to increase with redshift, as galaxies and their surrounding IHL subtend smaller angular scales and are therefore more strongly affected by the PSF. If left uncorrected, this effect could introduce a redshift-dependent bias in the inferred IHL fraction, highlighting the importance of applying an efficient PSF-correction technique.}

%no estoy diciendo nada que ademas nuestra tecnica recupera flujo perdido de las sources (ademas de borrar el flujo contaminante afuera)

 \textcolor{black}{While our results are obtained from controlled mock observations, the main conclusions are expected to be translated to real data. In our idealised mock setup, where both the IHL and point-sources are analytically defined and the PSF is \textcolor{black}{known and explicitly defined}, the impact of the PSF convolution can be isolated and corrected in a more controlled manner than in real observations. As a result, the difference between PSF-corrected and original measurements appears more clearly defined than it is expected to be in real data, where additional sources of systematic uncertainty (e.g. sky subtraction residuals, source crowding, and non-ideal light profiles) can partially mix with PSF-induced effects. Nevertheless, the PSF remains a dominant source of bias, and the qualitative improvement from applying the correction is expected to persist in real systems.}

%We find that the combination of both PSF-scattered flux removal and IHL estimation techniques is well suited to automatically analysing the IHL component in large datasets.

%esto no entiendo porque no pasa a low fractions?

%As expected, $\sigma_{\mathrm{Par_{fitted}}}$ is more precise around the fitted values also toward higher \textcolor{black}{$f_{\mathrm{IHL}}$}. For the weighted residuals, it is more clear now that the fitted flux values deviate, on average, by up to 2$\sigma$ from the injected values at higher \textcolor{black}{$f_{\mathrm{IHL}}$}, while the fitted effective radius deviates by $\sim$ 0.2$\sigma$. By analysing all of the distributions, we see how the two main factors mentioned above that affect our estimations, S/N of the IHL component and the PSF correction, dominate differently depending on the \textcolor{black}{$f_{\mathrm{IHL}}$} of the mock group. At lower \textcolor{black}{$f_{\mathrm{IHL}}$}, the fitting uncertainties dominate the overall uncertainties in our techniques to model the IHL, making our results unreliable in this range (\textcolor{black}{$f_{\mathrm{IHL}}$} = 0.01 - 0.05). From \textcolor{black}{$f_{\mathrm{IHL}}$} = 0.1 onward, the fitting uncertainties are significantly reduced, leaving space for the uncertainties of our PSF correction, causing a positive deviation in the flux estimations, but more precise size estimations. As we move toward higher \textcolor{black}{$f_{\mathrm{IHL}}$}, the PSF correction becomes the dominant factor, leading to systematic biases in the flux residuals. 

\textcolor{black}{In Appendix~\ref{B}, we provide a discussion of a small quantity of cases within our mock group dataset where \textsc{ProFit} struggles to recover reliable fits of the IHL component \textcolor{black}{due to issues arising from the source extractor process}\textcolor{black}{, even when its S/N is high.}}

%In particular cases where there are extended galaxy sources present in our mock group images, the fitted parameters can be biased by these segments and our IHL estimation technique does not perform well (especially at lower \textcolor{black}{$f_{\mathrm{IHL}}$}) even after applying the PSF correction. Additionally, due to the way we construct our mock dataset, large sources can be fragmented into smaller segments, which can further bias the results. Fortunately, these are only a few isolated cases in our dataset. See Appendix~\ref{A} for further details about these cases.

%it does not reach exactly one due to the PSF correction. While the 1$\sigma$-quantile posterior distributions become more precise around the fitted values (confirming that our IHL detection technique works properly), the weighted ratio distributions show that the true value is, on average, 2$\sigma$ away. \textcolor{black}{Correct after having good fluxratio/sigma}

%We also observe that for mocks groups with a lower fraction of IHL (\textcolor{black}{$f_{\mathrm{IHL}}$} $<$ 10\%), the recovered magnitude present high uncertainties. This is expected as the signal-to-noise ratio (S/N) is low and it is difficult to differentiate the IHL flux from the sky background ?? (plus the PSF light in the upper panel). However, for higher IHL fractions, we are able to recover the injected magnitude and effective radius with accuracy in both samples.

%\subsection{Caveats}
%\label{caveats results}

%esto no se si ponerlo !!!
%In particular cases where there are extended galaxy sources present in our mock group images, the fitted parameters can be biased by these segments and our IHL estimation technique does not perform well (especially at lower \textcolor{black}{$f_{\mathrm{IHL}}$}) even after applying the PSF correction. Additionally, due to the way we construct our mock dataset, large sources can be fragmented into smaller segments, which can further bias the results. Fortunately, these are only a few isolated cases in our dataset. See Appendix~\ref{A} for further details about these cases.

\section{Application to observations: IHL estimation using HSC-UD PDR3}

\label{sec:application}
\textcolor{black}{In this section, we estimate the IHL fraction of a GAMA group using real HSC-SSP PDR3 UD in the $\textit{g,r}$ and $\textit{i}$-bands. We apply two different techniques: the IHL estimation technique (Section~\ref{sec:IHL analysis}) and the SB limit method. In the latter, we slightly modify the method by including an extrapolation of the IHL towards the core of the system\textcolor{black}{, as defined in Section~\ref{sec:Intro}.}}

%apply our PSF-scattered flux removal and IHL estimation techniques on a GAMA group using real HSC-SSP PDR3 UD data in the $\textit{g,r}$ and $\textit{i}$-bands. 

In particular, we select the GAMA group identified by the ID G400138, which has 6 observed galaxy members, halo mass \textcolor{black}{$M_{\mathrm{halo}}$} $\sim$  9.4 $\times 10^{12}\,\mathrm{M_\odot}$, and redshift $z$ $\sim$ 0.21 (see left plot of Fig.\ref{Comparison techniques}). This group is located in the G02 GAMA field, which partially overlaps with the HSC-SSP PDR3 UD field SXDS (see Fig.~\ref{Map}). We select this group because its IHL component has already been analysed in \citealt{2023MNRAS.518.1195M}, where the authors estimated the IHL fraction in the $\textit{g,r}$ and $\textit{i}$-bands following two different methods and using HSC-SSP PDR2 UD data.

%A more detailed explanation of their methods is provided later. BORRE LA PARTE DONDE EXPLICO SUS METODOS

\textcolor{black}{In both approaches, we estimate \textcolor{black}{$f_{\mathrm{IHL}}$} following the next equation, \textcolor{black}{which is the same equation introduced in Eq.\ref{ecuacion}, but written here in a rearranged form}:}

\begin{equation}    
\\
\\
\\
\textcolor{black}{f_{\mathrm{IHL}}} = \frac{\textcolor{black}{F_{\mathrm{IHL}}}}{\textcolor{black}{F_{\mathrm{group}}} +\textcolor{black}{F_{\mathrm{IHL}}}}, \\
 \label{eq 3}
\end{equation}
\\
\noindent \textcolor{black}{where \textcolor{black}{$F_{\mathrm{IHL}}$} is the IHL flux and \textcolor{black}{$F_{\mathrm{group}}$} is the total group \textcolor{black}{flux} of G400138. We estimate \textcolor{black}{$F_{\mathrm{IHL}}$} following two different approaches: \textcolor{black}{one based on \textsc{ProFit}, where we fit the IHL component magnitude with a Sérsic model, and another where we mask all sources present in the image using an SB threshold value and fit a 1D exponential model to the SB profile of the IHL.} Regarding $F_\mathrm{{group}}$, we consider one version including only the confirmed GAMA galaxy members, and another version that also incorporates an extrapolation factor to account for faint members below the GAMA spectroscopic limit.}

\textcolor{black}{All estimations of \textcolor{black}{$f_{\mathrm{IHL}}$} are performed on both the original and the PSF-corrected images of G400138.}

\subsection{\textcolor{black}{Estimation of \textcolor{black}{$F_{\mathrm{IHL}}$}}}
\subsubsection{\textcolor{black}{\textsc{ProFit}}} 

\textcolor{black}{In our first approach, we estimate \textcolor{black}{$F_{\mathrm{IHL}}$} following our IHL estimation technique, which uses \textsc{ProFit} with a Sérsic exponential template to fit the IHL component of G400138. We use two different exponential models: one with only the magnitude and the effective radius as free parameters, and another where the axial ratio and position angle are also left as free parameters (as in App.~\ref{sec:non-circular}). This results in two different IHL models following the \textsc{ProFit} method, and therefore two values of \textcolor{black}{$F_{\mathrm{IHL}}$}.}

%(unlike the analysis in Sec.~\ref{sec:results}, the true IHL parameters are unknown in this case.)

%follow a similar methodology to that used in Sec.~\ref{sec:results}, which consists of applying the IHL estimation technique to both the original and the PSF-corrected images of G400138. The key difference from our analysis in Sec.~\ref{sec:results} is that, in this case, the true parameters of the IHL are unknown. Following our technique, we fit the IHL component of G400138 using two exponential models: one with only the magnitude and effective radius as free parameters, and another in which the axial ratio and position angle are also left as free parameters (as done in Sec.~\ref{caveats results}). We then estimate \textcolor{black}{$f_{\mathrm{IHL}}$} from the resulting models.

 %We then estimate the \textcolor{black}{$f_{\mathrm{IHL}}$} from the resulting models using the following equation: 

\subsubsection{\textcolor{black}{SB limit method with IHL core extrapolation}}

\textcolor{black}{In our second approach, we use the SB limit method to estimate \textcolor{black}{$F_{\mathrm{IHL}}$}.} This method consists of simply applying a threshold in the SB of the image and considering all light below this limit as IHL. We apply the same cut adopted by \citealt{2023MNRAS.518.1195M}, $\mu_{V}$ = 26.5 mag arcsec$^{-2}$, which we also correct for SB dimming and \textit{K}-correction in each band\footnote{We use the same corrections as \citealt{2023MNRAS.518.1195M}: dimming in the form (1+z)$^{-4}$ \citep{1930PNAS...16..511T,1934rtc..book.....T} and \textit{K}-correction:  \textit{K}-corr$_\mathrm{g}$ = 0.68 mag, \textit{K}-corr$_\mathrm{r}$ = 0.29 mag, and \textit{K}-corr$_\mathrm{i}$ = 0.13 mag \citep{2012MNRAS.420.1239L}.}. After masking all sources using this SB cut, we find the best exponential model that fits the \textcolor{black}{SB profile of the} IHL (\textcolor{black}{$\mathrm{\mu_{IHL,SB}(r)}$}) using the MCMC fitting package \textsc{hyper-fit}\footnote{\url{https://github.com/asgr/hyper.fit}. \textcolor{black}{\textsc{hyper-fit} is a Bayesian fitting tool for multi-dimensional linear models that incorporates the intrinsic scatter as part of the fit.}} \citep{2015PASA...32...33R}. We then extrapolate \textcolor{black}{$\mathrm{\mu_{IHL,SB}(r)}$} towards the core of the image in order to recover all the IHL that would otherwise be discarded due to the BCG mask. \textcolor{black}{We therefore assume that the IHL component extends into the central regions of the system, spatially overlapping with the BCG.} See Appendix~\ref{C} for further details about the fitting and extrapolation procedures in the SB limit method. We finally calculate \textcolor{black}{$F_{\mathrm{IHL}}$} by integrating \textcolor{black}{$\mathrm{\mu_{IHL,SB}(r)}$} up to a radius of $R$ = 275 kpc, \textcolor{black}{as indicated by the following equation}:

\begin{equation}    
\\
\\
\\
 F_\mathrm{IHL,SB} = \int_{0}^{R}
\mu_\mathrm{IHL,SB}(r)~2~\pi~r~dr. \\
\end{equation}
\\
\noindent  

The radius of $R$ = 275 kpc is the same as the semi-major axis of the elliptical aperture selected by \citealt{2023MNRAS.518.1195M}\textcolor{black}{, and is used here as a common reference scale, noting that our assumption of circular symmetry implies a different effective integration area. We further clarify that five out of the six member galaxies of the group lie within $R$, with only one object located outside it, as seen in Fig.~\ref{Comparison techniques}.}

\subsection{\textcolor{black}{Estimation of \textcolor{black}{$F_{\mathrm{group}}$}}}

\subsubsection{\textcolor{black}{Total group flux}}
\label{F_group}

\textcolor{black}{\textcolor{black}{$F_{\mathrm{group}}$} is calculated by first selecting the confirmed members of G400138 using the GAMA galaxy group catalogue G$^3$Cv10, and then summing the fluxes from the members. As G$^3$Cv10 only provides the magnitude of the galaxy members in the \textit{r}-band, we make use of the G02 input catalogue from the GAMA Data Release 3\footnote{\url{https://gama-survey.org/dr3/schema/table.php?id=9}} for the \textit{g} and \textit{i}-band estimations of \textcolor{black}{$F_{\mathrm{group}}$}.}

\subsubsection{\textcolor{black}{Extrapolated total group flux}}

\textcolor{black}{In addition, we estimate the extrapolated version of \textcolor{black}{$F_{\mathrm{group}}$}, rescaled by a factor that considers all the missing flux from galaxy members below the GAMA \textcolor{black}{spectroscopic} completeness limit.} This correction is already implemented in G$^3$Cv10, where \citealt{2011MNRAS.416.2640R} calculate the effective absolute magnitude limit of each group, measure the $\textit{r}_{AB}$-band luminosity contained within this limit, and then integrate the global GAMA galaxy luminosity function to a faint-end magnitude threshold to correct for the missing flux (See Sec. 4.4 of \citealt{2011MNRAS.416.2640R} for further details). The extrapolated total group light for G400138 is:

\begin{equation}    
\\
F_\mathrm{group, extrap.}  \sim F_\mathrm{group} \times 1.48, 
\\
\end{equation}  

\noindent where \textcolor{black}{$F_{\mathrm{group}}$} is obtained from Section \ref{F_group}.

%\vspace{1cm}

%In addition, we also re-estimate all of our \textcolor{black}{$f_{\mathrm{IHL}}$} by redefining \textcolor{black}{$F_{\mathrm{group}}$} to include not only the total luminosity of the observed group members, but also the extrapolated luminosity inferred from an arbitrarily faint absolute magnitude limit, in order to account for residual selection effects. This correction is already implemented in G$^3$Cv10 catalogue, where \citealt{2011MNRAS.416.2640R} calculate the effective absolute magnitude limit of each group, measure the $\textit{r}_{AB}$-band luminosity contained within this limit, and then integrate the global GAMA galaxy luminosity function to a faint-end magnitude threshold to correct for the missing flux (See Sec. 4.4 of \citealt{2011MNRAS.416.2640R} for further details). The extrapolated total group light for G400138 is \textcolor{black}{$F_{\mathrm{group,extrap.}}$} $\sim$ \textcolor{black}{$F_{\mathrm{group}}$} * 1.48. Following the same procedure as before, we re-estimate \textcolor{black}{$f_{\mathrm{IHL}}$} as: 

\textcolor{black}{It is worth highlighting that, in addition to the uncertainty in measuring \textcolor{black}{$F_{\mathrm{IHL}}$} using different techniques, there is also an uncertainty related to how we estimate \textcolor{black}{$F_{\mathrm{group}}$} in Section \ref{F_group}, which is the main contributor to the denominator of \textcolor{black}{$f_{\mathrm{IHL}}$}. To ensure that any discrepancy between our two methods to measure \textcolor{black}{$F_{\mathrm{IHL}}$} is due only to the IHL estimation, we use the same value of \textcolor{black}{$F_{\mathrm{group}}$} for both SB and \textsc{ProFit} approaches.}

\subsection{\textcolor{black}{Our IHL measurements}}

\textcolor{black}{By estimating the IHL flux using both the \textsc{ProFit} and the SB limit methods, and computing the total group flux with and without the extrapolation factor, we obtain four different values of \textcolor{black}{$f_{\mathrm{IHL}}$} per band. \textcolor{black}{Repeating this for both the original and PSF-corrected images of G400138 yields eight measurements in total per band.}}

\textcolor{black}{Our most robust \textcolor{black}{$f_{\mathrm{IHL}}$} estimations are those performed in the PSF-corrected image and calculated using \textcolor{black}{$F_{\mathrm{group,extrap.}}$}, shown in the upper panel of Table~\ref{tab:fihl}. The median IHL fractions per band considering these three measurements (two using \textsc{ProFit} and one using the SB limit method) are:
\textcolor{black}{$f_{\mathrm{g,IHL}}$} \textcolor{black}{$\approx$} 0.19$^{+0.09}_{-0.01}$, \textcolor{black}{$f_{\mathrm{r,IHL}}$} \textcolor{black}{$\approx$} 0.08$^{+0.06}_{-0.02}$, and \textcolor{black}{$f_{\mathrm{i,IHL}}$} \textcolor{black}{$\approx$} 0.06$^{+0.04}_{-0.02}$. To estimate these median \textcolor{black}{$f_{\mathrm{IHL}}$} values with only three measurements per band, we combine the individual measurements and their errors through a Monte Carlo approach. We generate $N = 1000$ random realisations of \textcolor{black}{$f_{\mathrm{IHL}}$}, assuming a Normal distribution centred on the measured value and with a standard deviation equal to its 1$\sigma$ uncertainty. We then merge the simulated samples \textcolor{black}{from all three bands} into a single distribution and compute the 16th, 50th, and 84th percentiles, which correspond to the median and its $\pm1\sigma$ confidence intervals.}

\subsection{\textcolor{black}{Comparison with \textcolor{black}{the literature}}}

\subsubsection{\textcolor{black}{Martínez-Lombilla et al. 2023}}

In this section, we compare our IHL fraction estimations of G400138 with the IHL fractions estimated by \citealt{2023MNRAS.518.1195M} based on their two methods. In short, their first method consists of modelling the brightest stars in the field using extended HSC-SSP PDR2 PSFs, and the three galaxies in the group core using 2D PSF-convolved models generated with \textsc{IMFIT} \citep{2015ApJ...799..226E}. After carefully subtracting the star and galaxy 2D models from the data and masking the undesirable background and foreground sources, the diffuse residual light is then interpreted as the IHL component. Their second method is the SB limit method, using the SB cut mentioned above ($\mu_{V}$ = 26.5 mag arcsec$^{-2}$) and an elliptical aperture with a semi-major axis of 275 kpc, ellipticity of 0.52, and position angle of 194$^\circ$. \textcolor{black}{These} authors calculated the total group light by summing the flux from all the members and diffuse light detected within this aperture. \textcolor{black}{Their IHL estimations are summarised in the lower panel of Table~\ref{tab:fihl}.}

%We summarise the \textcolor{black}{$f_{\mathrm{IHL}}$} from \citealt{2023MNRAS.518.1195M} in Table~\ref{tab:table ML}.

Our estimates of \textcolor{black}{$f_{\mathrm{IHL}}$}, along with those from \citealt{2023MNRAS.518.1195M}, are shown in the right plot of Fig.~\ref{Comparison techniques}. The values from \citealt{2023MNRAS.518.1195M} are shown as purple symbols, while our estimates are shown as green symbols. \textcolor{black}{Our $f_{\mathrm{IHL}}$} obtained using the \textsc{ProFit} modelling technique with fixed angle and axial ratio are shown as circles, those obtained with free angle and axial ratio as triangles, and those obtained using the SB limit method including the extrapolated IHL flux as squares. In all three cases, filled symbols represent estimations using the PSF-corrected images, while empty symbols correspond to those using the original versions. Solid colours indicate estimations that consider \textcolor{black}{$F_{\mathrm{group,extrap.}}$}, while pale colours consider \textcolor{black}{$F_{\mathrm{group}}$}. Regarding the estimations from \citealt{2023MNRAS.518.1195M}, those obtained using the 2D composite model are shown as diamonds, and those estimated using the SB limit method as empty squares with a cross inside. 

%\textcolor{black}{Our \textcolor{black}{$f_{\mathrm{IHL}}$} estimations are shown as solid dark symbols in Fig.~\ref{Comparison techniques}, while the medians are indicated with red crosses.}.

%This may be due to the increased sky background and fringing effects commonly present in the $\textit{i}$-band, which can affect LSB measurements. 

We find good agreement between all our methods in the three bands. Modelling the IHL with free angle and axial ratio results in a higher \textcolor{black}{$f_{\mathrm{IHL}}$} compared to when these parameters are fixed, as this model \textcolor{black}{is allowed to capture the ellipticity of the IHL.} As also expected, our PSF-corrected \textcolor{black}{$f_{\mathrm{IHL}}$} estimations are lower than those obtained from the original, non–PSF-corrected images. Regarding the estimations that consider \textcolor{black}{$F_{\mathrm{group,extrap.}}$}, the resulting \textcolor{black}{$f_{\mathrm{IHL}}$} values are consistently lower, since the numerator of \textcolor{black}{$f_{\mathrm{IHL}}$} remains fixed while the denominator increases.

When comparing with the estimations from \citealt{2023MNRAS.518.1195M}, our \textcolor{black}{$f_{\mathrm{IHL}}$} fall within their range of \textcolor{black}{$f_{\mathrm{IHL}}$}. In addition, we find the same decreasing trend as reported by the authors, with lower IHL fractions towards redder bands. This trend was also reported in \citealt{2018ApJ...857...79J}, who found evidence for higher \textcolor{black}{$f_{\mathrm{IHL}}$} in the bluer bands of merging clusters compared to redder bands. The 2D composite model \textcolor{black}{$f_{\mathrm{IHL}}$} from \citealt{2023MNRAS.518.1195M} are larger than our corrected estimations (solid dark symbols) across all bands. We \textcolor{black}{speculate} that the difference between our estimates and their 2D composite model estimates \textcolor{black}{may be} related to residual galaxy light in their IHL images, arising from sources in the field that are not modelled with analytical profiles. These residuals would make their IHL look brighter and consequently increase their \textcolor{black}{$f_{\mathrm{IHL}}$}. \textcolor{black}{In addition, even when we use the elliptical \textsc{ProFit} model, our analytical profiles may not fully capture the total extent of the IHL. This would bias the inferred IHL component towards fainter values and consequently lead to lower \textcolor{black}{$f_{\mathrm{IHL}}$} estimates in both the original and PSF-corrected cases.} Focusing on the SB limit technique,  we find higher \textcolor{black}{$f_{\mathrm{IHL}}$} values than those obtained by \citealt{2023MNRAS.518.1195M}. \textcolor{black}{In Appendix~\ref{C}, we explore the effect of extrapolating the IHL towards the core through an exponential fit as one possible explanation for the differences between our measurements and those reported by these authors. Our \textcolor{black}{$f_{\mathrm{IHL}}$} estimations that exclude the extrapolated IHL lie within the error bars of their results in the \textit{r} and \textit{i}-bands. However, our \textit{g}-band \textcolor{black}{$f_{\mathrm{IHL}}$} are still higher than their estimations.} It is worth noting that \citealt{2023MNRAS.518.1195M} used HSC-SSP PDR2 UD data, while in our work we use HSC-SSP PDR3 UD data. 
\textcolor{black}{Compared to PDR2, PDR3 implements an improved global sky subtraction algorithm. We also use our PDR3 PSF models. Both improvements better capture the light distribution of objects with extended wings. Since these extended wings are directly relevant for LSB measurements, this may introduce small systematic differences in the recovered diffuse emission and hence in $f_{\mathrm{IHL}}$.}

%The PDR3 sky subtraction algorithm and our PDR3 PSF models are designed to better capture the light distribution of objects with extended wings, which could have a minor impact on the IHL estimations. 

This comparison highlights that the measured IHL fraction can vary significantly even within a single group, depending on the method used \citep{2011ApJ...732...48R,2021ApJS..252...27K, 2024MNRAS.528..771B}.

%We attribute this difference to the extrapolated core IHL light incorporated in our analysis through the exponential fit. 

% Following our \textsc{ProFit} method, our nearly flat distribution across the different bands suggests that the spectral energy distribution (SED) of the IHL is broadly consistent with that of the average galaxy population within the group.

%\begin{table*}

%\centering
%\begin{tabular}{lcccccc}

%~~~~\textbf{Method} & \textcolor{black}{$f_{\mathrm{g,IHL}}$} & \textcolor{black}{$f_{\mathrm{r,IHL}}$} & \textcolor{black}{$f_{\mathrm{i,IHL}}$} & $\mathrm{mag_{g,IHL}}$ & $\mathrm{mag_{r,IHL}}$ & $\mathrm{mag_{i,IHL}}$\\
%\hline
%\textsc{ProFit} angle and axrat fixed & 0.180 $\pm$ 0.006 & 0.058 $\pm$ 0.002  & 0.044 $\pm$ 0.002 & 19.17 $\pm$ 0.03 & 19.19 $\pm$ 0.03  & 19.09 $\pm$ 0.04 \\
%\textsc{ProFit} angle and axrat free & 0.187 $\pm$ 0.012  & 0.083 $\pm$ 0.033 & 0.072 $\pm$ 0.029 & 19.13 $\pm$ 0.06  & 18.78 $\pm$ 0.41 & 18.53 $\pm$ 0.41 \\ SB extrap. & 0.273 $\pm$ 0.052  & 0.132 $\pm$ 0.036 & 0.089 $\pm$ 0.027 & 18.60 $\pm$ 0.30  & 18.21 $\pm$ 0.37 & 18.29 $\pm$ 0.39 \\

%\end{tabular}

%\caption{Our estimations of \textcolor{black}{$f_{\mathrm{IHL}}$} and $\mathrm{mag_{IHL}}$ for G400138 in the \textit{g},\textit{r}, and \textit{i}-bands derived using three different methods with HSC-UD data. All \textcolor{black}{$f_{\mathrm{IHL}}$} have been corrected for the PSF-scattered flux. The total group light is defined as the sum of the flux from confirmed members plus the extrapolated contribution from faint sources.}
%\label{tab:fihl}

%\end{table*}

\begin{table*}

\centering
\begin{tabular}{lcccccc}

~~~~\textbf{This work} & \textcolor{black}{$f_{\mathrm{g,IHL}}$} & \textcolor{black}{$f_{\mathrm{r,IHL}}$} & \textcolor{black}{$f_{\mathrm{i,IHL}}$} \\
\hline
\textsc{ProFit} angle and axrat fixed & 0.180 $\pm$ 0.006 & 0.058 $\pm$ 0.002  & 0.044 $\pm$ 0.002  \\
\textsc{ProFit} angle and axrat free & 0.187 $\pm$ 0.012  & 0.083 $\pm$ 0.033 & 0.072 $\pm$ 0.029 \\ SB extrap. & 0.273 $\pm$ 0.052  & 0.132 $\pm$ 0.036 & 0.089 $\pm$ 0.027 \\
\\
~~~~\textbf{Martínez-Lombilla et al. 2023} & &  &  \\
\hline
2D comp. model & 0.365 $\pm$ 0.022 & 0.305 $\pm$ 0.014  & 0.298 $\pm$ 0.011 \\
SB $\mu$$>$26.5 $\mathrm{\frac{mag}{arcsec^{2}}}$ & 0.146 $\pm$ 0.032  & 0.035 $\pm$ 0.037 & 0.016 $\pm$ 0.043  \\

\end{tabular}

\caption{IHL fraction estimations for G400138 in the \textit{g}, \textit{r}, and \textit{i}-bands obtained using the three methods developed in this work, as well as the two methods from \citealt{2023MNRAS.518.1195M}. All of our \textcolor{black}{$f_{\mathrm{IHL}}$} values shown in this table have been corrected for PSF-scattered flux.}
\label{tab:fihl}

\end{table*}

%The total group light is defined as the sum of the flux from confirmed members plus the extrapolated contribution from faint sources.

%\begin{table}

%\centering
%\begin{tabular}{lccc}

%~~~~\textbf{Method} &\textcolor{black}{$f_{\mathrm{g,IHL}}$} & \textcolor{black}{$f_{\mathrm{r,IHL}}$} & \textcolor{black}{$f_{\mathrm{i,IHL}}$} \\
%\hline
%2D comp. model & 0.365 $\pm$ 0.022 & 0.305 $\pm$ 0.014  & 0.298 $\pm$ 0.011 \\
%SB $\mu$$>$26.5 $\mathrm{\frac{mag}{arcsec^{2}}}$ & 0.146 $\pm$ 0.032  & 0.035 $\pm$ 0.037 & 0.016 $\pm$ 0.043  \\

%\end{tabular}

%\caption{\textcolor{black}{$f_{\mathrm{IHL}}$} values from %\citealt{2023MNRAS.518.1195M} in the \textit{g},\textit{r}, and \textit{i}-bands, derived using two different methods.}
%\label{tab:table ML}

%\end{table}

\subsubsection{\textcolor{black}{Other theoretical and observational works}}

\textcolor{black}{Our median \textcolor{black}{$f_{\mathrm{IHL}}$} (\textcolor{black}{\textcolor{black}{$f_{\mathrm{g,IHL}}$} \textcolor{black}{$\approx$} 0.19$^{+0.09}_{-0.01}$, \textcolor{black}{$f_{\mathrm{r,IHL}}$} \textcolor{black}{$\approx$} 0.08$^{+0.06}_{-0.02}$, and \textcolor{black}{$f_{\mathrm{i,IHL}}$} \textcolor{black}{$\approx$} 0.06$^{+0.04}_{-0.02}$}) are in good agreement with the IHL fractions predicted by \citealt{2011ApJ...732...48R} and \citealt{2024MNRAS.527.2624P}. \citealt{2011ApJ...732...48R} used N-body simulations of clusters \textcolor{black}{at $z$ = 0} \textcolor{black}{and found} \textcolor{black}{$0.1 \lesssim f_{\mathrm{IHL}} \lesssim 0.35$}. \textcolor{black}{Although these clusters are more massive than our analysed group ($\mathrm{M}_{\mathrm{200}} \gtrsim  0.8\times10^{14}~\mathrm{M}_\odot$), there is no clear consensus in the literature on whether galaxy clusters systematically host a higher fraction of IHL than groups \citep{2021Galax...9...60C,2022NatAs...6..308M}.} \textcolor{black}{On the other hand}, \citealt{2024MNRAS.527.2624P} used a kinematic approach to identify the IHL in the EAGLE simulation for haloes \textcolor{black}{at $z$ = 0, spanning the mass range $10^{11.2}~\mathrm{M}_\odot \lesssim \mathrm{M}_{\mathrm{200}} \lesssim  10^{15.4} ~\mathrm{M}_\odot$}. \textcolor{black}{In particular, for \textcolor{black}{$M_{\mathrm{halo}}$} $\sim$  9.4 $\times 10^{12}\,\mathrm{M_\odot}$, which is the halo mass of our analysed group, the authors found $0.1 \lesssim f_{\mathrm{IHL}} \lesssim 0.4$.} However, our median \textcolor{black}{$f_{\mathrm{IHL}}$} values are smaller than those reported by \citealt{2024A&A...683A..59C} and \citealt{2024MNRAS.528..771B}. \citealt{2024A&A...683A..59C} analysed a sample of $\sim$ 300 massive clusters \textcolor{black}{at $z$ = 0} from \textsc{the three hundred} data set, separating the BCG from the IHL using a 50 kpc aperture \textcolor{black}{and finding $0.3 \lesssim f_{\mathrm{IHL}} \lesssim 0.5$}. Interestingly, \citealt{2024MNRAS.528..771B} \textcolor{black}{compared intrinsic IHL fractions measured directly from cosmological simulations with those recovered using observational techniques applied to the same simulated data, finding that IHL fractions derived from cosmological hydrodynamical simulations tend to be higher than those inferred from observational studies. They further investigated possible sources of these offsets, including the adopted star formation model and projection effects, but found that neither can fully account for the observed discrepancies.} \textcolor{black}{In all of these theoretical works, since groups and clusters are analysed at $z = 0$, unlike our group located at $z \sim 0.21$, we do not expect significant evolution in the IHL component over this redshift interval, which would discard the possibility that systems at lower redshifts could have systematically higher $f_{\mathrm{IHL}}$.}}

%attributing this offset to several factors, including the limited simulation volume, the adopted star formation model, and projection effects.

\begin{figure*}
\centering  %\qquad
 \includegraphics[width=1\linewidth]{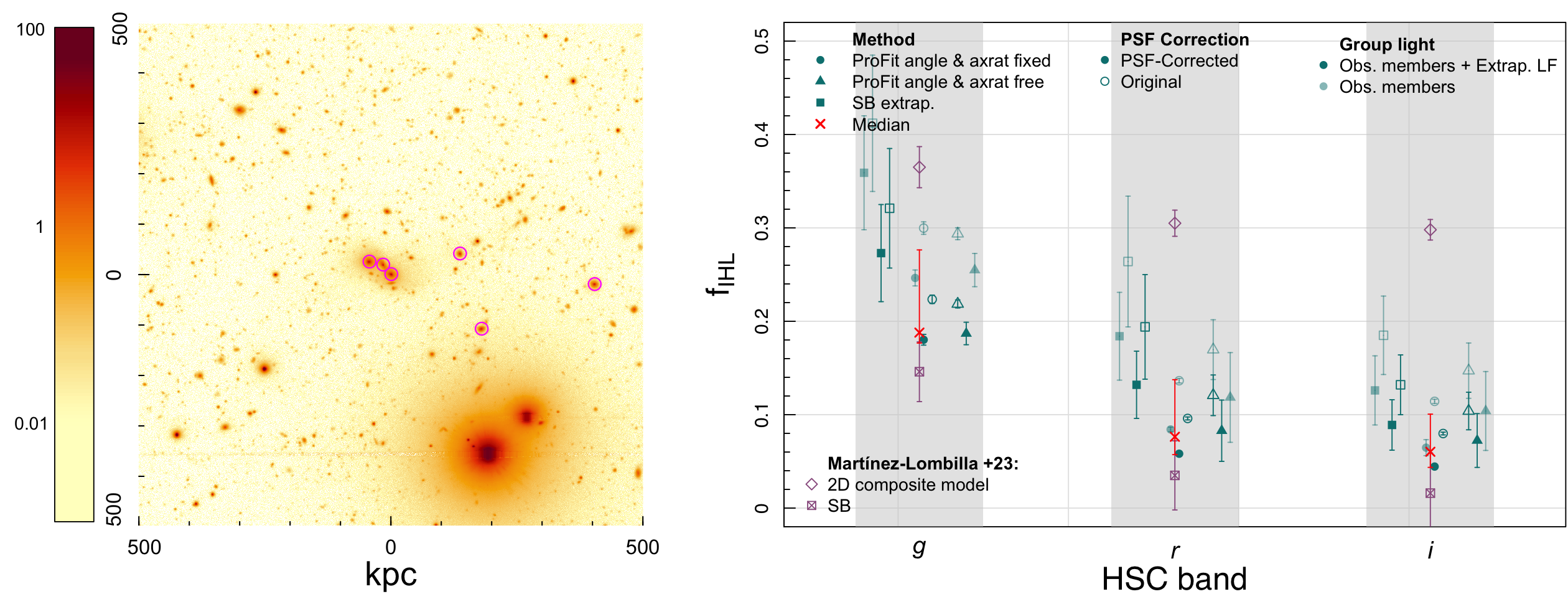}
    \caption{Estimations of the IHL fraction of the HSC UD GAMA group G400138. Left panel shows a 1 $\times$ 1 Mpc${^2}$ cutout of G400138 in the $\textit{g}$-band. \textcolor{black}{Galaxy members are highlighted with pink circles, while the two large objects in the lower right corner are bright stars.} Right panel shows the \textcolor{black}{$f_{\mathrm{IHL}}$} estimated in this work (green) and in \citealt{2023MNRAS.518.1195M} (purple). Our estimations obtained using \textsc{ProFit} with fixed angle and axial ratio are shown as circles, while those with angle and axial ratio as free parameters are shown as triangles. Those obtained using the SB limit method with the extrapolation towards the core are shown as squares. Filled symbols represent \textcolor{black}{$f_{\mathrm{IHL}}$} estimated using the PSF-corrected images, while empty symbols correspond to those estimated using the original versions. Solid colours represent estimations using \textcolor{black}{\textcolor{black}{$F_{\mathrm{group,extrap.}}$}}, while pale colours \textcolor{black}{correspond to estimations using \textcolor{black}{$F_{\mathrm{group}}$}}. The medians of our PSF-corrected \textcolor{black}{$f_{\mathrm{IHL}}$}, calculated using \textcolor{black}{\textcolor{black}{$F_{\mathrm{group,extrap.}}$}}, are shown as red crosses. Regarding the estimations from \citealt{2023MNRAS.518.1195M}, those obtained using the 2D composite model are shown as diamonds, while those from the SB limit method are shown as empty squares with a cross inside.
    } 
    \label{Comparison techniques}
\end{figure*}

\textcolor{black}{From an observational approach, our estimations are also consistent with several previous works. For example, \citealt{2004ApJ...609..617F} analysed the IHL component in four clusters at intermediate redshift using the SB limit method, finding \textcolor{black}{$f_{\mathrm{IHL}}$} \textcolor{black}{$\approx$} 0.05 -- 0.15 \textcolor{black}{with a Washington \textit{M} filter}. Using observations of six high-redshift clusters from the Hubble Frontier Fields (HFF) survey \textcolor{black}{and all the broad-band filters (from \textit{F435W} to \textit{F160W})}, \citealt{2018MNRAS.474..917M} measured an IHL fraction of $\mathrm{f_{IHL}} \sim 0.07$, consistent with earlier results for the same clusters by \citealt{2017ApJ...846..139M}. 
Other works have found larger \textcolor{black}{$f_{\mathrm{IHL}}$}, such as \citealt{2015MNRAS.449.2353B}, who found \textcolor{black}{$f_{\mathrm{IHL}}$} \textcolor{black}{$\approx$} 0.2 using \textcolor{black}{16-band photometry} from the Cluster Lensing And Supernova survey with Hubble (CLASH; \citealt{2012ApJS..199...25P}), and \citealt{2021MNRAS.502.2419F}, who measured the IHL fraction in HSC \textcolor{black}{\textit{i}-band} images of 18 clusters at intermediate redshift, finding \textcolor{black}{$f_{\mathrm{IHL}}$} \textcolor{black}{$\approx$} 0.24. \textcolor{black}{In some works, clusters are expected to have higher IHL fractions \citep{2007ApJ...666...20P,2018MNRAS.475..648P,2023MNRAS.518.3685A}, which could explain the higher fractions found. However, as mentioned above, there is still no clear consensus on a possible dependence of $f_{\mathrm{IHL}}$ on halo mass.} \textcolor{black}{It is important to note that only some of the observational studies discussed above explicitly accounted for the PSF in their analysis. For example, \citealt{2018MNRAS.474..917M} constructed extended PSF models and convolved the images with appropriate kernels to match their resolution across bands, while \citealt{2017ApJ...846..139M} incorporated the instrumental PSF into their \textsc{galfit} modelling and investigated the sensitivity of their IHL measurements to PSF variations. Similarly, \citealt{2021MNRAS.502.2419F} accounted for the PSF when fitting SB profiles, but explicitly stated that they did not deconvolve their data when computing the IHL. These studies therefore account for or mitigate PSF-related effects, although their approaches differ from ours, which explicitly corrects the IHL measurements for the extended PSF-scattered light.} As \textcolor{black}{shown} in Fig.~\ref{Comparison techniques}, \textcolor{black}{our} PSF-corrected values are \textcolor{black}{typically} $\sim$ 20\% smaller than the \textcolor{black}{corresponding values obtained without correcting for PSF-scattered light, indicating that differences in PSF treatment may partly contribute to the scatter among reported $f_{\mathrm{IHL}}$ values.} \textcolor{black}{It is \textcolor{black}{also} worth noting that our estimations correspond to a single system, whereas the works mentioned above typically report average IHL fractions derived from larger samples of groups or clusters.}}

%the non-corrected values, which could also explain the higher reported $f_{\mathrm{IHL}}$ values.} 

%In summary, \textcolor{black}{discrepancies in IHL fraction predictions largely arise from differences in definitions and measurement techniques}. It is important to note that significant progress in identifying realistic variations in IHL fractions between groups cannot be achieved until discrepancies due to differences in measurement techniques are resolved. As demonstrated in this work, the PSF effect can also significantly impact IHL measurements, meaning that different PSF treatments can lead to variations in the estimated IHL, particularly at lower \textcolor{black}{$f_{\mathrm{IHL}}$}. Furthermore, there is no clear agreement on how the IHL evolves with time or group/cluster mass. Statistically significant samples with the required depth across a wide range of cluster masses, redshifts and dynamical evolution stages are needed in order to understand how the IHL correlates with these properties.

In summary, \textcolor{black}{discrepancies in IHL fraction predictions largely arise from differences in definitions and measurement techniques}. It is important to note that significant progress in identifying realistic variations in IHL fractions between groups cannot be achieved until discrepancies due to differences in measurement techniques are resolved. \textcolor{black}{As demonstrated in this work, PSF treatment can also contribute to systematic differences in IHL measurements. In particular, Fig.~\ref{Comparison techniques} shows that the PSF-corrected \textcolor{black}{$f_{\mathrm{IHL}}$} values are typically $\sim$ 20\% smaller than the non-corrected values, indicating that different treatments of PSF scattering may partly contribute to the scatter between reported IHL fractions, particularly at lower \textcolor{black}{$f_{\mathrm{IHL}}$}.} \textcolor{black}{Statistically significant samples with sufficient SB sensitivity to robustly detect the faint IHL component are required. Such samples should span a wide range of cluster masses, redshifts, and dynamical stages to investigate how the IHL correlates with these properties.}

%\textcolor{black}{Statistically significant samples with sufficient surface-brightness sensitivity to robustly detect the faint IHL component across a wide range of cluster masses, redshifts, and dynamical stages are required to study how the IHL correlates with these properties.}

%Statistically significant samples with the required depth across a wide range of cluster masses, redshifts and dynamical evolution stages are needed in order to understand how the IHL correlates with these properties.

%Furthermore, there is no clear agreement on how the IHL evolves with time or group/cluster mass.

%It is known that the measured IHL fraction could vary significantly even within a single cluster, depending on the method used. \citep{2011ApJ...732...48R,2021ApJS..252...27K, 2024MNRAS.528..771B}.

%This allows us to analyse the IHL in the original image of G400138 while also analysing the IHL in a version where the PSF-scattered flux has been removed. This analysis aims to estimate the impact of the PSF-scattered flux when this is not removed from the image. Using the GAMA Galaxy Group Catalogue G$^3$Cv10, we estimate the IHL fraction of X before and after correcting for the PSF effect. 

\section{Conclusions}
\label{sec:Conclusions}
In this paper, we present a method to model the IHL component in a \textcolor{black}{galaxy group} or cluster, unbiased by the PSF-scattered flux of the telescope, \textcolor{black}{and use this methodology to quantify the impact of the PSF on IHL estimations.} This methodology consists of two techniques: \textcolor{black}{the first aims to correct most of the scattered light from the PSF effect in the image}, while the second models the IHL light following an exponential profile using \textcolor{black}{a 2-D galaxy profiling tool}. These two techniques can be applied independently to respectively correct for the PSF effect in any astronomical image or to accurately model the IHL component in any group or cluster image. However, if the goal of the user is to measure the IHL, we recommend applying the PSF correction prior to the modelling, as this will remove \textcolor{black}{most of} the PSF faint light that contributes with additional flux and \textcolor{black}{biases} the estimations.

\textcolor{black}{As the goal of this paper is to measure the impact of PSF-scattered flux on IHL estimations, we generated a dataset of 5440 HSC mock groups with an analytically injected circular exponential IHL component, and modelled the IHL using a circular exponential profile with \textsc{ProFit}. This idealised scenario, in which we fit a circular IHL component with a circular model, allows us to better isolate the effect of the PSF and quantify its impact on the IHL properties. Nevertheless, we also demonstrate the robustness of \textsc{ProFit} in fitting non-circular and more realistic IHL components.}

\textcolor{black}{The galaxies in our mock dataset were generated using real \textcolor{black}{\textit{r}-band} HSC data and the actual GAMA spatial configuration of 544 groups. However, we analytically injected the IHL component, sampling each group image ten times with a circular exponential IHL profile at varying IHL fractions ranging from 0.01 to 0.5. This results in our total dataset of 5440 \textit{r}-band HSC-SSP PDR3 mock groups. This approach allows us to capture the diversity of morphologies, SB profiles, and clustering properties using actual data while having control over the IHL properties (flux and size) that we later attempt to model. We apply our PSF-scattered flux removal technique to the mock dataset, generating a PSF-corrected dataset.} In parallel, we preserve the original dataset without applying the PSF correction, \textcolor{black}{while simply applying a masking technique,} serving as a control sample. We then apply the IHL estimation techniques to both datasets, allowing us to compare the resulting IHL fitted models with and without the PSF correction applied, and thereby quantifying the impact of PSF-scattered light on IHL measurements. Our results show that removing the PSF-scattered flux significantly improves the accuracy of the size and flux estimates of the IHL component in the mock groups. \textcolor{black}{Without the PSF correction, the IHL model substantially overestimates the true flux and effective radius. For the faintest IHL component ($f_{\mathrm{IHL}} = 0.01$), the recovered flux and effective radius are larger than their true values by factors of $\sim$84.4 and $\sim$9.1, respectively. For the most prominent component ($f_{\mathrm{IHL}} = 0.5$), the recovered flux is $\sim$0.87 times the true flux, while the recovered effective radius is $\sim$2 times the true value. More generally, our analysis shows that the PSF correction is not required for recovering the flux when $f_{\mathrm{IHL}} > 0.4$, whereas it is required for the effective radius across the full range of IHL fractions considered. After applying our PSF correction, the recovered flux and effective radius are much closer to their true values. For $f_{\mathrm{IHL}} = 0.01$, both quantities are recovered at $\sim$0.89 times their true values, while for $f_{\mathrm{IHL}} = 0.5$, the recovered flux and effective radius are $\sim$0.95 and $\sim$1 times their true values, respectively.}

%This way we are able to compare the injected parameters versus the recovered parameters when we do not apply the PSF correction (PSF-corrected sample) and the recovered parameters when we apply it (original sample). 

%\textcolor{mycolor}{In addition, we demonstrate the robustness of \textsc{ProFit} in fitting non-circular IHL components. To do so, we generated a small sample of HSC mock groups with an injected elliptical exponential IHL.} Our results show that the IHL detection technique also performs effectively in this scenario, achieving good agreement in capturing both the flux and the shape of the elliptical IHL.

%we investigate the flexibility of \textsc{ProFit} to fit elliptical exponential IHL components with a small sample of HSC mock observations \textcolor{mycolor}{with an injected elliptical exponential IHL}. 

Finally, we apply our IHL estimation technique to measure the PSF-corrected IHL fraction of the G400138 GAMA group using HSC-SSP PDR3 UD data in the \textit{g,r}, and \textit{i}-bands. \textcolor{black}{Consistent with the flexibility of \textsc{ProFit} in modelling elliptical extended sources,} we allow \textsc{ProFit} to model not only the flux and size, but also the shape and orientation of the IHL. We also estimate \textcolor{black}{$f_{\mathrm{IHL}}$} of G400138 using the surface brightness limit method, with the additional step of accounting for the IHL light beneath the core masks through an exponential fit. We compare our IHL fractions with those obtained in \citealt{2023MNRAS.518.1195M}, finding the same decreasing trend towards redder wavelengths as the authors. In addition, our \textcolor{black}{$f_{\mathrm{IHL}}$} values fall within their range of \textcolor{black}{$f_{\mathrm{IHL}}$}. Differences in the techniques applied likely account for the remaining discrepancies between the two works. On average, our IHL fractions of G400138 are: \textcolor{black}{$f_{\mathrm{g,IHL}}$} \textcolor{black}{$\approx$} 0.19$^{+0.09}_{-0.01}$, \textcolor{black}{$f_{\mathrm{r,IHL}}$} \textcolor{black}{$\approx$} 0.08$^{+0.06}_{-0.02}$, and \textcolor{black}{$f_{\mathrm{i,IHL}}$} \textcolor{black}{$\approx$} 0.06$^{+0.04}_{-0.02}$.

%Our results show that the elliptical IHL model finds a \textcolor{black}{$f_{\mathrm{IHL}}$} $\sim$ 58\% larger than the \textcolor{black}{$f_{\mathrm{IHL}}$} from the circular IHL model. Also, when we do not correct for the PSF effect, the original \textcolor{black}{$f_{\mathrm{IHL}}$} is up to $\sim$ 73\% larger than the PSF-corrected \textcolor{black}{$f_{\mathrm{IHL}}$}.

In this paper, we have demonstrated the importance of removing the PSF flux when measuring the IHL component of a galaxy group or cluster, especially when \textcolor{black}{the IHL fraction is low.} \textcolor{black}{By following our methodology, we are able to determine the IHL with and without the PSF correction, with the true IHL value expected to be better represented by the PSF-corrected estimation. Additionally, we demonstrated that our methodology can be applied to large mock datasets at low computational cost, suggesting that the same approach could be extended to the analysis of IHL components in large observational surveys.}

%Through this analysis, we have also demonstrated that this methodology \textcolor{mycolor}{can be applied to analyse the IHL component of large datasets. }

%is well-suited for the automated IHL analysis of large datasets.

\section*{Acknowledgements}
LPGN, ASGR, and SB acknowledge support from the ARC Future Fellowship scheme (FT200100375). SB acknowledges funding by the Australian Research Council (ARC) Laureate Fellowship scheme (FL220100191). LJMD acknowledges support from the ARC Future Fellowship scheme (FT200100055). Parts of this research were supported by the Australian Research Council Centre of Excellence for All Sky Astrophysics in 3 Dimensions (ASTRO 3D), through project number CE170100013.

This paper is based on data collected at the Subaru Telescope and retrieved from the Hyper Suprime-Cam (HSC) data archive system. The Hyper Suprime-Cam (HSC) collaboration includes the astronomical communities of Japan and Taiwan, and Princeton University. The HSC instrumentation and software were developed by the National Astronomical Observatory of Japan (NAOJ), the Kavli Institute for the Physics and Mathematics of the Universe (Kavli IPMU), the University of Tokyo, the High Energy Accelerator Research Organization (KEK), the Academia Sinica Institute for Astronomy and Astrophysics in Taiwan (ASIAA), and Princeton University. Funding was contributed by the FIRST program from the Japanese Cabinet Office, the Ministry of Education, Culture, Sports, Science and Technology (MEXT), the Japan Society for the Promotion of Science (JSPS), Japan Science and Technology Agency (JST), the Toray Science Foundation, NAOJ, Kavli IPMU, KEK, ASIAA, and Princeton University. This paper makes use of software developed for Vera C. Rubin Observatory. We thank the Rubin Observatory for making their code available as free software at \,\url{http://pipelines.lsst.io/}. This paper is based on data collected at the Subaru Telescope and retrieved from the HSC data archive system, which is operated by the Subaru Telescope and Astronomy Data Center (ADC) at NAOJ. Data analysis was in part carried out with the cooperation of Center for Computational Astrophysics (CfCA), NAOJ. We are honored and grateful for the opportunity of observing the Universe from Maunakea, which has the cultural, historical and natural significance in Hawaii. 

We have used catalogues from GAMA, a joint European-Australasian project based around a spectroscopic campaign using the Anglo-Australian Telescope. The GAMA input catalogue is based on data taken from the Sloan Digital Sky Survey and the UKIRT Infrared Deep Sky Survey. Complementary imaging of the GAMA regions is being obtained by a number of independent survey programmes including GALEX MIS, VST KiDS, VISTA VIKING, WISE, Herschel-ATLAS, GMRT and ASKAP providing UV to radio coverage. GAMA is funded by the STFC (UK), the ARC (Australia), the AAO, and the participating institutions. The GAMA website is \,\url{https://www.gama-survey.org/}.

We have also used catalogues from WAVES, a joint European-Australian project based around a spectroscopic campaign using the 4MOST. The WAVES input catalogue is based on data taken from the European Southern Observatory’s VST and VISTA telescopes. Complementary imaging of the WAVES regions is being obtained by a number of independent survey programmes including GALEX MIS, VST, WISE, HerschelATLAS, and ASKAP providing UV to radio coverage. WAVES is funded by the ARC (Australia) and the participating institutions. The WAVES website is \,\url{https://wavesurvey.org}. Based on observations made with ESO Telescopes at the La Silla Paranal Observatory under
programme ID 179.A-2004 and ID 177.A-3016.

This work was supported by resources provided by the Pawsey Supercomputing Research Centre’s Setonix Supercomputer (\,\url{https://doi.org/10.48569/18sb-8s43}), with funding from the Australian Government and the Government of Western Australia.

All of the work presented here was made possible by the free and open R software environment \citep{R-core}. All figures in this paper were made using the R \textsc{magicaxis} package \citep{2016ascl.soft04004R}. This work also makes use of the \textsc{celestial} package \citep{2016ascl.soft02011R}. The software tools used are: \textsc{ProFit}: \url{https://github.com/ICRAR/ProFit}, \citealt{2017MNRAS.466.1513R} (LGPL-3); \textsc{ProFound}: \url{https://github.com/asgr/ProFound}, \citealt{2018MNRAS.476.3137R} (LGPL-3); \textsc{ProPane}: \url{https://github.com/asgr/ProPane}, \citealt{2024MNRAS.528.5046R} (LGPL-3); \textsc{Rfits}: \url{https://github.com/asgr/Rfits} (LGPL-3); \textsc{RWCS}: \url{https://github.com/asgr/Rwcs} (LGPL-3). All associated \textsc{protools} software (v0.2) is available on GitHub for use immediately: \url{https://github.com/asgr/ProTools}; \citealt{robotham2023a}.

%%%%%%%%%%%%%%%%%%%%%%%%%%%%%%%%%%%%%%%%%%%%%%%%%%
\section*{Data Availability}
The mock observations were recreated using images from the Hyper Suprime-Cam Public Data Release 3 (HSC-PDR3; \citealt{2022PASJ...74..247A}). The catalogue information was extracted from the Galaxy And Mass Assembly (GAMA) survey Galaxy Group Catalogue (G$^3$Cv10; \citealt{2011MNRAS.413..971D}). The WAVES data will be made available in an upcoming data release.



\bibliographystyle{mnras}
\bibliography{paper} % if your bibtex file is called example.bib

%\begin{thebibliography}{99}
%\end{thebibliography}

%%%%%%%%%%%%%%%%%%%%%%%%%%%%%%%%%%%%%%%%%%%%%%%%%%

%%%%%%%%%%%%%%%%% APPENDICES %%%%%%%%%%%%%%%%%%%%%

\appendix

\section{ProFit: IHL non-circular profile fitting}
\label{sec:non-circular}

\textcolor{black}{The IHL estimation technique uses an exponential model to fit this component. In Sec.~\ref{case-study}, we focus on the performance of \textsc{ProFit} when fitting circular IHL components, while in Sec.~\ref{full dataset} we \textcolor{black}{again model the IHL of mock group images with a circular profile, using the same profile adopted for the injection in order to isolate the impact of PSF-scattered light.} In this section, we instead focus on the performance of \textsc{ProFit} when fitting more realistic, non-circular IHL components. This extension is motivated by the fact that real IHL distributions are generally not perfectly axisymmetric, and allowing ellipticity provides a more flexible smooth description of their global structure. We note that additional non-smooth substructures (e.g. tidal streams and plumes) may also be present in real systems, but these are not expected to be well captured by either circular or elliptical parametric models and are therefore beyond the scope of the present analysis.}

%As mentioned in Sec.~\ref{sec:IHL analysis}, systems that have suffered recent galactic interactions where the IHL formation is an ongoing process, might exhibit features, such as tidal streams and plumes, that are not properly modelled with a circular IHL component.}

%\textcolor{black}{In Sec.~\ref{case-study}, we focus on the performance of \textsc{ProFit} when fitting circular IHL components, and in Sec.~\ref{full dataset} we isolate the impact of PSF-scattered light by constructing an idealised setup in which the IHL is modelled with the same profile used for injection. In this section, we instead focus on the performance of \textsc{ProFit} when fitting more realistic, non-circular IHL components.}

 %\textcolor{mycolor}{In this section, we analyse the flexibility of \textsc{ProFit} to fit non-circular IHL components.}
 
  %Regarding the flexibility of our IHL estimation technique, as mentioned in Sec.~\ref{sec:IHL analysis}, s

 \textcolor{black}{To do this, we generate new mock observations with an inclined and elongated IHL component and fit this component with an elliptical model. For completeness, we also analyse the results of fitting the same component with a circular model.} \textcolor{black}{To generate the new set of mocks with an inclined and elongated IHL, we use the same methodology presented in Section \ref{sec:Data}. That is, we take one of the mock observations that has the real configuration of a GAMA group with real HSC data, but instead of adding an analytical circular IHL component, we inject an analytical IHL component with an inclined elongated shape.} We create a set of 10 new mocks, each with a different IHL fraction (ranging from 0.01 to 0.5). To inject this component, we again use the 2D image-generation mode of \textsc{ProFit} with a Sérsic template. In this case, however, we adopt a minor-to-major axis ratio of \textcolor{black}{$A_{\mathrm{rat}}$} = 0.5, instead of \textcolor{black}{$A_{\mathrm{rat}}$} = 1, and a position angle of $\textit{$\theta$}$ = 45$^\circ$, instead of $\textit{$\theta$}$ = 0$^\circ$. We use the configuration of the GAMA group identified by ID G200043 (\textcolor{black}{$N_{\mathrm{fof}}$} = 19, \textcolor{black}{$M_{\mathrm{halo}}$} $\sim$  3.4 $\times 10^{14}\,\mathrm{M_\odot}$ and $z$ $\sim$ 0.24), resulting in a set of 10 G200043 mocks with the new IHL component.

 As mentioned above, \textcolor{black}{we use \textsc{ProFit} in the fitting mode with} two different templates to model the IHL in this set of mocks. The first is the circular exponential template model with \textcolor{black}{$A_{\mathrm{rat}}$} = 1, $\textit{$\theta$}$ = 0$^\circ$ and free $\textit{mag}$ and \textcolor{black}{$R_{\mathrm{eff}}$}. The second one is the elliptical exponential template, where \textcolor{black}{$A_{\mathrm{rat}}$} and $\textit{$\theta$}$ are also left as free parameters.

\begin{figure*}
\centering  %\qquad
 \includegraphics[width=1\linewidth]{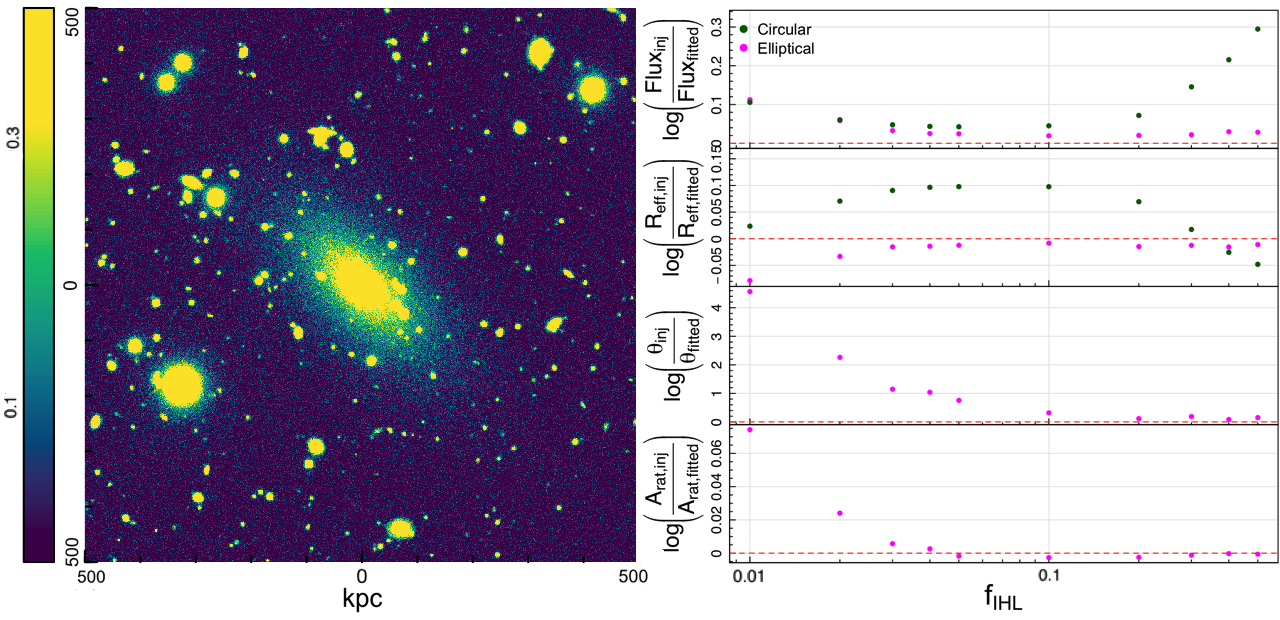}
    \caption{Comparison between fitting the
    inclined, elongated IHL of G200043 mock (left panel) with \textcolor{black}{\textsc{ProFit} using} an elliptical model  versus a circular model. Right panels show the ratios between the injected and fitted values for the flux (first), effective radius (second), position angle (third) and axial ratio (fourth) when fitting with an elliptical IHL model (pink) versus a circular IHL model (green). \textcolor{black}{The red dashed lines mark the zero point in each panel.} Our results are more accurate for the elliptical IHL model, especially towards higher fractions.}
    \label{caveat}
\end{figure*}

 The left panel of Fig.~\ref{caveat} shows the G200043 mock with an IHL of \textcolor{black}{$A_{\mathrm{rat}}$} = 0.5, $\textit{$\theta$}$ = 45$^\circ$,  and \textcolor{black}{$f_{\mathrm{IHL}}$} = 0.5. In the right panel, we compare the ratio between the injected and fitted parameters for the circular fitted IHL model in green versus the elliptical fitted IHL model in pink. The first panel shows the comparison for the flux, the second for the effective radius, the third for the position angle, and the fourth for the axial ratio. The elliptical model performs better as the IHL fraction increases because it is more flexible in capturing the elongated shape of the IHL, especially as the structure becomes more prominent. In contrast, the circular model struggles to capture the full flux of the IHL at higher fractions, leading to an increasing discrepancy between the flux ratios. At \textcolor{black}{$f_{\mathrm{IHL}}$} = 0.5, the injected flux is $\sim$ 100\% larger than the fitted flux, meaning that the circular model recovers only about half of the true flux. In the case of the effective radius, the injected value is $\sim$ 25\% larger than the fitted value. \textcolor{black}{We do not include the ratio comparison for the position angle and axial ratio for the circular model as these parameters are fixed in this model. From this test, we conclude that \textsc{ProFit} also performs well when fitting non-circular IHL components.}

%Top panels show the GAMA mock group 200043 with circular injected IHL (left) and elongated injected IHL (right).

%The model being simplistic will have effects too though, yes. You could comment on that by generating a pretty extreme model (axrat of 0.5 say) and seeing what bias that has on the implied IHL flux etc.

%Aaron: I more meant that we can use our technique to state what the implied IHL is with and without PSF correction, and the truth likely sits somewhere in between in general (probably much nearer to the corrected version though). The model being simplistic will have effects too though, yes. You could comment on that by generating a pretty extreme model (axrat of 0.5 say) and seeing what bias that has on the implied IHL flux etc.

%comparar en algun momento como cambia la fitted mag y radio sin aplicar la psf correction y luego de aplicar y resaltar la importance de descartar esta luz contaminants

%capaz estaria bueno hacer tabla para el group ID que meustra en la image donde dice para cada fraction de ihl, el effective radius y magnitude injectado versus el eff radius y magnitude fitteados. ams el dibujo!

\section{Case studies with IHL estimation limitations}
\label{B}
\begin{figure*}
\centering  %\qquad
 \includegraphics[width=1\linewidth]{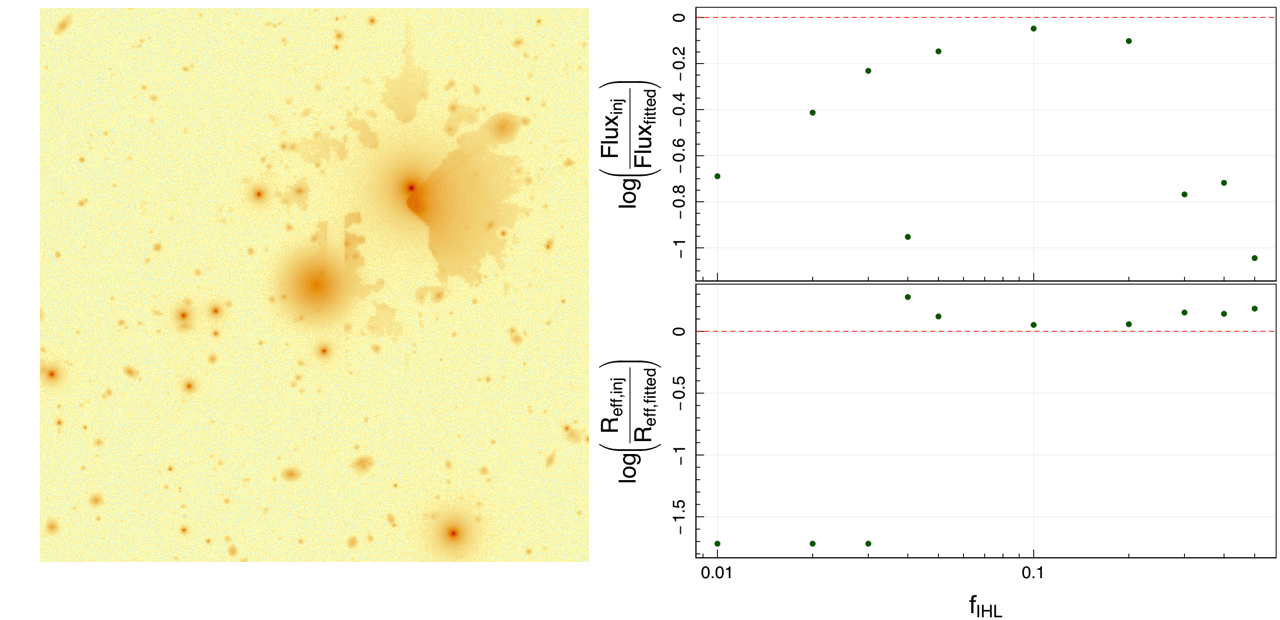}
    \caption{\textcolor{black}{Case study of the mock constructed using the GAMA group configuration from G100224. The left panel shows the G100224 mock with an injected IHL component of $f_\mathrm{IHL}$ = 0.5. The right panel presents the results of fitting the IHL component to the ten G100224 mocks, each with a different injected IHL fraction. The upper plot compares the injected and fitted flux values, while the lower plot compares the injected and fitted effective radii. \textsc{ProFit} struggles to accurately fit the IHL component in the G100224 mocks at both low and high \textcolor{black}{$f_{\mathrm{IHL}}$}, primarily due to the presence of the fragmented bright star in the upper right side of the image.}}
    \label{G100224}
\end{figure*}

\textcolor{black}{In a small subset of our mock group image dataset ($\sim $5\% of our 544 GAMA groups) where extended objects occupy a large fraction of the group image, \textsc{ProFit} struggles to accurately fit the flux and size of the IHL model even when the S/N of the IHL component is high. \textcolor{black}{For instance, bright stars are sometimes fragmented in the WAVES star/galaxy catalogue \textcolor{black}{(generated using \textsc{ProFound})} and consequently classified as ambiguous objects\footnote{\textcolor{black}{The star/galaxy separation used here corresponds to the initial WAVES classification, based on the procedure developed for GAMA \citep{2020MNRAS.496.3235B}. This early version had known limitations, including photometry difficulties near bright stars, which motivated the development of a more sophisticated classification method to be presented in Bellstedt et al. in prep.}}. In our mock-generation process, we treat all ambiguous sources as galaxies, which causes these fragmented star components to appear as extended objects in the mocks. This occurs because galaxies are generated using real HSC cutouts, while stars are modelled analytically as point sources. As a result, any bright-star fragments treated as galaxies introduce stellar contamination in the mocks, which can bias the recovered IHL properties.} This is the case, for example, of the mocks generated using the configuration of the GAMA group G100224. We show the G100224 mock with an injected IHL component of $f_\mathrm{IHL} = 0.5$ in the left panel of Fig.~\ref{G100224}, where the bright star in the upper-right region of the group image is misclassified as multiple galaxies.  In the right panel, we compare the injected and fitted values of the flux (upper plot) and effective radius (lower plot) for the ten mocks generated from G100224. The fitted flux and size are accurately recovered only for intermediate \textcolor{black}{$f_{\mathrm{IHL}}$} values. Fortunately, these are only a few cases in our dataset of mocks and do not significantly affect the median IHL estimations presented in Section \ref{sec:results}.}

\section{SB limit method with extrapolation towards the core}
\label{C}
\textcolor{black}{In Section \ref{sec:application}, we fit the IHL component of the G400138 group using real HSC-UD data and applying the SB limit method in the \textit{g}, \textit{r}, and \textit{i}-bands.} By definition, this method does not account for all the IHL light that lies beneath the masked regions. In particular, as predicted in the literature \citep{2024A&A...683A..59C, 2024MNRAS.527.2624P}, the quantity of IHL light beneath the masks of the core galaxies can represent a significant percentage, which would have an impact in the estimation of the \textcolor{black}{$f_{\mathrm{IHL}}$}. To compensate for the unaccounted IHL flux of G400138 \textcolor{black}{when applying the SB limit method}, \textcolor{black}{in Section \ref{sec:application}} we fit an exponential profile to the unmasked flux of group image (which is assumed to be the IHL) and extrapolate this towards the center of the image. \textcolor{black}{This approach allows us to include the masked IHL contribution in the \textcolor{black}{$f_{\mathrm{IHL}}$} estimation.}

\begin{figure}
\centering  %\qquad
 \includegraphics[width=1\linewidth]{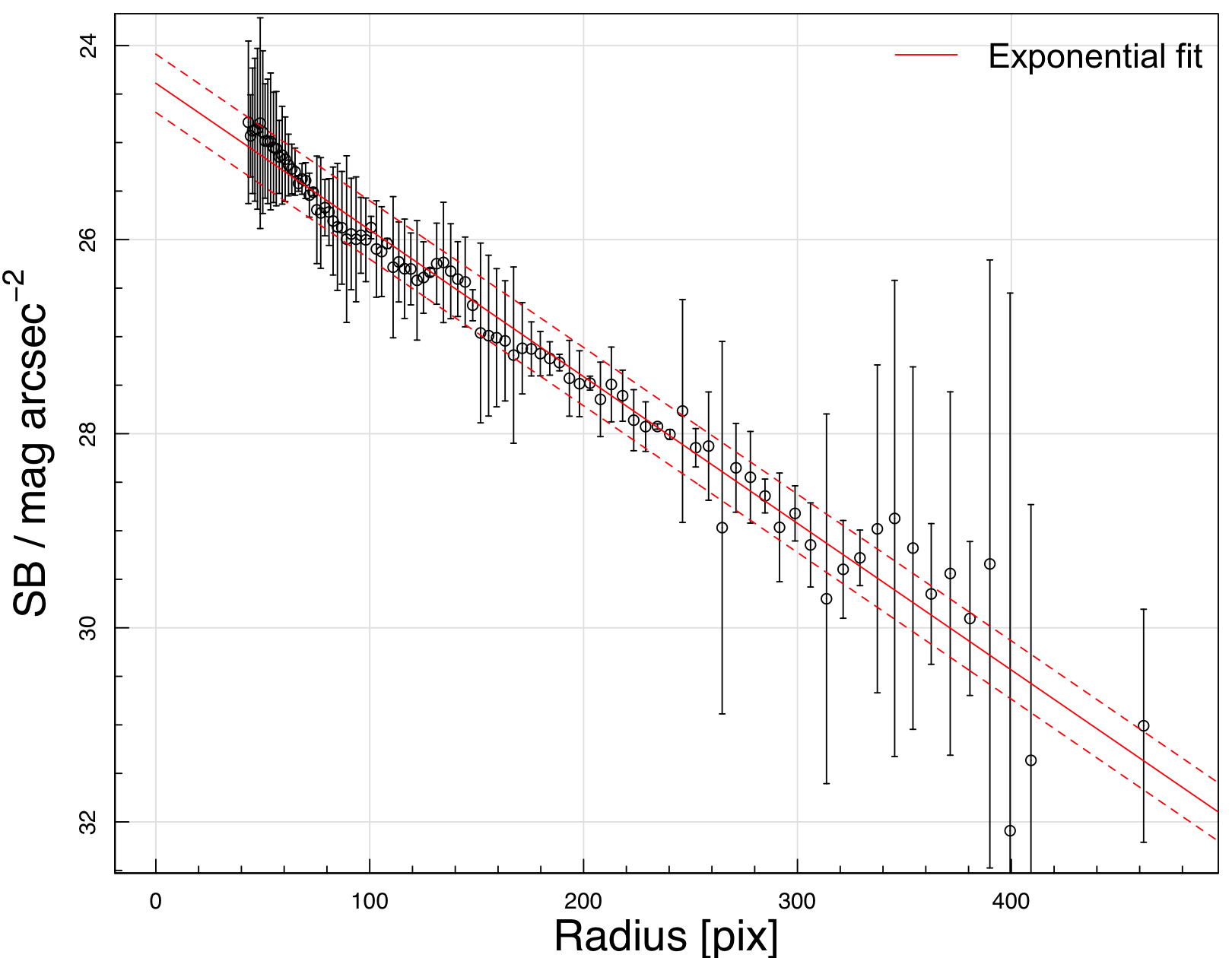}
    \caption{\textit{g}-band SB profile of the IHL flux of G400138 using HSC-PDR3 UD data. Each circle represents the median IHL value within its corresponding aperture.  As we follow the SB limit method, the IHL counts do not extend towards the center of the group. The red line shows the exponential fit to these points, allowing us to account for the IHL flux that lies beneath the core masks. The dashed red lines indicate the fitting uncertainties. This profile is corrected for the PSF effect, galactic extinction, SB dimming, and \textit{K}-corrected.}
    \label{hyperfit}
\end{figure}

\textcolor{black}{Here, we explain how we estimate the extrapolated IHL light towards the core of the system. After masking all sources in the group image of G400138, we generate the SB profile of this component. As an example, we show the \textit{g}-band SB profile of G400138 in Fig.~\ref{hyperfit}. All IHL SB profiles in the \textit{g,r,i}-bands have been corrected for the PSF effect, galactic extinction, SB dimming, and \textit{K}-corrected. We use the same corrections as \citealt{2023MNRAS.518.1195M}: dimming in the form (1+z)$^{-4}$ \citep{1930PNAS...16..511T,1934rtc..book.....T}; \textit{K}-correction:  \textit{K}-corr$_{g}$ = 0.68 mag, \textit{K}-corr$_{r}$ = 0.29 mag, and \textit{K}-corr$_{i}$ = 0.13 mag \citep{2012MNRAS.420.1239L}; and Galactic absorption: \textit{A}$_{g}$ = 0.092 mag, \textit{A}$_{r}$ = 0.066 mag, and \textit{A}$_{i}$ = 0.050 mag \citep{2018MNRAS.474.3875B}. We then use \textsc{hyper-fit} to fit an exponential model to this component}, which via Bayesian inference uses a likelihood analysis to minimise the scatter in the fitted function. The extrapolated best-fitting model is shown in red in Fig.~\ref{hyperfit}, and the associated fitting uncertainties are indicated by dashed black lines. Given how well the \textcolor{black}{red model fits the IHL SB profile of G400138}, we suggest that an exponential model is a reasonable \textcolor{black}{choice for fitting this component with \textsc{ProFit} in our IHL estimation technique.}

% approximation for the IHL component of our dataset of HSC-GAMA mocks. 

%In addition, we show the exponential models of the IHL SB profiles of G400138 in the $\textit{g,r}$ and $\textit{i}$-bands in Fig.~\ref{SB profiles}. 

%\begin{figure}
%\centering  %\qquad
 %\includegraphics[width=1\linewidth]%{Paper2/Figures/SB_extrapolated_profiles_comp.png}
 %   \caption{$\textit{g,r}$ and $\textit{i}$-bands exponential models of the IHL component of G400138 using HSC-PDR3 UD data. We use an exponential fit to model the IHL flux and compensate for the missing flux beneath the core masks. All profiles have been corrected for the PSF effect, galactic extinction, surface brightness dimming, and \textit{K}-corrected.}
 %   \label{SB profiles}
%\end{figure}

\begin{figure}
\centering  %\qquad
 \includegraphics[width=1\linewidth]{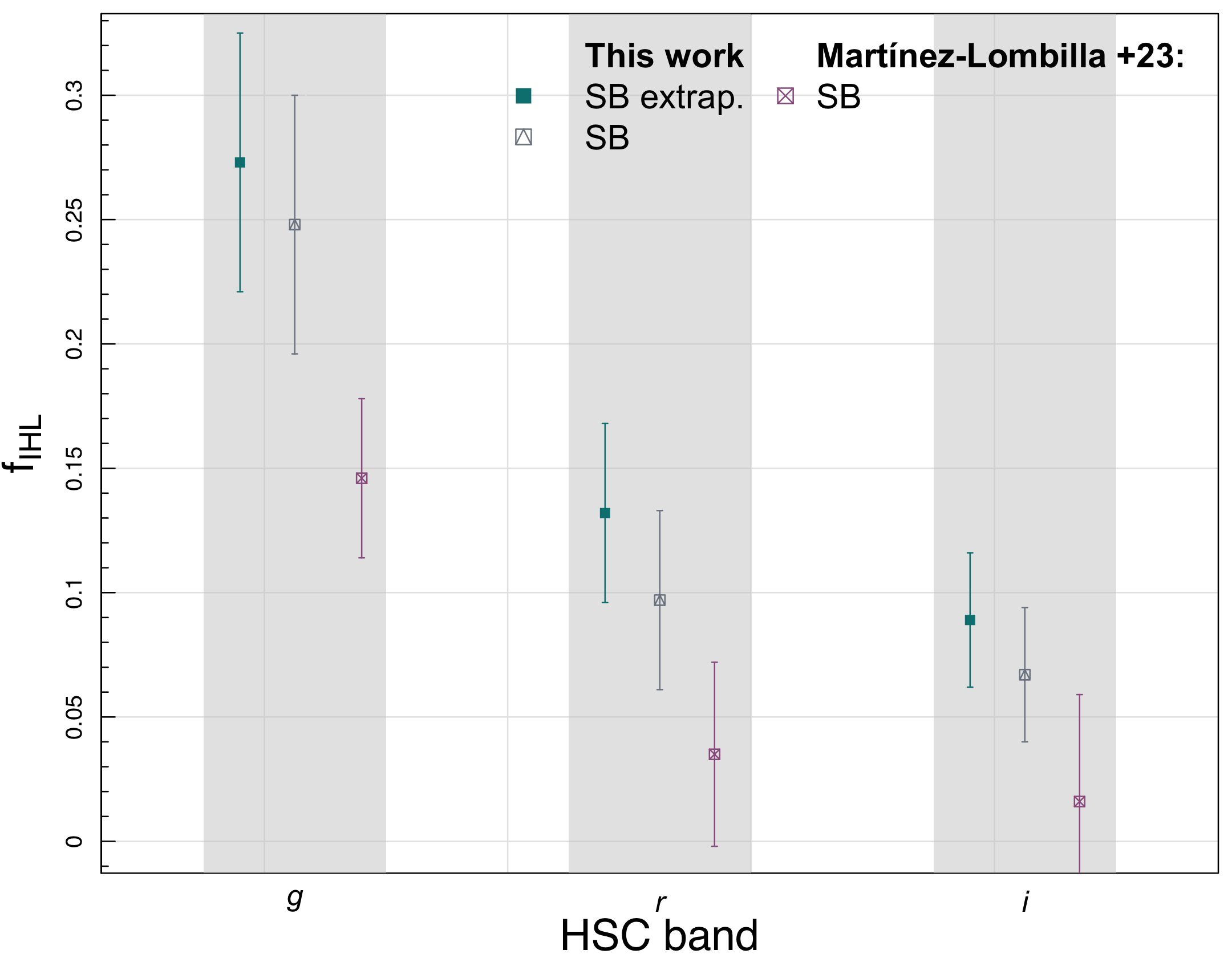}
    \caption{\textcolor{black}{$f_{\mathrm{IHL}}$} of G400138 obtained with the SB limit method. Green and gray are measurements from this work, while purple are from \citealt{2023MNRAS.518.1195M}. Green \textcolor{black}{$f_{\mathrm{IHL}}$} include an estimation of the extrapolated IHL that lies beneath the core masks, while gray values do not, resulting in systematically lower measurements.}
    \label{fraction extrapolated}
\end{figure}

When the extrapolated IHL component is included in our \textcolor{black}{$f_{\mathrm{IHL}}$} estimations, the resulting values increase, as shown in Fig.~\ref{fraction extrapolated}, where we compare \textcolor{black}{$f_{\mathrm{IHL}}$} derived using the SB limit method without the extrapolated light (gray) and with it included (green). All measurements have been PSF-corrected and consider the total group light as the sum of the light from confirmed members plus the extrapolated contribution from faint sources \textcolor{black}{(i.e. $F_\mathrm{{group,extrap.}})$}. For comparison, we include the \textcolor{black}{$f_{\mathrm{IHL}}$} from \citealt{2023MNRAS.518.1195M} derived using the same method in purple. Our \textcolor{black}{$f_{\mathrm{IHL}}$} estimations that exclude the extrapolated IHL lie within the error bars of their results in the \textit{r} and \textit{i}-bands. However, our \textit{g}-band \textcolor{black}{$f_{\mathrm{IHL}}$} are still higher than their estimations, which can be related to intrinsic differences in the SB technique, PSF treatment plus differences between HSC-UD PDR2 and PDR3 data.

\section{Comparison of our PSF correction method with Wiener deconvolution}
\label{D}
\textcolor{black}{In Sec.~\ref{Comparison with existing PSF-correction methods}, we discuss the differences between our PSF-correction method and other approaches for mitigating PSF-scattered flux in an image, including various deconvolution algorithms. In this section, we focus on the Wiener deconvolution method to quantitatively compare the results obtained by applying our PSF-scattered flux removal technique and the Wiener algorithm to the same mock group image. This comparison allows us to assess the robustness of our method relative to a deconvolution approach, which may introduce artefacts and can be highly sensitive to the adopted PSF model \citep{Bredell_2023_WACV}.}

\begin{figure}
\centering  %\qquad
 \includegraphics[width=1\linewidth]{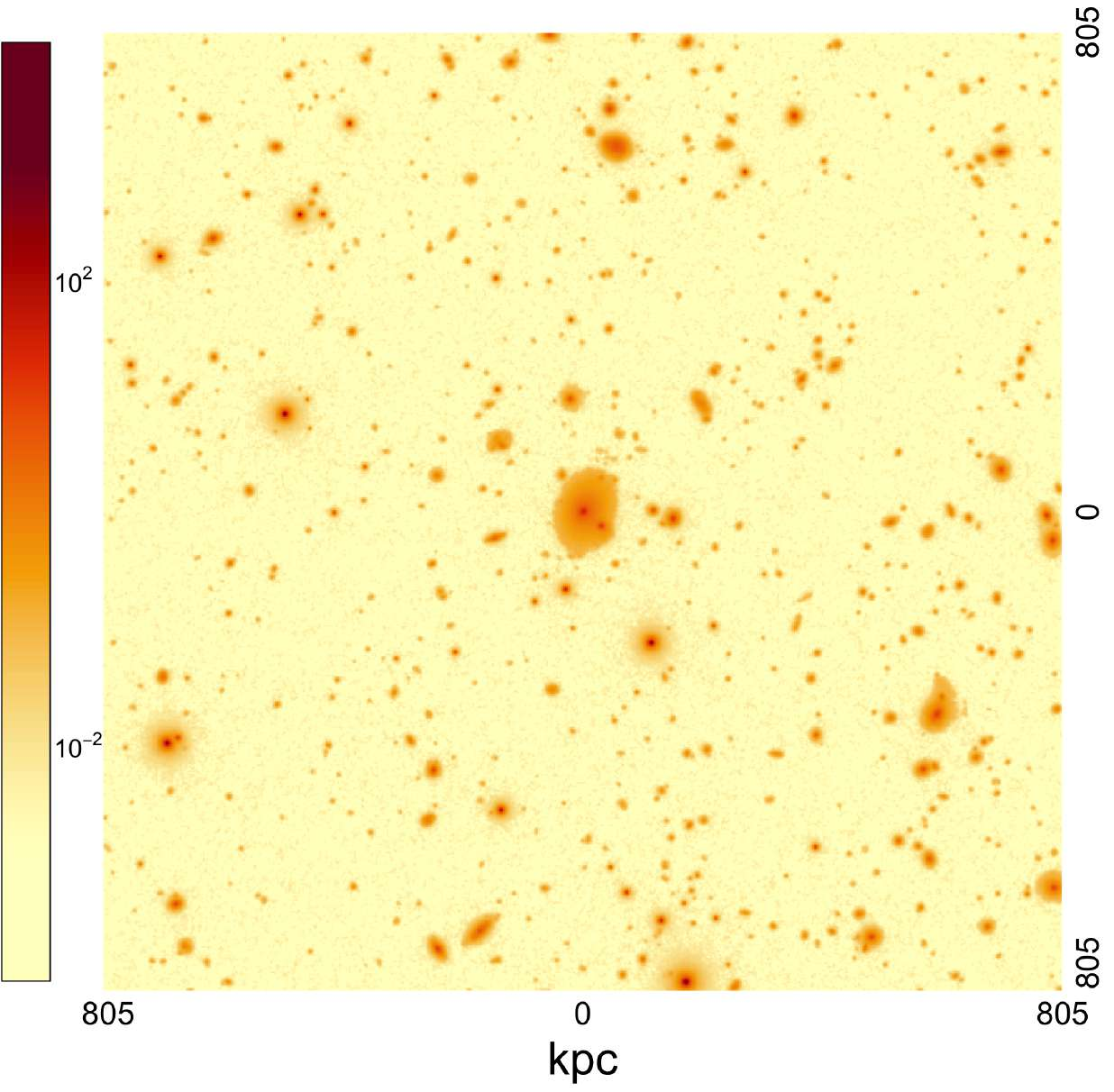}
\caption{\textcolor{black}{HSC mock group convolved image created using data from the GAMA group identified by ID 200133.}}
 \label{deconv group}
 \end{figure}

\textcolor{black}{For this comparison, we use an \textit{r}-band HSC mock image of a GAMA galaxy group. To do this, we first take the mock intrinsic version of the image and convolve it with the \textit{r}-band HSC PSF model used in this work and introduced in \citealt{2024MNRAS.531.2517G}. This mock is constructed using the configuration of the GAMA group identified by the ID 200133 ({$N_{\mathrm{fof}}$} = 11, \textcolor{black}{$M_{\mathrm{halo}}$} $\sim$  5.8 $\times 10^{13}\,\mathrm{M_\odot}$, and $z$ $\sim$ 0.13). Fig. \ref{deconv group} shows the PSF-convolved version of the G200133 mock group image, which has a size of 2048 pixels per side, corresponding to 805 kpc per side in physical distance. We then follow two different approaches to remove the PSF effect from the convolved group image: one in which we apply our PSF-scattered flux removal technique, introduced in Sec. \ref{sec:PSF analysis}, and another in which we apply the Wiener deconvolution algorithm \citep{wiener1949}. We repeat both PSF removal approaches using two different PSF models: the original one (\textit{PSF}), which was used to convolve the mock group intrinsic image, and a slightly different model, which we create by modifying \textit{PSF} as follows:}

\begin{equation}
\\
PSF_\mathrm{mod} = PSF^{0.7}.
\\
\end{equation}

\begin{figure*}
\centering  %\qquad
 \includegraphics[width=1\linewidth]{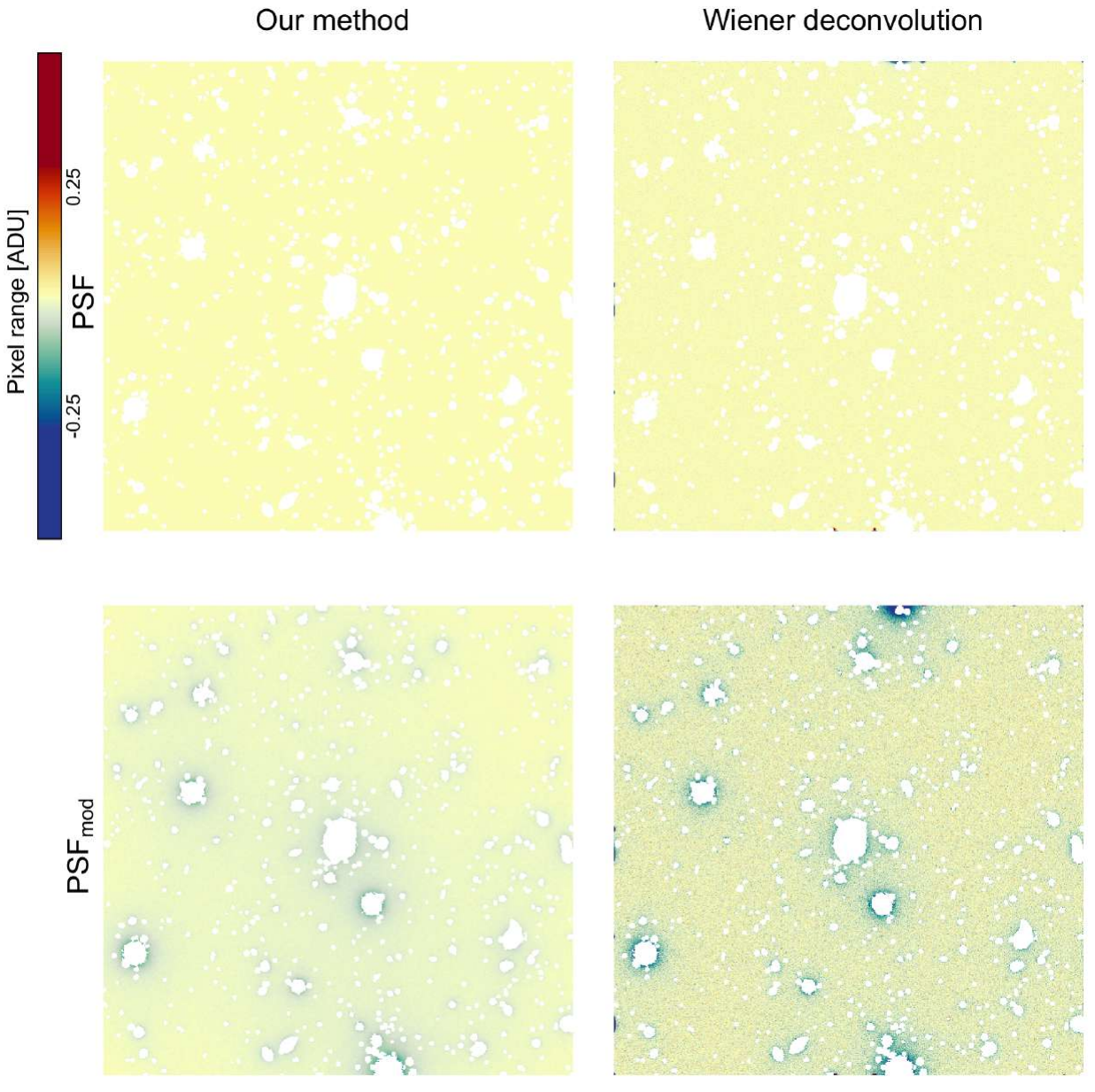}

\caption{\textcolor{black}{PSF-corrected versions of the G200133 mock obtained using two different methods (the PSF-scattered flux removal technique and Wiener deconvolution) and two different PSF models ($PSF$ and $PSF_\mathrm{mod}$). The first column shows the versions corrected using our technique, while the second column shows the versions deconvolved using the Wiener method. In the first row, the PSF model used is $PSF$, while in the second row it is $PSF_\mathrm{mod}$. All four versions have been normalised to the same colour scale. In the first row, the background of the deconvolved image is more heterogeneous than that of the image corrected using our method, and artefacts are visible at the image borders. Both effects are more pronounced when $PSF_\mathrm{mod}$ is used.}}
 \label{deconv 4 group image}
 \end{figure*}

\begin{figure*}
\centering  %\qquad
 \includegraphics[width=1\linewidth]{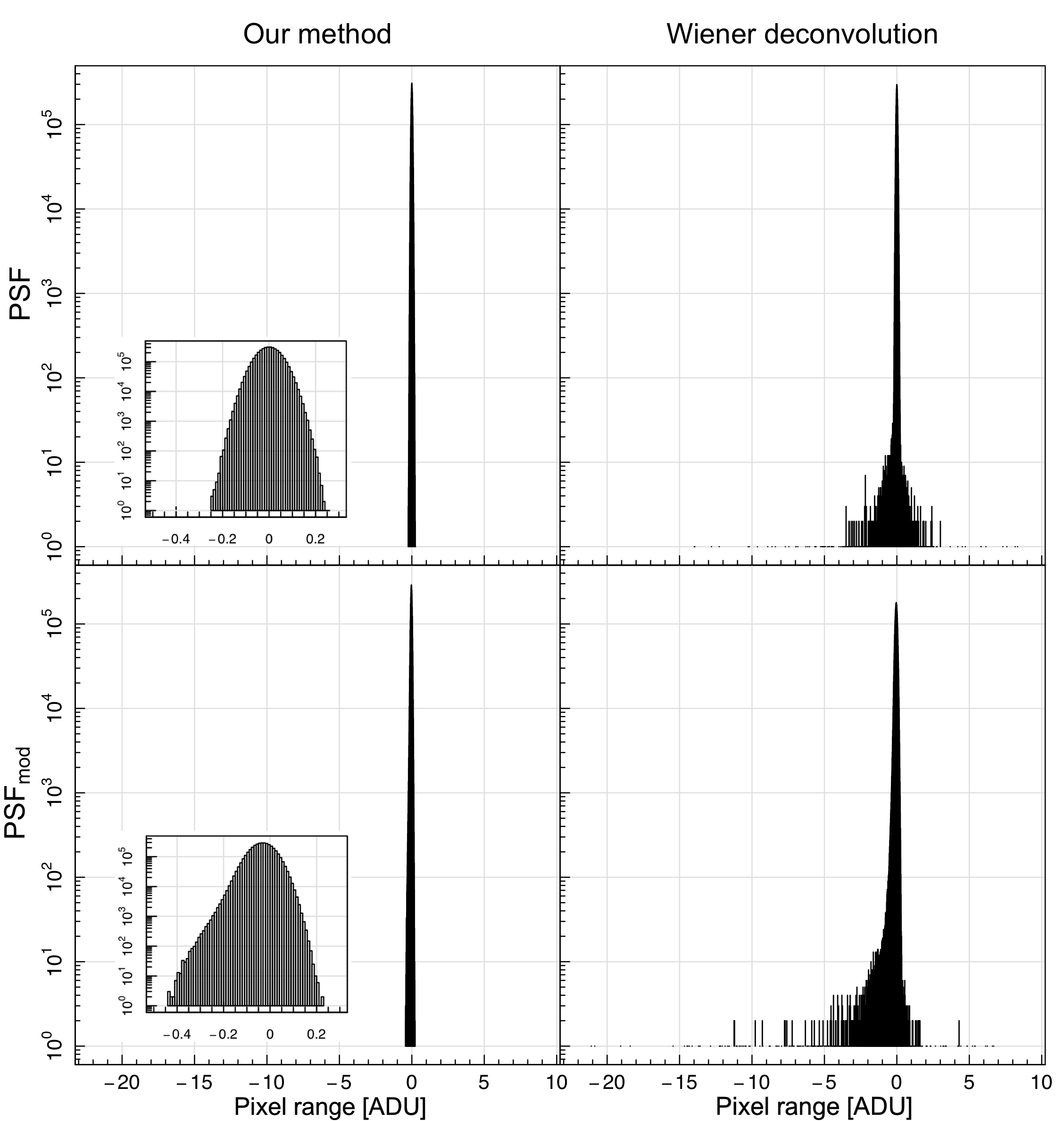}

\caption{\textcolor{black}{Histograms of the residuals between the four PSF-corrected versions of the G200133 mock and its intrinsic version. The structure of the plot is the same as in Fig. \ref{deconv 4 group image}: the first column shows the versions corrected using our technique, while the second column shows the versions deconvolved using the Wiener method, and in the first row, the PSF model used is $PSF$, while in the second row it is $PSF_\mathrm{mod}$. We have added zoomed-in histograms of the results from our method to better visualise the expanded pixel range in both cases. The pixel range is substantially larger in the deconvolution case, ranging from approximately $-15$ to $10$, compared to $-0.3$ to $0.3$ for our method. When using $PSF_\mathrm{mod}$, the pixel range for our method increases slightly, reaching $-0.4$, whereas for the deconvolution method, the minimum reaches values as low as $-20$, highlighting its greater sensitivity to small changes in the PSF.}}

 \label{residuals deconv group}

 \end{figure*}

\textcolor{black}{This test compares the response of the two methods to numerical variations in the assumed PSF model, showing that our method is more robust to these variations than the deconvolution method. In Fig. \ref{deconv 4 group image}, we show four versions of the G200133 image obtained using the two different PSF-correction methods and the two different PSF models. The first column shows the group images corrected using our PSF-scattered flux removal technique, while the second column shows the group images PSF-deconvolved using the Wiener algorithm. For both methods, the images in the first row use \textit{PSF}, whereas those in the second row use $PSF_\mathrm{mod}$. All four images use the same colour scale, ranging from -0.5 to 0.5. Focusing on our method, the background is relatively smooth in the upper-left panel, whereas the background in the upper-right panel is more structured, with some artefacts visible near the edges of the image. These features are introduced by the deconvolution process. When using $PSF_\mathrm{mod}$ to correct the group image with our method (lower-left panel), the oversubtraction issue discussed in Sec. \ref{cavetas psf} around each segmentation region becomes more noticeable. However, the sky background remains relatively smooth, in contrast to the lower-right panel, where the background is even more heterogeneous and the edge artefacts are more pronounced. This illustrates that even small numerical differences in the PSF can propagate into visible artefacts in the deconvolved image. We conclude that our method has a smaller impact on the image background and is less sensitive to changes in the PSF model used for the correction than the Wiener deconvolution method. In Fig. \ref{residuals deconv group} we show histograms of the residuals between the four PSF-corrected versions of the G200133 mock and its intrinsic version (i.e. without PSF convolution). All panels have the same x- and y-axis scales. We have also added zoomed-in histograms of the results from our method to better visualise the expanded pixel range in both cases. The residuals obtained with the deconvolution method span a substantially larger pixel range, from approximately $-15$ to $10$, compared to $-0.3$ to $0.3$ for our technique. This indicates that the deconvolution introduces larger deviations from the intrinsic image, resulting in stronger residual structures and more pronounced pixel-level deviations from the expected intrinsic flux. In contrast, the residuals from our method remain concentrated closer to zero. When using $PSF_\mathrm{mod}$, the pixel range for our method increases slightly, reaching $-0.4$, whereas for the deconvolution method, the minimum reaches values as low as $-20$. This highlights the sensitivity of the deconvolution method to small changes in the PSF, which can introduce more pronounced negative residuals.}

%Moreover, the pixel range does not increase when using $PSF_\mathrm{mod}$ in the deconvolution case, suggesting that these larger residual variations are mainly associated with the deconvolution process itself rather than with the choice of PSF model.

\textcolor{black}{Regarding the computational cost, neither approach is computationally demanding for the image size considered here. Applying the Wiener deconvolution method to a 2048-pixel-per-side image using a PSF of the same size takes approximately 2 seconds, while our technique takes approximately 3 seconds.}

%%%%%%%%%%%%%%%%%%%%%%%%%%%%%%%%%%%%%%%%%%%%%%%%%%

% Don't change these lines
\bsp	% typesetting comment
\label{lastpage}
\end{document}